\documentclass{article}
\usepackage{amssymb}
\usepackage{amsfonts}
\usepackage{amsmath}

\input{tcilatex}

\begin{document}
%\title{{\small UCRHEP-T362}\\
\title{
\textbf{Solutions to the Renormalization Group}\\
\textbf{Equations for Yukawa Matrices as an}\\
\textbf{Answer to the Quark and Lepton Mass }\\
\textbf{Problem}}
\author{Bipin R. Desai and Alexander R. Vaucher  \
\  \and \ \ \ \ \ \ \ \ \ \ \ \ \ \ \ \ \ \ \ \ \ \ \ \ \
\ \ \ \ \ \ \ \ \ \ \ \ \ \ \ \ \ \ \  \\
%EndAName
University of California, Riverside, California 92507}
\maketitle

\begin{abstract}
If the scale dependence of a Yukawa matrix is assumed to be determined
entirely by the dominant 33-element, then the renormalization group
equations can be expressed in terms of two separate equations: a
differential equation for the 33-coupling, and, an algebraic equation for
the scale-independent 3x3 matrix that is found to have only two non-trivial,
hierarchical, solutions with eigenvalues (0,0,1) and (0,1,1). The mass
matrices are constructed from these solutions by rotating them first by the
experimentally known mixing matrices-the CKM for quarks and charged leptons,
and the CKM-analog for the seesaw generated Majorana neutrinos-and then
incorporating the appropriate texture zeros. A uniform, hierarchical,
description for the mass matrices of quarks and leptons is thus achieved, in
terms of the mixing paramters, that give mass eigenvalues consistent with
experiments as well as reproduce the input mixing angles. Inverted hierarchy
in neutrinos is also discussed. Only a single scale $\left( \thickapprox
10^{13}\text{ }GeV\right) $for the seesaw neutrinos is involved rather than
their mass distribution. No new particles are otherwise invoked.
\end{abstract}
\vspace{1cm}
\newpage

\textbf{I. Introduction}

The Yukawa matrices which describe the mass values in the quark sector of
the standard model show a pronounced hierarchical pattern, as do the mixing
angles of the CKM matrix [1, 2],

In the lepton sector the charged lepton masses show a similar pattern of
hierarchy. The neutrinos, on the other hand, are massless in the minimal
version of the standard model, as no right-handed neutrinos are assumed to
exist. The neutrino oscillation data [3,4,5,6,7,8] indicate, however, that
neutrinos have masses which are extremely small and have a pattern which can
be consistent with hierarchy or, alternatively, inverted hierarchy [9]$.$

The mixing angles, the CKM-analog [10], for leptons for a diagonal charged
lepton mass matrix are quite different. While for CKM, all three rotation
angles are small, here two of the angles are found to be quite
large[3,4,5,6,7,8]\textbf{.}

We would like to address all these facts together and find a uniform,
underlying explanation by solving the renormalization group equations in
MSSM (minimal supersymmetric standard model) [11,12] supplemented by two
assumptions which reflect physics beyond the standard model. We assume the
well known seesaw mechanism [13] for generating the small (Majorana)
neutrino masses in which the standard model left-handed neutrinos couple to
large mass right-handed neutrinos. And we take account of the fact that
texture zeros may be present in the Yukawa matrices of quarks, charged
leptons and neutrinos [1,11,14,15,16,17,18,19].

We begin first in, sections II and III, with discussing the renormalization
group equations (RGE) of the Yukawa matrices.The RGEs, to one loop in MSSM,
for up-quark $(U)$ and down-quark $(D)$ Yukawa matrices are given as follows
[11]

\begin{equation}
\dfrac{dU}{dt}=\frac{1}{16\pi ^{2}}\left[ -%
\sum_{i}c_{i}g_{i}^{2}+3UU^{+}+DD^{+}+Tr(3UU^{+})\mathbf{1}\right] U 
\tag{1.1}  \label{1.1}
\end{equation}%
\ \ \ \ \ \ \ \ \ \ \ \ \ \ \ \ \ \ \ \ \ \ \ \ \ \ \ \ \ \ \ \ \ \ \ \ \ \
\ \ \ \ \ \ \ \ \ \ \ \ \ \ \ \ \ \ \ \ \ \ \ \ \ \ \ \ \ \ \ \ \ \ \ \ \ \
\ \ \ \ \ \ \ \ \ \ \ \ \ \ \ \ \ \ $\ \ \ \ \ \ $

\begin{equation}
\dfrac{dD}{dt}=\frac{1}{16\pi ^{2}}\left[ -\sum_{i}c_{i}^{\prime
}g_{i}^{2}+3DD^{+}+UU^{+}+Tr(3DD^{+})\mathbf{1}\right] D  \tag{1.2}
\label{1.2}
\end{equation}%
where $t=ln(\mu $ $/1Gev),$ $\mu $ being the energy variable, $c_{i}$ and $%
c_{i}^{\prime }$ are known constants and $g_{i}^{\prime }s$ are the gauge
couplings.\ \ \ \ \ \ \ \ \ \ \ \ \ \ \ \ \ \ \ \ \ \ \ \ \ \ \ \ \ \ \ \ \
\ \ \ \ \ \ \ \ \ \ \ \ \ \ \ \ \ \ \ \ \ \ \ \ \ \ \ \ \ \ \ \ \ \ \ \ \ \
\ \ \ \ \ \ \ \ \ \ \ \ \ \ \ \ \ \ \ \ \ \ \ \ \ \ \ \ \ \ \ \ \ \ \ \ \ \
\ \ \ \ \ \ \ \ \ \ \ \ \ \ \ \ \ \ \ \ \ \ \ \ \ \ \ \ \ \ \ \ \ \ \ \ \ \
\ \ \ \ \ \ \ \ \ \ \ \ \ \ \ \ \ \ \ \ \ \ \ \ \ \ \ \ \ \ \ \ \ \ \ \ \ \
\ \ \ \ 

In the lepton sector, we consider the charged lepton Yukawa matrix, $E,$
and, for the neutrinos we assume a seesaw mechanism, as mentioned earlier,
and consider the (Dirac) neutrino matrix, $N$, that couples the standard
model left-handed neutrinos to the large-mass right-handed seesaw neutrinos
given by the mass distribution, $M_{R}.$To one loup in MSSM the RGEs are
given by [11,12]

\begin{equation}
\dfrac{dE}{dt}=\frac{1}{16\pi ^{2}}\left[ -%
\sum_{i}d_{i}g_{i}^{2}+3EE^{+}+NN^{+}+Tr(EE^{+})\mathbf{1+}Tr(3DD^{+})%
\mathbf{1}\right] E  \tag{1.3}  \label{1.3}
\end{equation}%
\ \ \ \ \ \ \ \ \ \ \ \ \ \ \ \ \ \ \ \ \ \ \ \ \ \ \ \ \ \ \ \ \ \ \ \ \ \
\ \ \ \ \ \ \ \ \ \ \ \ \ \ \ \ \ \ \ \ \ \ \ \ \ \ 

\begin{equation}
\dfrac{dN}{dt}=\frac{1}{16\pi ^{2}}\left[ -\sum_{i}d_{i}^{\prime
}g_{i}^{2}+3NN^{+}+EE^{+}+Tr(NN^{+})\mathbf{1+}Tr(3UU^{+})\mathbf{1}\right] U
\tag{1.4}  \label{1.4}
\end{equation}%
where $d_{i}$and $d_{i}^{\prime }$ are known constants. The Majorana
neutrino mass matrix, $\kappa ,$\ is then given by the seesaw formula
[12,13]\ \ \ \ \ \ \ \ \ \ \ \ \ \ \ \ \ \ \ \ \ \ \ \ \ \ \ \ \ \ \ \ \ \ \
\ \ \ \ \ \ \ \ \ \ \ \ \ \ \ \ \ \ \ \ \ \ \ \ \ \ \ \ \ \ \ \ \ \ \ \ \ \
\ \ \ \ \ \ \ \ \ \ \ \ \ \ \ \ \ \ \ \ \ \ \ \ \ \ \ \ \ \ \ \ \ \ \ \ \ \
\ \ \ \ \ \ \ \ \ \ \ \ \ \ \ \ \ \ \ \ \ \ \ \ \ \ \ \ \ \ \ \ \ \ \ \ \ \
\ \ \ \ \ \ \ \ \ \ \ \ \ \ \ \ \ \ \ \ \ \ \ \ \ \ \ \ \ \ \ \ \ \ \ \ \ \
\ \ \ \ \ \ \ \ \ \ \ \ \ \ \ \ \ \ \ \ \ \ \ \ \ \ \ \ \ \ \ \ \ \ \ \ \ \
\ \ \ \ \ \ \ \ \ \ \ \ \ \ \ \ \ \ \ \ \ \ \ \ \ \ \ \ \ \ \ \ \ \ \ \ 

\begin{equation}
\kappa =N^{T}M_{R}^{-1}N  \tag{1.5}  \label{1.5}
\end{equation}%
\ \ \ \ \ \ \ \ \ \ \ \ \ \ \ \ \ \ \ \ \ \ \ \ \ \ \ \ \ \ \ \ \ \ \ \ \ \
\ \ \ \ \ \ \ \ \ \ \ \ \ \ \ \ \ \ \ \ \ \ \ \ \ \ \ \ \ \ \ \ \ \ \ \ \ \
\ \ \ \ \ \ \ \ \ \ \ \ \ \ \ \ \ \ \ \ \ \ \ \ \ \ \ \ \ \ \ \ \ \ \ \ \ \
\ \ \ \ \ \ \ \ \ \ \ \ \ \ \ \ \ \ \ \ \ \ \ \ \ \ \ \ \ \ \ 

Renormalization group equations, used as a tool to determine the properties
of the quark Yukawa matrices have been considered previously, specfically
for the case in which the top coupling is assumed dominanat[1,14,20]. \
Recently, the RGEs have also been used to reconcile with the neutrino data
and to investigate the differences between CKM and the CKM-analog mixing
angles [10,12,21]

Let us consider the $U$ matix given by (1.1). Assuming the 33-matrix element
to be dominant and equal to the top-quark Yukawa coupling, $\lambda _{t},$we
obtain the following equation ignoring $D$\ and the $g_{i}^{\prime }s$\ as
they are small compared to $\lambda _{t}$%
\begin{equation}
\dfrac{d\lambda _{t}}{dt}=\dfrac{3\lambda _{t}^{3}}{8\pi ^{2}}  \tag{1.6}
\label{1.6}
\end{equation}%
\ \ \ \ \ \ \ \ \ \ \ \ \ \ \ \ \ \ \ \ \ \ \ \ \ \ \ \ \ \ \ \ \ \ \ \ \ \
\ \ \ \ \ \ \ \ \ \ \ \ \ \ \ \ \ \ \ \ \ \ \ \ \ \ \ \ \ \ \ \ \ \ \ \ \ \
\ \ \ \ \ \ \ \ \ \ \ \ \ \ \ \ \ \ \ \ \ \ \ \ \ \ $\ $\ $\ \ \ \ \ \ \ \ \
\ \ \ \ \ \ \ \ \ \ \ \ \ \ \ \ \ \ \ \ \ \ \ \ \ \ \ \ \ \ \ \ \ $\ \ \ \ \
\ \ \ \ \ \ \ \ \ \ \ \ \ \ \ \ \ \ \ \ \ \ \ \ \ \ \ \ \ \ \ \ \ \ \ \ \ \
\ \ \ \ \ \ \ \ \ \ \ \ \ \ \ \ \ \ \ \ \ \ \ \ \ \ \ \ \ \ \ \ \ \ \ \ \ \
\ \ \ \ \ \ \ \ \ \ \ \ \ \ \ \ \ \ \ \ \ \ \ \ \ \ \ \ \ \ \ \ \ \ \ \ \ \
\ \ \ The solution is given by,

\begin{equation}
\lambda _{t}\left( t\right) =\lambda _{0t}\left[ 1+\frac{3}{4\pi ^{2}}\left(
t_{0}-t\right) \lambda _{0t}^{2}\right] ^{-\frac{1}{2}}  \tag{1.7}
\label{1.7}
\end{equation}%
\ \ \ \ \ \ \ \ \ \ \ \ \ \ \ \ \ \ \ \ \ \ \ \ \ \ \ \ \ \ \ \ \ \ \ \ \ \
\ \ \ \ \ \ \ \ \ \ \ \ \ \ \ \ \ \ \ \ \ \ \ \ \ \ $\ \ \ \ \ \ \ \ \ \ \ \
\ \ \ \ \ \ \ \ \ \ \ \ \ \ \ \ \ \ \ \ \ \ \ \ \ \ \ \ \ \ \ \ \ \ \ \ \ $
\ \ \ \ \ \ \ \ \ \ \ \ \ \ \ \ \ \ \ \ \ \ \ \ \ \ \ \ \ \ \ \ \ \ \ \ \ \
\ \ \ \ \ \ \ \ \ \ \ \ \ \ \ \ \ \ \ \ \ \ \ \ \ \ \ \ \ \ \ \ \ \ \ \ \ \
\ \ \ \ \ \ \ \ \ \ \ \ \ \ \ \ \ \ \ \ \ \ \ \ \ \ \ \ \ \ \ \ \ \ \ \ \ \
where we have taken $\lambda _{t}\left( t_{0}\right) =\lambda _{0t}$ at a
convenient scale parameter $t_{0}$.

The question we now wish to explore is this: just as the magnitude of $%
\lambda _{t}$\ dominates the U-matrix, are there solutions to (1.1)$\ $such
that the scale-dependence, $\lambda _{t}\left( t\right) $\ given by (1.7)\
also describes the scale-dependence of the entire U- matrix ? In other
words, does equation (1.1)$\ $\ allow a factorizable solution of the type

\begin{equation}
U=U_{0}\lambda _{t}\left( t\right)  \tag{1.8}  \label{1.8}
\end{equation}%
\ \ \ \ \ \ \ \ \ \ \ \ \ \ \ \ \ \ \ \ \ \ \ \ \ \ \ \ \ \ \ \ \ \ \ \ \ \
\ \ \ \ \ \ \ \ \ \ \ \ \ \ \ \ \ \ \ \ \ \ \ \ \ \ \ \ \ \ \ \ \ \ \ \ \ \
\ \ \ \ \ \ \ \ \ \ \ \ \ \ \ \ \ \ \ \ \ \ \ \ \ \ \ \ \ \ \ \ \ \ \ \ \ \
\ \ \ \ \ \ \ \ \ \ \ \ \ \ \ \ \ \ \ \ \ \ \ \ \ \ \ \ \ \ \ \ \ \ \ \ \ \
\ \ \ \ \ \ \ \ \ \ \ \ \ \ \ \ \ \ \ \ \ \ \ \ \ \ \ \ \ \ \ \ \ \ \ \ \ \
\ \ \ \ \ \ \ \ \ \ \ \ \ \ \ \ \ \ \ \ \ \ \ \ \ \ \ \ \ \ \ \ \ \ \ \ \ \
\ \ \ \ \ \ \ \ \ \ \ \ \ \ \ \ \ \ \ \ \ \ \ \ \ \ \ \ \ \ \ \ \ \ \ \ \ \
\ \ \ \ \ \ \ \ \ \ \ \ \ \ \ \ \ \ \ \ \ \ \ \ \ \ \ \ \ \ \ \ \ \ \ \ \ \
\ \ \ \ \ \ \ \ \ \ \ \ \ \ \ \ \ \ \ \ \ \ \ \ \ \ \ \ \ \ \ \ \ \ \ \ \ \
\ \ \ \ \ \ \ \ \ \ so that the $t$-dependence, as given by $\lambda
_{t}\left( t\right) ,$\ can be factored out, leaving behind a matrix of
coefficients $U_{0}$\ that is independent of the scale, $t$. \ Similarly, we
wish to explore the possibility that the solutions to (1.2), (1.3) and (1.4)
can also be expressed in a factorizable form with the scale-dependence
residing entirely in the dominant matrix element of each, given by the
b-quark, $\tau -$lepton, and the largest mass (Dirac) neutrino, $m_{3}$\ (or 
$m_{1}$for inverted hierarchy) e.g.

\begin{equation}
D=D_{0}\lambda _{b}\left( t\right)  \tag{1.9a}  \label{1.9a}
\end{equation}

\begin{equation}
E=E_{0}\lambda _{\tau }\left( t\right)  \tag{1.9b}  \label{1.9b}
\end{equation}

\begin{equation}
N=N_{0}\lambda _{3v}\left( t\right)  \tag{1.9c}  \label{1.9c}
\end{equation}%
\ \ \ \ \ \ \ \ \ \ \ \ \ \ \ \ \ \ \ \ \ \ \ \ \ \ \ \ \ \ \ \ \ \ \ \ \ \
\ \ \ \ \ \ \ \ \ \ \ \ \ \ \ \ \ \ \ \ \ \ \ \ \ \ \ \ \ \ \ \ \ \ \ \ \ \
\ \ \ \ \ \ \ \ \ \ \ \ \ \ \ \ \ \ \ \ \ \ \ \ \ \ \ \ \ \ \ \ \ \ \ \ \ \
\ \ \ \ \ \ \ \ \ \ \ \ \ \ \ \ \ \ \ \ \ \ \ \ \ \ \ \ \ \ \ \ \ 

Such solutions, indeed, exist as we will discuss below, in which the RGEs
split into two equations, one a differential equation for the dominant
(scale-dependent) Yukawa coupling and the other an algebraic equation for
the (scale-independent) matrix of the coefficients. It is found that the
scale-independent matrices (e.g. $U_{0},D_{0}$\ etc.) can be classified in
terms of their eigenvalues in simple, diagonal --"primordial"-- forms.

In deriving these results we assume $U,D,N$ and $\kappa $ to be real and
symmetric.

There are, actually only two non-trivial solutions for the scale-independent
matrices : one we designate as "hierarchical", and the other
"semi-hierarchical", of the forms

\begin{equation}
\left[ 
\begin{array}{ccc}
0 & 0 & 0 \\ 
0 & 0 & 0 \\ 
0 & 0 & 1%
\end{array}%
\right] \ \ and\ \ \left[ 
\begin{array}{ccc}
0 & 0 & 0 \\ 
0 & 1 & 0 \\ 
0 & 0 & 1%
\end{array}%
\right]  \tag{1.10}  \label{1.10}
\end{equation}%
respectively. \ In particular, $U_{0}$\ is identified as the former, $D_{0}$%
\ as the mixture of the two and similarly for the leptons. A general, real,
symmetic, solution can be obtained from this "primordial" matrix by rotating
it through arbitrary angles by a unitary matrix. That is, if $U_{diag}$\ and 
$D_{diag}$\ are related to the "primordial" matrices then one can write the
general matrices as

\begin{equation}
U_{0}=VU_{diag}V^{\dagger }  \tag{1.11}  \label{1.11}
\end{equation}

\begin{equation}
D_{0}=VD_{diag}V^{\dagger }  \tag{1.12}  \label{1.12}
\end{equation}%
where\ $V$ is a unitary matrix. A similar situation exits for the leptons.\
\ \ \ \ \ \ \ \ \ \ \ \ \ \ \ \ \ \ \ \ \ \ \ \ \ \ \ \ \ \ \ \ \ \ \ \ \ \
\ \ \ \ \ \ \ \ \ \ \ \ \ \ \ \ \ \ \ \ \ \ \ \ \ \ \ \ \ \ \ \ \ \ \ \ \ \
\ \ \ \ \ \ \ \ \ \ \ \ \ \ \ \ \ \ \ \ \ \ \ \ \ \ \ \ \ \ \ \ \ \ \ \ \ \
\ \ \ \ \ \ \ \ \ \ \ \ \ \ \ \ \ \ \ \ \ \ \ \ \ \ \ \ \ \ \ \ \ \ \ \ \ \
\ \ \ \ \ \ \ \ \ \ \ \ \ \ \ \ \ \ \ \ \ \ \ \ \ \ \ \ \ \ \ \ \ \ \ \ \ \
\ \ \ \ \ \ \ \ \ \ \ \ \ \ \ \ \ \ \ \ \ \ \ \ \ \ \ \ \ \ \ \ \ \ \ \ \ \
\ \ \ \ \ \ \ \ \ \ \ \ \ \ \ \ \ \ \ \ \ \ \ \ \ \ \ \ \ \ \ \ \ \ \ \ \ \
\ \ \ \ \ \ \ \ \ \ \ \ \ \ \ \ \ \ \ \ \ \ \ \ \ \ 

Thus the Yukawa matrix elements will depend directly on the rotation
parameters of $V.$

In section IV, confining first to the quark sector, what we find most
remarkable is that, if we assume the unitary matrix $V$ \ to be the same as
the CKM matrix, we obtain Yukawa matrices $U_{0}$ and $D_{0}$ that are
exactly what one generally expects when expressed in powers of the Cabibbo
parameter, $\lambda $ [1]. Namely,

\begin{equation}
U_{0}\thickapprox 
\begin{bmatrix}
\lambda ^{8} & \lambda ^{6} & \lambda ^{4} \\ 
\lambda ^{6} & \lambda ^{4} & \lambda ^{2} \\ 
\lambda ^{4} & \lambda ^{2} & 1%
\end{bmatrix}
\tag{1.13}  \label{1.13}
\end{equation}%
Similarly for $D_{0}$ where the hierarchy is found to be less pronounced,
just as expected [1].\ \ \ \ \ \ \ \ \ \ \ \ \ \ \ \ \ \ \ \ \ \ \ \ \ \ \ \
\ \ \ \ \ \ \ \ \ \ \ \ \ \ \ \ \ \ \ \ \ \ \ \ \ \ \ \ \ \ \ \ \ \ \ \ \ \
\ \ \ \ \ \ \ \ \ \ \ \ \ \ \ \ \ \ \ \ \ \ \ \ \ \ \ \ \ \ \ \ \ \ \ \ \ \
\ \ \ \ \ \ \ \ \ \ \ \ \ \ \ \ \ \ \ \ \ \ \ \ \ \ \ \ \ \ \ \ \ \ \ \ \ \
\ \ \ \ \ \ \ \ \ \ \ \ \ \ \ \ \ \ \ \ \ \ \ \ \ \ \ \ \ \ \ \ \ \ \ \ \ \
\ \ \ \ \ \ \ \ \ \ \ \ \ \ \ \ \ \ \ \ \ \ \ \ \ \ \ \ \ \ \ \ \ \ \ \ \ \
\ \ \ \ \ \ \ \ \ \ \ \ \ \ \ \ \ \ \ \ \ \ \ \ \ \ \ \ \ \ \ \ \ \ \ \ \ \
\ \ \ \ \ \ \ \ \ \ \ \ \ \ \ \ \ \ \ \ \ \ \ \ \ \ \ \ \ \ \ \ \ \ \ \ \ \
\ \ 

In the lepton sector, discussed in V, the solutions for the charged leptons
exhibit a similar hierarchy through the small angles of the CKM matrix. For
the neutrinos, we argue that for the mass matrix, $\kappa $, defined in
(1.5) the candidate for the appropriate unitary matrix $V$ is the
large-angle lepton mixing matrix, the CKM-analog [10].

We thus show that for each of the four systems of particles, the mass matrix
can be described in terms of the appropriate mixing parameters through an
expression that is very simple and transparent.

Although our model reproduces the expected hierarchical pattern of the mass
matrix elements, it is, at this stage, incomplete, because, for example, the
eigenvalues of $U_{0}$ and $D_{0}$ in equation (1.11) and (1.12) are still
given by their "primordial" representatios i.e. $(0,0,1)$ and $(0,1,1)$.
Furthermore, the mixing matrix turns out to be a unit matrix, since the same
matrix, $V$, that diagonalizes $U_{0},$ also diagonalizes $D_{0}.$

All that is remedied, however, once we incorporate texture zeros, as we do
in section V, whereby, in $U_{0}$ and $D_{0}$, constructed through equations
(1.11)\textbf{\ }and (1.12), some of the entries are replaced by zeros.

Confining ourselves, briefly, to the quark sector we note that the
(symmetric) Yukawa matrices $U_{0}$ and $D_{0}$ that are consistent with
experiments and allow the maximum number of texture zeros, are found to be
of the following type,

\begin{equation}
U_{0}=\left[ 
\begin{array}{ccc}
0 & 0 & X \\ 
0 & X & 0 \\ 
X & 0 & X%
\end{array}%
\right] ,\ \ \ \ D_{0}=\left[ 
\begin{array}{ccc}
0 & X & 0 \\ 
X & X & X \\ 
0 & X & X%
\end{array}%
\right]  \tag{1.14}  \label{1.14}
\end{equation}%
\ \ \ \ \ \ \ \ \ \ \ \ \ \ \ \ \ \ \ \ \ \ \ \ \ \ \ \ \ \ \ \ \ \ \ \ \ \
\ \ \ \ \ \ \ \ \ \ \ \ \ \ \ \ \ \ \ \ \ \ \ \ \ \ \ \ \ \ \ \ \ \ \ \ \ \
\ \ \ \ \ \ \ \ \ \ \ \ \ \ \ \ \ \ \ \ \ \ \ \ \ \ \ \ \ \ \ \ \ \ \ \ \ \
\ \ \ \ \ \ \ \ \ \ \ \ \ \ \ \ \ \ \ \ \ \ \ \ \ \ \ \ \ \ \ \ \ \ \ \ \ \
\ \ \ \ \ \ \ \ \ \ \ \ \ \ \ \ \ \ \ \ \ \ \ \ \ \ \ \ \ \ \ \ \ \ \ \ \ \
\ \ \ \ \ \ \ \ \ \ \ \ \ \ \ \ \ \ \ \ \ \ \ \ \ \ \ \ \ \ \ \ \ \ \ \ \ \
\ \ \ \ \ \ \ \ \ \ \ \ \ \ \ \ \ \ \ \ \ \ \ \ \ \ \ \ \ \ \ \ \ \ \ \ \ \
\ \ \ \ \ \ with a total of five texture zeros, where $X$\ denotes non-zero
entries[16.17]. We assume this to be the pattern of the zeros in our case.

Two things happen as soon as texture zeros are incorporated. First, the
eigenvalues values of (the newly textured) $U_{0}$ and $D_{0}$ will no
longer be given simply by the primordial values, but will be proportional to
the matrix elements of $V$ \ that were involved in (1.11) and (1.12).
Second, the mixing matrix will not necessarily be a unit matrix since there
will now be a mismatch between matrices that diagonalize $U_{0}$ and $D_{0},$
as is evident from the structures in (1.14)

We need then to define two types of mixing matrices :$V_{CKM}^{(in)}$ and $%
V_{CKM}^{(out)}.$

The "input" mixing matrix, $V_{CKM}^{(in)}$, represents $V$ which generates
Yukawa matrices $U_{0}$ and $D_{0}$ through equations (1.11) and (1.12). The
entries in these matrices are the experimentally known CKM parameters.

After texture zeros are incorporated, according to the structures (1.14), an
"output" CKM, $V_{CKM}^{(out)}$ is defined which is a mixing matrix between
(the newly textured) $U_{0}$ and $D_{0}$. That is, if $V_{u}$ and $V_{d}$
diagonalize the two (newly textured) matrices $U_{0}$ and $D_{0},$
respectively, then,

\begin{equation}
V_{CKM}^{(out)}=V_{u}^{\dagger }V_{d}  \tag{1.15}  \label{1.15}
\end{equation}

If $V_{CKM}^{(out)}$ turns out to be the same as the experimentally known
CKM matrix (i.e. if $V_{CKM}^{(out)}=$ $V_{CKM}^{(in)}$) then it is a
success of the model, justifying the choice of $V_{CKM}^{(in)}$ as the
unitary matrix $V$ in (1.11) and (1.12). That is, indeed, found to be the
case when we carry out our calculations.

In other words, what was put in to generate the mass matrices, came out, in
a self-consistent manner, when we calculated the mixing angles.

At the same time, we also find that the eigenvalues, or rather, the ratios
of the eigenvalues to the 33-elements are consistent with the experimentally
determined mass ratios.

For the leptons, discussed in VI and VII, we take the charged lepton
structure for $E_{0}$\ to be the same as $D_{0}$. \ And for the neutrinos we
refer to an extensive analysis of the texture in the neutrino mass matrix
which points to two possible types of structures that are consistent with
experiments[18]. The so called $A$-type structure given by

\begin{equation}
\kappa _{0}=\left[ 
\begin{array}{ccc}
0 & 0 & X \\ 
0 & X & X \\ 
X & X & X%
\end{array}%
\right] \ \ or\ \kappa _{0}=\left[ 
\begin{array}{ccc}
0 & X & 0 \\ 
X & X & X \\ 
0 & X & X%
\end{array}%
\right]  \tag{1.16}  \label{1.16}
\end{equation}%
\ \ \ \ \ \ \ \ \ \ \ \ \ \ \ \ \ \ \ \ \ \ \ \ \ \ \ \ \ \ \ \ \ \ \ \ \ \
\ \ \ \ \ \ \ \ \ \ \ \ \ \ \ \ \ \ \ \ \ \ \ \ \ \ \ \ \ \ \ \ \ \ \ \ \ \
\ \ \ \ \ \ \ \ \ \ \ \ \ \ \ \ \ \ \ \ \ \ \ \ \ \ \ \ \ \ \ \ \ \ \ \ \ \
\ \ \ \ \ \ \ \ \ \ \ \ \ \ \ \ \ \ \ \ \ \ \ \ \ \ \ \ \ \ \ \ \ \ \ \ \ \
\ \ \ \ \ \ \ \ \ \ \ \ \ \ \ \ \ \ \ \ \ \ \ \ \ \ \ \ \ \ \ \ \ \ \ \ \ \
\ \ \ \ \ \ \ \ \ \ \ \ \ \ \ \ \ \ \ \ \ \ \ \ \ \ \ \ \ \ \ \ \ \ \ \ \ \
\ \ \ \ \ \ \ \ \ \ \ \ \ \ \ \ \ \ \ \ \ \ \ \ \ \ \ \ \ \ \ \ \ \ \ \ \ \
\ \ \ \ \ \ \ \ \ \ \ which gives hierarchical neutrino mass values, and the
C-type\ 
\begin{equation}
\kappa _{0}=\left[ 
\begin{array}{ccc}
X & X & X \\ 
X & 0 & X \\ 
X & X & 0%
\end{array}%
\right]  \tag{1.17}  \label{1.17}
\end{equation}%
\ \ \ \ \ \ \ \ \ \ \ \ \ \ \ \ \ \ \ \ \ \ \ \ \ \ \ \ \ \ \ \ \ \ \ \ \ \
\ \ \ \ \ \ \ \ \ \ \ \ \ \ \ \ \ \ \ \ \ \ \ \ \ \ \ \ \ \ \ \ \ \ \ \ \ \
\ \ \ \ \ \ \ \ \ \ \ \ \ \ \ \ \ \ \ \ \ \ \ \ \ \ \ \ \ \ \ \ \ \ \ \ \ \
\ \ \ \ \ \ \ \ \ \ \ \ \ \ \ \ \ \ \ \ \ \ \ \ \ \ \ \ \ \ \ \ \ \ \ \ \ \
\ \ \ \ \ \ \ \ \ \ \ \ \ \ \ \ \ \ \ \ \ \ \ \ \ \ \ \ \ \ \ \ \ \ \ \ \ \
\ \ \ \ \ \ \ \ \ \ \ \ \ \ \ \ \ \ \ \ \ \ \ \ \ \ \ \ \ \ \ \ \ \ \ \ \ \
\ \ \ \ \ \ \ \ \ \ \ \ \ \ \ \ \ \ \ \ \ \ \ \ \ \ \ \ \ \ \ \ \ \ \ \ \ \
\ \ \ \ \ \ \ \ \ \ \ \ \ \ \ \ \ \ \ \ \ \ \ \ \ \ \ \ \ \ \ \ \ which
gives an inverted hierarchy [18,19]$.$

As in the case of quarks, as we discuss in section V, we have $%
V_{CKM}^{(in)} $ that generates charged leptons Yukawa matrices, $E_{0},$%
while for $\kappa _{0}$, that role, we argue, will be played by $%
V_{large}^{(in)}$ which is the large angle mixing matrix obtained from the
neutrino oscillation data [3,4,5,6,7,8,10] in the basis where the charged
leptons are diagonal. And, $V_{large}^{(out)}$ will be an appropriate
product of the matrix that diagonalizes $\kappa _{0}$, and the one which
diagonalizes $E_{0}$ since the $E_{0}$ matrix we obtain from our
prescription will not necessarily be diagonal.

In other words, the process is the same as for $U_{0}$ and $D_{0}$. \ We
first create $E_{0}$ and $\kappa _{0}$ by the appropriately designated $V$.
Then we replace some of the entries by zeros as given above and examine the
newly textured matrices.

Once again we find that, with the A-type structure, $V_{large}^{(out)}$ is
the same as $V_{large}^{(in)}$, except for a small discrepancy in the the
angle $s_{2}$\ for the 1-3 sub matrix. The mass eigenvalues are found to be
consistent with experiments.

We discuss the C-type structure for $\kappa _{0}$ separately, in section
VIII, since it produces an inverted hierarchy which can not be accomodated
within the framwork of the "primordial" matrices we have pursued which are
basically hierarchical in nature. Instead we point out that, for this case,
the appropriate basis is provided by

\begin{equation}
\left[ 
\begin{array}{ccc}
1 & 0 & 0 \\ 
0 & 0 & 0 \\ 
0 & 0 & 0%
\end{array}%
\right]  \tag{1.18}  \label{1.18}
\end{equation}

We find that, as a consequence, $V_{large}^{(out)}$ is essentially identical
to $V_{large}^{(in)}$ with a maximal atmospheric angle mixing (2-3
submatrix), and $s_{2}=0$. \ The neutrino mass eigenvalues are also
consistent with experiments.

\bigskip

\textbf{II.The RG Equations}

First let us define the following more convenient variable

\begin{equation}
x=\frac{3}{4\pi ^{2}}\left( t_{0}-t\right)  \tag{2.1}  \label{2.1}
\end{equation}%
\ \ \ \ \ \ \ \ \ \ \ \ \ \ \ \ \ \ \ \ \ \ \ \ \ \ \ \ \ \ \ \ \ \ \ \ \ \
\ \ \ \ \ \ \ \ \ \ \ \ \ \ \ \ \ \ \ \ \ \ \ \ \ \ \ \ \ \ \ \ \ \ \ \ \ \
\ \ \ \ \ \ \ \ \ \ \ \ \ \ \ \ \ \ \ \ \ $\ \ \ \ $\ $\ \ \ \ \ \ \ \ \ \ \
\ \ \ \ \ \ \ \ \ \ \ \ \ \ \ \ \ \ \ \ \ \ \ \ \ \ $\ \ \ \ \ \ \ \ \ \ \ \
\ \ \ \ \ \ \ \ \ \ \ \ \ \ \ \ \ \ \ \ \ \ \ \ \ \ \ \ \ \ \ \ \ \ \ \ \ \
\ \ \ \ \ \ \ \ \ \ \ \ \ \ \ \ \ \ \ \ \ \ \ \ \ \ \ \ \ \ \ \ \ \ \ \ \ \
\ \ \ \ \ \ \ \ \ \ \ \ \ \ \ \ \ \ \ \ \ \ \ \ then, (1.6) and (1.7) can be
expressed as

\begin{equation}
-\dfrac{d\lambda _{t}}{dx}=\frac{1}{2}\lambda _{t}^{3}  \tag{2.2}
\label{2.2}
\end{equation}%
\ \ \ \ \ \ \ \ \ \ \ \ \ \ \ \ \ \ \ \ \ \ \ \ \ \ \ \ \ \ \ \ \ \ \ \ \ \
\ \ \ \ \ \ \ \ \ \ \ \ \ \ \ \ \ \ \ \ \ \ \ \ \ \ \ \ \ \ \ \ \ \ \ \ \ \
\ \ \ \ \ \ \ \ \ \ \ \ \ \ \ \ \ \ \ \ \ \ \ \ \ \ \ $\ \ \ \ $\ $\ \ \ \ \
\ \ \ \ \ \ \ \ \ \ \ \ \ \ \ \ \ \ \ \ \ \ \ \ \ \ $

\begin{equation}
\lambda _{t}\left( x\right) =\lambda _{0t}\left[ 1+x\lambda _{0t}^{2}\right]
^{-\frac{1}{2}}  \tag{2.3}  \label{2.3}
\end{equation}%
\ \ \ \ \ \ \ \ \ \ \ \ \ \ \ \ \ \ \ \ \ \ \ \ \ \ \ \ \ \ \ \ \ \ \ \ \ \
\ \ \ \ \ \ \ \ \ \ \ \ \ \ \ \ \ \ \ \ \ \ \ \ \ \ \ \ \ \ \ \ \ \ \ \ \ \
\ \ \ \ \ \ $\ \ \ \ \ \ \ \ \ \ \ \ \ \ \ \ $\ $\ \ \ \ \ \ \ \ \ \ \ \ \ \
\ \ \ \ \ \ \ \ \ $

Secondly, ignoring the gauge terms, which are small, we re-arrange the
remaining terms in (1.1) and (1.2), as follows

\begin{equation}
-\dfrac{dU}{dx}=\frac{1}{4}\left[ UU^{+}+Tr(UU^{+})\mathbf{1}\right] U+\frac{%
1}{12}DD^{+}U  \tag{2.4}  \label{2.4}
\end{equation}%
\ \ \ \ \ \ \ \ \ \ \ \ \ \ \ \ \ \ \ \ \ \ \ \ \ \ \ \ \ \ \ \ \ \ \ \ \ \
\ \ \ \ \ \ \ $\ \ \ \ \ \ \ \ \ \ \ \ \ $\ $\ \ \ \ \ \ \ \ \ \ \ \ \ \ \ \
\ \ \ $

\begin{equation}
-\dfrac{dD}{dx}=\frac{1}{4}\left[ DD^{+}+Tr(DD^{+})\mathbf{1}\right] D+\frac{%
1}{12}UU^{+}D  \tag{2.5}  \label{2.5}
\end{equation}%
\ \ \ \ \ \ \ \ \ \ \ \ \ \ \ \ \ \ \ \ \ \ \ \ \ \ \ \ \ \ \ \ \ \ \ \ \ \
\ \ \ \ \ \ \ \ $\ \ \ \ \ \ \ \ \ \ \ \ \ \ $\ $\ \ \ \ \ \ \ \ \ \ \ \ \ \
\ \ \ \ \ \ \ \ \ \ \ \ \ \ \ \ $\ \ \ \ \ \ \ \ \ \ \ \ \ \ \ \ \ \ \ \ \ \
\ \ \ \ \ \ \ \ \ \ \ \ \ \ \ \ \ \ \ \ \ \ \ \ \ \ \ \ \ \ \ \ \ \ \ \ \ \
\ \ \ \ \ \ \ \ \ \ \ \ \ \ \ \ \ \ \ \ \ \ \ \ \ \ \ \ \ \ \ \ \ \ \ \ \ \
\ \ \ \ \ \ \ \ \ \ \ \ \ \ \ \ \ \ \ \ \ \ \ \ \ \ \ \ \ \ \ \ \ \ \ \ \ \
\ \ \ \ \ \ \ where the first term, in the square brackets in each, involves
"uncoupled" terms, and the second involves coupling between $U$ and $D$.

In a similar fashion we write, $E$ and $N$ in the following form

\begin{equation}
-\dfrac{dE}{dx}=\frac{1}{12}\left[ 3EE^{+}+Tr(EE^{+})\mathbf{1}\right] E\ +%
\frac{1}{12}NN^{+}E\ \ +\frac{1}{4}Tr(DD^{+})\mathbf{1}E\ \   \tag{2.6}
\label{2.6}
\end{equation}%
\ \ \ \ \ \ \ \ \ \ \ \ \ \ \ \ \ \ \ \ \ \ \ \ \ \ \ \ \ \ \ \ \ \ \ \ \ \
\ \ \ \ \ \ \ \ \ \ \ \ \ \ \ \ \ \ \ \ \ \ \ \ \ \ \ \ \ \ \ \ \ \ \ 

\begin{equation}
-\dfrac{dN}{dx}=\frac{1}{12}\left[ 3NN^{+}+Tr(NN^{+})\mathbf{1}\right] N\ +%
\frac{1}{12}EE^{+}N\ \ +\frac{1}{4}Tr(UU^{+})\mathbf{1}N\   \tag{2.7}
\label{2.7}
\end{equation}%
where, besides the uncoupled and coupling terms, we have a term, in each,
which contains a trace in the quark sector that basically normalizes
(re-scales) the solutions.\ \ \ \ \ \ \ \ \ \ \ \ \ \ \ \ \ \ \ \ \ \ \ \ \
\ \ \ \ \ \ \ \ \ \ \ \ \ \ \ \ \ \ \ \ \ \ \ \ \ \ \ \ \ \ \ \ \ \ \ \ \ \
\ \ \ \ \ \ \ \ \ \ \ \ \ \ \ \ \ \ \ \ \ \ \ \ \ \ \ \ \ \ \ \ \ \ \ \ \ \
\ \ \ \ \ \ \ \ \ \ \ \ \ \ \ \ \ \ \ \ \ \ \ \ \ \ \ \ \ \ \ \ \ \ \ \ \ \
\ \ \ \ \ \ \ \ \ \ \ \ \ \ \ \ \ \ \ \ \ \ \ \ \ \ \ \ \ \ \ \ \ \ \ \ \ \
\ \ \ \ \ \ \ \ \ \ \ \ \ \ \ \ \ \ \ \ \ \ \ \ \ \ \ \ \ \ \ \ \ \ \ \ \ \
\ \ \ \ \ \ \ \ \ \ \ \ \ \ \ \ \ \ \ \ \ \ \ \ \ \ \ \ \ \ \ \ \ \ \ \ \ \
\ \ \ \ \ \ 

\bigskip

\textbf{III. Solutions to the RGEs and the eigenvalues in the quark sector}

The equations involving only the uncoupled terms in both (2.4) and (2.5)
look like

\begin{equation}
-\dfrac{dM}{dx}=\frac{1}{4}\left[ MM^{+}+Tr(MM^{+})\mathbf{1}\right] M 
\tag{3.1}  \label{3.1}
\end{equation}%
\ \ \ \ \ \ \ \ \ \ \ \ \ \ \ \ \ \ \ \ \ \ \ \ \ \ \ \ \ \ \ \ \ \ \ \ \ \
\ \ \ \ \ \ \ \ \ \ \ \ \ \ \ \ \ \ \ \ \ \ \ \ \ $\ \ \ \ \ \ \ \ \ \ \ \ \ 
$\ $\ \ \ \ \ \ \ \ \ \ \ \ \ \ \ \ \ \ \ \ \ \ \ \ \ \ \ \ \ \ \ \ \ \ \ $\
\ \ \ \ \ \ \ \ \ \ \ \ \ \ \ \ \ \ \ \ \ \ \ \ \ \ \ \ \ \ \ \ \ \ \ \ \ \
\ \ \ \ \ \ \ \ \ \ \ \ \ \ \ \ \ \ \ \ \ \ \ \ \ \ \ \ \ \ \ \ \ \ \ \ \ \
\ \ \ \ \ \ \ \ \ \ \ \ \ \ \ \ \ \ \ \ \ \ \ \ \ \ \ \ \ \ \ \ \ \ \ \ \ \
\ \ \ \ \ \ \ \ \ \ \ \ \ \ \ \ \ \ \ \ \ \ \ \ \ \ \ \ \ \ \ \ \ \ \ \ \ \
\ \ \ \ \ \ \ \ \ \ \ \ \ \ \ \ \ \ \ \ \ \ \ \ \ \ \ \ \ \ \ \ For a
solution of the type (1.8) and (1.9a)\ 
\begin{equation}
M=M_{0}\lambda _{m}\left( x\right)  \tag{3.2}  \label{3.2}
\end{equation}%
\ \ \ \ \ \ \ \ \ \ \ \ \ \ \ \ \ \ \ \ \ \ \ \ \ \ \ \ \ \ \ \ \ \ \ \ \ \
\ \ \ \ \ \ \ \ \ \ \ \ \ \ \ \ \ \ \ \ \ \ \ \ \ \ \ \ \ \ \ \ \ \ \ \ \ \
\ \ \ \ \ \ \ \ \ \ \ \ \ \ \ \ \ \ \ \ \ \ \ \ \ \ \ \ \ \ \ \ \ \ \ \ \ \
\ \ \ \ \ \ \ \ \ \ \ \ \ \ \ \ \ \ \ \ \ \ \ \ \ \ \ \ \ \ \ \ \ \ \ \ \ \
\ \ \ \ \ \ \ \ \ \ \ \ \ \ \ \ \ \ \ \ \ \ \ \ \ \ \ \ \ \ \ \ \ \ \ \ \ \
\ \ \ \ \ \ \ \ \ \ \ \ \ \ \ \ \ \ \ \ \ \ \ \ \ \ \ \ \ \ \ \ \ \ \ \ \ \
\ \ \ \ \ \ \ \ \ \ \ \ \ \ \ \ \ \ \ \ \ \ \ \ \ \ \ \ \ \ \ \ \ \ \ \ \ \
\ \ \ \ \ \ \ \ \ \ \ \ \ \ \ \ \ \ \ \ \ \ \ \ \ \ \ \ \ \ \ \ \ \ \ \ \ \
\ \ \ \ \ \ \ \ \ \ \ \ \ \ \ with $\lambda _{m}\left( x\right) $
representing the dominant 33-matrix element, we have two equations

\begin{equation}
-\dfrac{d\lambda _{m}\left( x\right) }{dx}=\frac{1}{2}\lambda _{m}^{3}\left(
x\right)  \tag{3.3}  \label{3.3}
\end{equation}%
and,\ \ \ \ \ \ \ \ \ \ \ \ \ \ \ \ \ \ \ \ \ \ \ \ \ \ \ \ \ \ \ \ \ \ \ \
\ \ \ \ \ \ \ \ \ \ \ \ \ \ \ \ \ \ \ \ \ \ \ \ \ \ \ \ \ \ \ \ \ \ \ \ \ \
\ \ \ \ \ \ \ \ \ \ \ \ \ \ \ \ \ \ \ \ \ \ \ \ \ \ \ \ \ \ \ \ \ \ \ \ \ \
\ \ \ \ \ \ \ \ \ \ \ \ \ \ \ \ \ \ \ \ \ \ \ \ \ \ \ \ \ \ \ \ \ \ \ \ \ \
\ \ \ \ \ \ \ \ \ \ \ \ \ \ \ \ \ \ \ \ \ \ \ \ \ \ \ \ \ \ \ \ \ \ \ \ \ \
\ \ \ \ \ \ \ \ \ \ \ \ \ \ \ \ \ 

\begin{equation}
M_{0}=\frac{1}{2}\left[ M_{0}M_{0}^{\dagger }+trM_{0}M_{0}^{\dagger }\mathbf{%
1}\right] M_{0}  \tag{3.4}  \label{3.4}
\end{equation}%
\ \ \ \ \ \ \ \ \ \ \ \ \ \ \ \ \ \ \ \ \ \ \ \ \ \ \ \ \ \ \ \ \ \ \ \ \ \
\ \ \ \ \ \ \ \ \ \ \ \ \ \ \ \ \ \ \ \ \ \ \ \ \ \ \ \ \ \ \ \ \ \ \ \ \ \ $%
\ \ \ \ \ \ \ \ \ \ \ \ \ \ \ \ \ \ \ \ \ \ \ \ \ \ \ \ \ \ \ \ \ \ \ \ \ \
\ \ \ $

Let us now consider equation (3.4)\textbf{\ }and assume $M_{0}$ to be real.
We diagonalize the equation through unitary matrices $V_{1}$ and $V_{2}$\ to
obtain

\begin{equation}
M_{diag}=V_{1}^{\dagger }M_{0}V_{2}  \tag{3.5}  \label{3.5}
\end{equation}%
\ \ \ \ \ \ \ \ \ \ \ \ \ \ \ \ \ \ \ \ \ \ \ \ \ \ \ \ \ \ \ \ \ \ \ \ \ \
\ \ \ \ \ \ \ \ \ \ \ \ \ \ \ \ \ \ \ \ \ \ \ \ \ \ \ \ \ \ \ \ \ \ \ \ \ \
\ \ \ \ \ \ \ \ \ \ \ \ \ \ \ \ \ \ \ \ \ \ \ \ \ \ \ \ \ \ \ \ \ \ \ \ \ \
\ \ \ \ \ \ \ \ \ \ \ \ \ \ \ \ \ \ \ \ \ \ \ \ \ \ \ \ \ \ \ \ \ \ \ \ \ \
\ \ \ \ \ \ \ \ \ \ \ \ \ \ \ \ \ \ \ \ \ \ \ \ \ \ \ \ \ \ \ \ \ \ \ \ \ \
\ \ \ \ \ \ \ \ \ \ \ \ \ \ \ \ \ \ \ \ \ \ \ \ \ \ \ \ \ \ \ \ \ \ \ \ \ \
\ \ \ \ \ \ \ \ \ \ \ \ \ \ \ \ \ \ \ \ \ \ \ \ \ \ \ \ \ \ \ \ \ \ \ \ \ \
\ \ \ \ \ \ \ \ \ \ \ \ \ \ \ \ \ \ \ \ \ \ \ \ \ \ \ \ \ \ \ \ \ \ \ \ \ \
\ \ \ \ \ \ \ \ \ \ \ \ \ \ \ \ \ \ \ \ \ and

\begin{equation}
M_{diag}=\frac{1}{2}\left[ M_{diag}^{2}+trM_{diag}^{2}\mathbf{1}\right]
M_{diag}  \tag{3.6}  \label{3.6}
\end{equation}%
\ \ \ \ \ \ \ \ \ \ \ \ \ \ \ \ \ \ \ \ \ \ \ \ \ \ \ \ \ \ \ \ \ \ \ \ \ \
\ \ \ \ \ \ \ \ \ \ \ \ \ \ \ \ \ \ \ \ \ \ \ \ \ \ \ \ \ \ \ \ \ \ \ \ \ \
\ \ \ \ \ \ \ \ \ \ \ \ \ \ \ \ \ \ \ \ \ \ \ \ \ \ \ \ \ \ \ \ \ \ \ \ \ \
\ \ \ \ \ \ \ \ \ \ \ \ \ \ \ \ \ \ \ \ \ \ \ \ \ \ \ \ \ \ \ \ \ \ \ \ \ \
\ \ \ \ \ \ \ \ \ \ \ \ \ \ \ \ \ \ \ \ \ \ \ \ \ \ \ \ \ \ \ \ \ \ \ \ \ \
\ \ \ \ \ \ \ \ \ \ \ \ \ \ \ \ \ \ \ \ \ \ \ \ \ \ \ \ \ \ \ \ \ \ \ \ \ \
\ \ \ \ \ \ \ \ \ \ \ \ \ \ \ \ \ \ \ \ \ \ \ \ \ \ \ \ \ \ \ \ \ \ \ \ \ \
\ \ \ \ \ \ \ \ \ \ \ \ \ \ \ \ \ \ \ \ \ \ \ \ \ If we express the above
relation in terms of the eigenvalues

\begin{equation}
M_{diag}=%
\begin{bmatrix}
\lambda _{1} & 0 & 0 \\ 
0 & \lambda _{2} & 0 \\ 
0 & 0 & \lambda _{3}%
\end{bmatrix}
\tag{3.7}  \label{3.7}
\end{equation}%
\ \ \ \ \ \ \ \ \ \ \ \ \ \ \ \ \ \ \ \ \ \ \ \ \ \ \ \ \ \ \ \ \ \ \ \ \ \
\ \ \ \ \ \ \ \ \ \ \ \ \ \ \ \ \ \ \ \ \ \ \ \ \ \ \ \ \ \ \ \ \ \ \ \ \ \ $%
\ \ \ \ \ \ \ \ \ \ \ \ \ \ \ \ \ \ \ \ \ \ \ \ \ \ \ \ \ \ \ \ \ \ \ \ \ \
\ \ $\ \ \ \ \ \ \ \ \ \ \ \ \ \ \ \ \ \ \ \ \ \ \ \ \ \ \ \ \ \ \ \ \ \ \ \
\ \ \ \ \ \ \ \ \ \ \ \ \ \ \ \ \ \ \ \ \ \ \ \ \ \ \ \ \ \ \ \ \ \ \ \ \ \
\ \ \ \ \ \ \ \ \ \ \ \ \ \ \ \ \ \ \ \ \ \ \ \ \ \ \ \ \ \ \ \ \ \ \ \ \ \
\ \ \ \ \ \ \ \ \ \ \ \ \ \ \ then we obtain the following relations

\begin{equation}
\lambda _{1}=\frac{1}{2}\left[ 2\lambda _{1}^{3}+\lambda _{1}\left( \lambda
_{2}^{2}+\lambda _{3}^{2}\right) \right]  \tag{3.8a}  \label{3.8a}
\end{equation}

\begin{equation}
\lambda _{2}=\frac{1}{2}\left[ 2\lambda _{2}^{3}+\lambda _{2}\left( \lambda
_{3}^{2}+\lambda _{1}^{2}\right) \right]  \tag{3.8b}  \label{3.8b}
\end{equation}

\begin{equation}
\lambda _{3}=\frac{1}{2}\left[ 2\lambda _{3}^{3}+\lambda _{3}\left( \lambda
_{1}^{2}+\lambda _{2}^{2}\right) \right]  \tag{3.8c}  \label{3.8c}
\end{equation}%
\ 

We assume the eigenvalues to be real and order them in the sequence $\lambda
_{1}\leq \lambda _{2}\leq \lambda _{3},$and consider only positive values
for these fermions. There are then the following four possibilities

(i) All the $\lambda _{i}$s are zero. This is a trivial solution

(ii) All the $\lambda _{i}$s are non-zero in which case they must all be
equal. This is also a trivial solution if $V_{1}=V_{2}$, as is the case for $%
U_{0}$ and $D_{0}$\ defined in (1.11) and (1.12) and, when $M_{0}$ becomes a
unit matrix

(iii) $\lambda _{1}=0,$ in which case $\lambda _{2}=$\ \ $\lambda _{3}$

(iv)\ \ $\lambda _{1}=\lambda _{2}=0$, in which case $\lambda _{3}=1$\ \ 

Therefore, there are only two non-trivial solution, which we describe as
follows\ \ \ \ \ \ \ \ \ \ \ \ \ \ \ \ \ \ \ \ \ \ \ \ \ \ \ \ \ \ \ \ \ \ \
\ \ \ \ \ \ \ \ \ \ \ \ \ \ \ \ \ \ \ \ \ \ \ \ \ \ \ $\ $\ $\ \ \ \ \ \ \ \
\ \ \ \ \ \ \ \ \ \ \ \ \ \ \ \ \ \ \ \ \ \ \ \ \ \ \ \ \ \ \ \ \ $\ \ \ \ \
\ \ \ \ \ \ \ \ \ \ \ \ \ \ \ \ \ \ \ \ \ \ \ \ \ \ \ \ \ \ \ \ \ \ \ \ \ \
\ \ \ \ \ \ \ \ \ \ \ \ \ \ \ \ \ \ \ \ \ \ \ \ \ \ \ \ \ \ \ \ \ \ \ \ \ \
\ \ \ \ \ \ \ \ \ \ \ \ \ \ \ \ \ 

\bigskip (i) "Hierarchical" solution with $\lambda _{1}=\lambda _{2}=0$, $%
\lambda _{3}=1$%
\begin{equation}
M_{diag}^{\left( 1\right) }=%
\begin{bmatrix}
0 & 0 & 0 \\ 
0 & 0 & 0 \\ 
0 & 0 & 1%
\end{bmatrix}
\tag{3.9}  \label{3.9}
\end{equation}%
\ \ \ \ \ \ \ \ \ \ \ \ \ \ \ \ \ \ \ \ \ \ \ \ \ \ \ \ \ \ \ \ \ \ \ \ \ \
\ \ \ \ \ \ \ \ \ \ \ \ \ \ \ \ \ \ \ \ \ \ \ \ \ \ \ \ \ \ \ \ \ \ \ \ \ \
\ \ \ \ \ \ \ \ \ \ \ \ \ \ \ \ \ \ \ $\ \ \ \ \ \ \ \ \ \ \ \ \ \ \ \ \ \ \
\ \ \ \ \ \ \ \ \ \ \ \ \ \ \ \ \ \ $%
\begin{equation}
M_{0}^{\left( 1\right) }=V_{1}M_{diag}^{\left( 1\right) }V_{2}^{\dagger } 
\tag{3.10}  \label{3.10}
\end{equation}%
\ 

(ii) "Semi-hierarchical" solution with $\lambda _{1}=0,\lambda _{2}=\lambda
_{3}=\sqrt{\frac{2}{3}}$

\begin{equation}
M_{diag}^{\left( 2\right) }=\sqrt{\frac{2}{3}}%
\begin{bmatrix}
0 & 0 & 0 \\ 
0 & 1 & 0 \\ 
0 & 0 & 1%
\end{bmatrix}
\tag{3.11}  \label{3.11}
\end{equation}%
\ \ \ \ \ \ \ \ \ \ \ \ \ \ \ \ \ \ \ \ \ \ \ \ \ \ \ \ \ \ \ \ \ \ \ \ \ \
\ \ \ \ \ \ \ \ \ \ \ \ \ \ \ \ \ \ \ \ \ \ \ \ \ \ \ \ \ \ \ \ \ \ \ \ \ \
\ \ \ \ \ \ \ \ \ \ \ \ \ \ $\ \ \ \ \ \ \ \ \ \ \ \ \ \ \ \ \ \ \ \ \ \ \ \
\ \ \ \ \ \ \ \ \ \ \ $

\begin{equation}
M_{0}^{\left( 2\right) }=V_{1}M_{diag}^{\left( 2\right) }V_{2}^{\dagger } 
\tag{3.12}  \label{3.12}
\end{equation}%
\ 

\bigskip

\textbf{A. Hierarchical solutions}

In Appendix I a general form of $M_{0}^{\left( 1\right) }$ is obtained by
using equation (3.4) directly.

Below we obtain $M_{0}^{\left( 1\right) },$assumed real and symmetric, from
relation (3.5) by taking

\begin{equation}
V_{1}=V_{2}=V  \tag{3.13}  \label{3.13}
\end{equation}

where we will take $V$ as the product of three rotations that diagonalizes,
in succession, the 2-3, 1-3 and 1-2 sub-matrices by angles $\theta
_{1,}\theta _{2,}$and\ $\theta _{3,}$respectively i.e.

\begin{equation}
V=%
\begin{bmatrix}
c_{2}c_{3} & c_{2}s_{3} & s_{2} \\ 
-c_{1}s_{3}-s_{1}s_{2}c_{3} & c_{1}c_{3}-s_{1}s_{2}s_{3} & s_{1}c_{2} \\ 
s_{1}s_{3}-c_{1}s_{2}c_{3} & -s_{1}c_{3}-c_{1}s_{2}s_{3} & c_{1}c_{2}%
\end{bmatrix}
\tag{3.14}  \label{3.14}
\end{equation}%
where $s_{i}$ and $c_{i\text{ }}$are the corresponding sine and
cosine-values. With (3.13) and (3.14) we then have the following simple
form\ \ \ \ \ \ \ \ \ \ \ \ \ \ \ \ \ \ \ \ \ \ \ \ \ \ \ \ \ \ \ \ \ \ \ \
\ \ \ \ \ \ \ \ \ \ \ \ \ \ \ \ \ \ \ \ \ \ \ \ \ \ \ \ \ \ \ \ \ \ \ \ \ \
\ \ \ \ \ \ \ \ \ \ \ \ \ \ \ \ \ \ \ \ \ \ \ \ \ \ \ \ \ \ \ \ \ \ \ \ \ \
\ \ \ \ \ \ \ \ \ \ \ \ \ \ \ \ \ \ \ \ \ \ \ \ \ \ \ \ \ \ \ \ \ \ \ \ \ \
\ \ \ \ \ \ \ \ \ \ \ \ \ \ \ \ \ \ \ \ \ \ \ \ \ \ \ \ \ \ \ \ \ \ \ \ \ \
\ \ \ \ \ \ \ \ \ \ \ \ \ \ \ \ \ \ \ \ \ \ \ \ \ \ \ \ \ \ \ \ \ \ \ \ \ \
\ \ \ \ \ \ \ \ \ \ \ \ \ \ \ \ \ \ \ \ \ \ \ \ \ \ \ \ \ \ \ \ \ \ 

\begin{equation}
M_{0}^{\left( 1\right) }=%
\begin{bmatrix}
s_{2}^{2} & c_{2}s_{1}s_{2} & c_{1}c_{2}s_{2} \\ 
c_{2}s_{1}s_{2} & c_{2}^{2}s_{1}^{2} & c_{1}c_{2}^{2}s_{1} \\ 
c_{1}c_{2}s_{2} & c_{1}c_{2}^{2}s_{1} & c_{1}^{2}c_{2}^{2}%
\end{bmatrix}
\tag{3.15}  \label{3.15}
\end{equation}%
$\ \ \ \ \ \ \ \ \ \ \ \ \ \ \ \ \ \ \ \ \ \ \ \ \ \ \ \ \ \ \ \ \ \ \ \ \ \
\ \ \ \ \ \ \ \ \ \ \ \ \ \ \ \ \ \ \ \ \ \ \ \ \ \ \ \ \ \ \ \ \ \ \ \ \ \
\ \ \ \ \ \ \ \ \ \ \ \ \ \ \ \ \ \ \ \ \ \ \ \ \ \ \ \ \ \ \allowbreak $\ \
\ \ \ \ \ \ \ \ \ \ \ \ \ \ \ \ \ \ \ \ \ \ \ \ \ \ \ \ \ \ \ \ \ \ \ \ \ \
\ \ \ \ \ \ \ \ \ \ \ \ \ \ \ \ \ \ \ \ \ \ \ \ \ \ \ \ \ \ \ \ \ \ \ \ \ \
\ \ \ \ \ \ \ \ \ \ \ \ \ \ \ \ \ \ \ \ \ \ \ \ \ \ \ \ \ \ \ \ \ \ \ \ \ \
\ \ \ \ \ \ \ \ \ \ \ \ \ \ \ \ \ \ \ \ \ \ \ \ \ \ \ \ \ \ \ \ \ \ \ \ \ \
\ \ \ \ \ \ \ \ \ \ \ \ \ \ \ \ \ \ \ \ \ \ \ \ \ There is thus a direct
connection between the mass matrix and the rotation parameters.\ This shows
a classsic hierarchy pattern, particularly for the case of small angles, in
which case one can write $\left( c_{i}\thickapprox 1\right) $

\begin{equation}
M_{0}^{\left( 1\right) }=%
\begin{bmatrix}
s_{2}^{2} & s_{1}s_{2} & s_{2} \\ 
s_{1}s_{2} & s_{1}^{2} & s_{1} \\ 
s_{2} & s_{1} & 1%
\end{bmatrix}
\tag{3.16}  \label{3.16}
\end{equation}

\ \ \ \ \ \ \ \ \ \ \ \ \ \ \ \ \ \ \ \ \ \ \ \ \ \ \ \ \ \ \ \ \ \ \ \ \ \
\ \ \ \ \ \ \ \ \ \ \ \ \ \ \ \ \ \ \ \ \ \ \ \ \ \ \ \ \ \ \ \ \ \ \ \ \ \
\ \ \ \ \ \ \ \ \ \ \ \ \ \ \ \ \ \ \ \ \ \ \ \ \ \ \ \ \ \ \ \ \ \ \ \ \ \
\ \ \ \ \ \ \ \ \ \ \ \ \ \ \ \ \ \ \ \ \ \ \ \ \ \ \ \ \ \ \ \ \ \ \ \ \ \
\ \ \ \ \ \ \ \ \ \ \ \ \ \ \ \ \ \ \ \ \ \ \ \ \ \ \ \ \ \ \ \ \ \ \ \ \ \
\ \ \ \ \ \ \ \ \ \ \ \ \ \ \ \ \ \ \ \ \ \ \ \ \ \ \ \ \ \ \ \ \ \ \ \ \ \
\ \ \ \ \ \ \ \ \ \ \ \ \ \ \ \ \ \ \ \ \ \ \ \ \ \ \ \ \ \ \ \ \ \ \ \ \ \
\ \ \ \ \ \ \ \ \ \ \ \ \ \ \ \ \ \ \ \ \ \ \ \ \ 

\textbf{B. Semi-hierarchical solutions}

Following the same procedure as above, the semi-hierarchical patterns are
created by writing

\begin{equation}
M_{0}^{\left( 2\right) }=VM_{diag}^{\left( 2\right) }V^{\dagger }  \tag{3.17}
\label{3.17}
\end{equation}%
With (3.13) and (3.14), $M_{0}^{\left( 2\right) }$ in (3.12) becomes, for
small angles $\left( c_{i}\thickapprox 1\right) $\ \ \ \ \ \ \ \ \ \ \ \ \ \
\ \ \ \ \ \ \ \ \ \ \ \ \ \ \ \ \ \ \ \ \ \ \ \ \ \ \ \ \ \ \ \ \ \ \ \ \ \
\ \ \ \ \ \ \ \ \ \ \ \ \ \ \ \ \ \ \ \ \ \ \ \ \ \ \ \ \ \ \ \ \ \ \ \ \ \
\ \ \ \ \ \ \ \ \ \ \ \ \ \ \ \ \ \ \ \ \ \ \ \ \ \ \ \ \ \ \ \ \ \ \ \ \ \
\ \ \ \ \ \ \ \ \ \ \ \ \ 

\begin{equation}
M_{0}^{\left( 2\right) }=\sqrt{\frac{2}{3}}%
\begin{bmatrix}
s_{3}^{2} & s_{3} & s_{2}-s_{3}s_{1} \\ 
s_{3} & s_{1}^{2}+1 & -s_{2}s_{3} \\ 
s_{2}-s_{3}s_{1} & -s_{2}s_{3} & s_{1}^{2}+1%
\end{bmatrix}
\tag{3.18}  \label{3.18}
\end{equation}%
\ \ \ \ \ \ \ \ \ \ \ \ \ \ \ \ \ \ \ \ \ \ \ \ \ \ \ \ \ \ \ \ \ \ \ \ \ \
\ \ \ \ \ \ \ \ \ \ \ \ \ \ \ \ \ \ \ \ \ \ \ \ \ \ \ \ \ \ \ \ \ \ \ \ \ \
\ \ \ \ \ \ \ \ \ \ \ \ \ \ \ \ \ \ \ \ \ \ \ 

\bigskip

\textbf{IV. U and D matrices, and the CKM angles}

\textbf{A. U-matrix}

If we assume $\lambda _{b}<<\lambda _{t}$ then the contribution of the
second term in (2.4) given by

\begin{equation}
\frac{1}{12}DD^{+}U  \tag{4.1}  \label{4.1}
\end{equation}%
will be of the order $\frac{1}{12}\lambda _{b}^{2}\lambda _{t}$ which is
much smaller than the first term in (2.4) which is of the order$\frac{1}{2}%
\lambda _{t}^{3}$ . It can, therefore, be safely neglected. The U-matrix
will satisfy, to an excellent approximation, the following

\begin{equation}
-\dfrac{dU}{dx}=\frac{1}{4}\left[ UU^{+}+Tr(UU^{+})\mathbf{1}\right] U 
\tag{4.2}  \label{4.2}
\end{equation}%
which is the same equation as (3.4) for $M_{0}$ and will have the solutions
already discussed.\ \ \ \ \ \ \ \ \ \ \ \ \ \ \ \ \ \ \ \ \ \ \ \ \ \ \ \ \
\ \ \ \ \ \ \ \ \ \ \ \ \ \ \ \ \ \ \ \ \ \ \ \ \ \ \ \ \ \ \ \ \ \ \ \ \ \
\ \ \ \ \ \ \ \ \ \ \ \ \ \ \ \ \ \ \ \ \ \ \ \ \ \ \ \ \ \ \ \ \ \ \ \ \ \
\ \ \ \ \ \ \ \ \ \ \ \ \ \ \ \ \ \ \ \ \ \ \ \ \ \ \ \ \ \ \ \ \ \ \ \ \ \
\ \ \ \ \ \ \ \ \ \ \ \ \ \ \ \ 

We now assume that \ $U_{0}$\ \ is of the "hierarchical form" so that

\begin{equation}
U_{diag}=M_{diag}^{\left( 1\right) }=\ 
\begin{bmatrix}
0 & 0 & 0 \\ 
0 & 0 & 0 \\ 
0 & 0 & 1%
\end{bmatrix}
\tag{4.3}  \label{4.3}
\end{equation}%
and\ \ \ \ \ \ \ \ \ \ \ \ \ \ \ \ \ \ \ \ \ \ \ \ \ \ \ \ \ \ \ \ \ \ \ \ \
\ \ \ \ \ \ \ \ \ \ \ \ \ \ \ \ \ \ \ \ \ \ \ \ \ \ \ \ \ \ \ \ \ \ \ \ \ \
\ \ \ \ \ \ \ \ \ \ \ \ \ \ \ \ \ \ \ \ \ \ \ \ \ \ \ \ \ \ \ \ \ \ \ \ \ \
\ \ \ \ \ \ \ \ \ \ \ \ \ \ \ \ \ \ \ \ \ \ \ \ \ \ \ \ \ \ \ \ \ \ \ \ \ \
\ \ \ \ \ \ \ \ \ \ \ \ \ \ \ \ \ \ \ \ \ \ \ \ \ \ \ \ \ \ \ \ \ \ \ \ \ \
\ \ \ \ \ \ \ \ \ \ \ \ \ \ \ \ \ \ \ \ \ \ \ \ \ \ \ \ \ \ \ \ \ \ \ \ \ \
\ \ \ \ \ \ \ \ \ \ \ \ \ \ \ \ \ \ \ \ \ \ \ \ \ \ \ \ \ \ \ \ \ \ \ \ \ \
\ \ \ \ \ \ \ \ \ \ \ \ \ \ \ \ \ \ \ \ \ \ \ \ \ \ \ \ \ \ \ \ \ \ \ \ \ \
\ \ \ \ \ \ \ \ \ \ \ \ \ \ \ \ \ \ \ \ \ \ \ \ \ \ \ \ \ \ \ \ \ \ \ \ \ \
\ \ \ \ \ \ \ \ \ \ \ \ \ \ \ \ \ \ \ \ \ \ \ \ \ \ \ \ \ \ \ \ \ \ \ \ \ 

\begin{equation}
U_{0}=V%
\begin{bmatrix}
0 & 0 & 0 \\ 
0 & 0 & 0 \\ 
0 & 0 & 1%
\end{bmatrix}%
V^{\dagger }\   \tag{4.4}  \label{4.4}
\end{equation}%
If we take \ $V$ to be the same as the CKM matrix\ \ \ \ \ \ \ \ \ \ \ \ \ \
\ \ \ \ \ \ \ \ \ \ \ \ \ \ \ \ \ \ \ \ \ \ \ \ \ \ \ \ \ \ \ \ \ \ \ \ \ \
\ \ \ \ \ \ \ \ \ \ \ \ \ \ \ \ \ \ \ \ \ \ \ \ \ \ \ \ \ \ \ \ \ \ \ \ \ \
\ \ \ \ \ \ \ \ \ \ \ \ \ \ \ \ \ \ \ \ \ \ \ \ \ \ \ \ \ \ \ \ \ \ \ \ \ \
\ \ \ \ \ \ \ \ \ \ \ \ \ \ \ \ \ \ \ \ \ \ \ \ \ \ \ \ \ \ \ \ \ \ \ \ \ \
\ \ \ \ \ \ \ \ \ \ \ \ \ \ \ \ \ \ \ \ \ \ \ \ \ \ \ \ \ \ \ \ \ \ \ \ \ \
\ \ \ \ \ \ \ \ \ \ \ \ \ \ \ \ \ \ \ \ \ \ \ \ \ \ \ \ \ \ \ \ \ \ \ \ \ \
\ \ \ \ \ \ \ \ \ \ 

\begin{equation}
V\ \ =\ V_{CKM}  \tag{4.5}  \label{4.5}
\end{equation}%
then, as discussed in III A, equations (3.15), we have\ \ \ \ \ \ \ \ \ \ \
\ \ \ \ \ \ \ \ \ \ \ \ \ \ \ \ \ \ \ \ \ \ \ \ \ \ \ \ \ \ \ \ \ \ \ \ \ \
\ \ \ \ \ \ \ \ \ \ \ \ \ \ \ \ \ \ \ \ \ \ \ \ \ \ \ \ \ \ \ \ \ \ \ \ \ \
\ \ \ \ \ \ \ \ \ \ \ \ \ \ \ \ \ \ \ \ \ \ \ \ \ \ \ \ \ \ \ \ \ \ \ \ \ \
\ \ \ \ \ \ \ \ \ \ \ \ \ \ \ \ \ \ \ \ \ \ \ \ \ \ \ \ \ \ \ \ \ \ \ \ \ \
\ 

\begin{equation}
U_{0}=M_{0}^{\left( 1\right) }=%
\begin{bmatrix}
s_{2}^{2} & c_{2}s_{1}s_{2} & c_{1}c_{2}s_{2} \\ 
c_{2}s_{1}s_{2} & c_{2}^{2}s_{1}^{2} & c_{1}c_{2}^{2}s_{1} \\ 
c_{1}c_{2}s_{2} & c_{1}c_{2}^{2}s_{1} & c_{1}^{2}c_{2}^{2}%
\end{bmatrix}
\tag{4.6}  \label{4.6}
\end{equation}%
where the angles now are the same as given by the CKM matrix\ , we call it
"input" CKM, giving us the expression (with $c_{i\text{ }}\thickapprox 1$)\
\ \ \ \ \ \ \ \ \ \ \ \ \ \ \ \ \ \ \ \ \ \ \ \ \ \ \ \ \ \ \ \ \ \ \ \ \ \
\ \ \ \ \ \ \ \ \ \ \ \ \ \ \ \ \ \ \ \ \ \ \ \ \ \ \ \ \ \ \ \ \ \ \ \ \ \
\ \ \ \ \ \ \ \ \ \ \ \ \ \ \ \ \ \ \ \ \ \ \ \ \ \ \ \ \ \ \ \ \ \ \ \ \ \
\ \ \ \ \ \ \ \ \ \ \ \ \ \ \ \ \ \ \ \ \ \ \ \ \ \ \ \ \ \ \ \ \ \ \ \ \ \
\ \ \ \ \ \ \ \ \ \ \ \ \ \ \ \ \ \ \ \ \ \ \ \ \ \ \ \ \ \ \ \ \ \ \ \ \ \
\ \ \ \ \ \ \ \ \ \ \ \ \ \ \ \ \ \ \ \ \ \ \ \ \ \ \ \ \ \ \ \ \ \ \ \ \ \ $%
\ $

\begin{equation}
U_{0}=%
\begin{bmatrix}
s_{2}^{2} & s_{1}s_{2} & s_{2} \\ 
s_{1}s_{2} & s_{1}^{2} & s_{1} \\ 
s_{2} & s_{1} & 1%
\end{bmatrix}
\tag{4.7}  \label{4.7}
\end{equation}%
\ \ \ \ \ \ \ \ \ \ \ \ \ \ \ \ \ \ \ \ \ \ \ \ \ \ \ \ \ \ \ \ \ \ \ \ \ \
\ \ \ \ \ \ \ \ \ \ \ \ \ \ \ \ \ \ \ \ \ \ \ \ \ \ \ \ \ \ \ \ \ \ \ \ \ \
\ \ \ \ \ \ \ \ \ \ \ \ \ \ \ \ \ \ \ \ \ \ \ \ \ \ \ \ \ \ \ \ \ \ \ \ \ \
\ \ \ \ \ \ \ \ \ \ \ \ \ \ \ \ \ \ \ \ \ \ \ \ \ \ \ \ \ \ \ \ \ \ \ \ \ \
\ \ \ \ \ \ \ \ \ \ \ \ \ \ \ \ \ \ \ \ \ \ \ \ \ \ \ \ \ \ \ \ \ \ \ \ \ \
\ \ \ \ \ \ \ \ \ \ \ \ \ \ \ \ \ \ \ \ \ \ \ \ \ \ \ \ \ \ \ \ \ \ \ \ \ \
\ \ \ \ \ \ \ \ \ \ \ \ \ \ \ \ \ \ \ \ \ \ \ \ \ \ \ \ \ \ \ \ 

\bigskip with%
\begin{equation}
V_{CKM}^{(in)}=%
\begin{bmatrix}
1 & s_{3} & s_{2} \\ 
-s_{3} & 1 & s_{1} \\ 
s_{1}s_{3}-s_{2} & -s_{1} & 1%
\end{bmatrix}
\tag{4.8}  \label{4.8}
\end{equation}

Let us elaborate further on the above result by expressing the CKM matrix in
the Wolfenstein representation [2], keeping only the leading terms

\begin{equation}
V_{CKM}\ =\left[ 
\begin{array}{ccc}
1 & \lambda & A\lambda ^{3}\left( \rho -i\eta \right) \\ 
-\lambda & 1 & A\lambda ^{2} \\ 
A\lambda ^{3}\left( 1-\rho -i\eta \right) & -A\lambda ^{2} & 1%
\end{array}%
\right]  \tag{4.9}  \label{4.9}
\end{equation}

Comparing this with the expression for $V_{CKM}^{(in)}$ above we can extract 
$s_{1},s_{2},$ and $s_{3}$ as follows considering only the magnitudes of the
Wolfenstein parameters (ignoring phases).\ 
\begin{equation}
s_{1}\approx A\lambda ^{2}  \tag{4.10a}  \label{4.10a}
\end{equation}

\begin{equation}
s_{2}\approx A\lambda ^{3}\sqrt{\rho ^{2}+\eta ^{2}}  \tag{4.10b}
\label{4.10b}
\end{equation}

\begin{equation}
s_{3}\approx \lambda  \tag{4.10c}  \label{4.10c}
\end{equation}%
\ \ \ \ \ \ \ \ \ \ \ \ \ \ \ \ \ \ \ \ \ \ \ \ \ \ \ \ \ \ \ \ \ \ \ \ \ \
\ \ \ \ \ \ \ \ \ \ \ \ \ \ \ \ \ \ \ \ \ \ \ \ \ \ \ \ \ \ \ \ \ \ \ \ \ \
\ \ \ \ \ \ \ \ \ \ \ \ \ \ \ \ \ \ \ \ \ \ \ \ \ \ \ \ \ \ \ \ \ \ \ \ \ \
\ \ \ \ \ \ \ \ \ \ \ \ \ \ \ \ \ \ \ \ \ \ \ \ \ \ \ \ \ \ \ \ \ \ \ \ \ \
\ \ \ \ \ \ 

Substituting the above values in the expression for $U_{0}$ given by we
obtain

\begin{equation}
U_{0}=%
\begin{bmatrix}
A^{2}\lambda ^{6}\left( \rho ^{2}+\eta ^{2}\right) & A^{2}\lambda ^{5}\sqrt{%
\rho ^{2}+\eta ^{2}} & A\lambda ^{3}\sqrt{\rho ^{2}+\eta ^{2}} \\ 
A^{2}\lambda ^{5}\sqrt{\rho ^{2}+\eta ^{2}} & A^{2}\lambda ^{4} & A\lambda
^{2} \\ 
A\lambda ^{3}\sqrt{\rho ^{2}+\eta ^{2}} & A\lambda ^{2} & 1%
\end{bmatrix}
\tag{4.11}  \label{4.11}
\end{equation}%
Experimentally [22], one finds, in order of magnitude terms,$\allowbreak $ \
\ \ \ \ \ \ \ \ \ \ \ \ \ \ \ \ \ \ \ \ \ \ \ \ \ \ \ \ \ \ \ \ \ \ \ \ \ \
\ \ \ \ \ \ \ \ \ \ \ \ \ \ \ \ \ \ \ \ \ \ \ \ \ \ \ \ \ \ \ \ \ \ \ \ \ \
\ \ \ \ \ \ \ \ \ \ \ \ \ \ \ \ \ \ \ \ \ \ \ \ \ \ \ \ \ \ \ \ \ \ \ \ \ \
\ \ \ \ \ \ \ \ \ \ \ \ \ \ \ \ \ \ \ \ \ \ \ \ \ \ \ \ \ \ \ \ \ \ \ \ \ \
\ \ \ \ \ \ \ \ \ \ \ \ \ 

\begin{equation}
A\thickapprox 1  \tag{4.12a}  \label{4.12a}
\end{equation}

\begin{equation}
\sqrt{\rho ^{2}+\eta ^{2}}\thickapprox \lambda  \tag{4.12b}  \label{4.12b}
\end{equation}%
therefore, the above expression simplifies to\ \ \ \ \ \ \ \ \ \ \ \ \ \ \ \
\ \ \ \ \ \ \ \ \ \ \ \ \ \ \ \ \ \ \ \ \ \ \ \ \ \ \ \ \ \ \ \ \ \ \ \ \ \
\ \ \ \ \ \ \ \ \ \ \ \ \ \ \ \ \ \ \ \ \ \ \ \ \ \ \ \ \ \ \ \ \ \ \ \ \ \
\ \ \ \ \ \ \ \ \ \ \ \ \ \ \ \ \ \ \ \ \ \ \ \ \ \ \ \ \ \ \ \ \ \ \ \ \ \
\ \ \ \ \ \ \ \ \ \ \ \ \ \ \ \ \ \ \ \ \ \ \ \ \ \ \ \ \ \ \ \ \ \ \ \ \ \
\ \ \ \ \ \ \ \ \ \ \ \ \ \ \ \ \ \ \ \ \ \ \ \ \ \ \ \ \ \ \ \ \ \ \ \ \ \
\ \ \ \ \ \ \ \ \ \ \ \ \ \ \ \ \ \ \ \ \ \ \ \ \ \ \ \ \ \ \ \ \ \ \ \ \ \
\ \ \ \ \ \ \ \ \ \ \ \ \ \ \ \ \ \ \ \ \ \ \ \ \ \ \ \ \ \ \ \ \ \ \ \ \ \
\ \ \ \ \ \ \ \ \ \ \ \ \ \ \ \ \ \ \ \ \ \ \ \ \ \ 

\begin{equation}
U_{0}\thickapprox 
\begin{bmatrix}
\lambda ^{8} & \lambda ^{6} & \lambda ^{4} \\ 
\lambda ^{6} & \lambda ^{4} & \lambda ^{2} \\ 
\lambda ^{4} & \lambda ^{2} & 1%
\end{bmatrix}
\tag{4.13}  \label{4.13}
\end{equation}%
which is in excellent agreement with the behavior one expects for $U_{0}$
[1].\ \ \ \ \ \ \ \ \ \ \ \ \ \ \ \ \ \ \ \ \ \ \ \ \ \ \ \ \ \ \ \ \ \ \ \
\ \ \ \ \ \ \ \ \ \ \ \ \ \ \ \ \ \ \ \ \ \ \ \ \ \ \ \ \ \ \ \ \ \ \ \ \ \
\ \ \ \ \ \ \ \ \ \ \ \ \ \ \ \ \ \ \ \ \ \ \ \ \ \ \ \ \ \ \ \ \ \ \ \ \ \
\ \ \ \ \ \ \ \ \ \ \ \ \ \ \ \ \ \ \ \ \ \ \ \ \ \ \ \ \ \ \ \ \ \ \ \ \ \
\ \ \ \ \ \ \ \ \ \ \ \ \ \ \ \ \ \ \ \ \ \ \ \ \ \ \ \ \ \ \ \ \ \ \ \ \ \
\ \ \ \ \ \ \ \ \ \ \ \ \ \ \ \ \ \ \ \ \ \ \ \ \ \ \ \ \ \ \ \ \ \ \ \ \ \
\ \ \ \ \ \ \ \ \ \ \ \ \ \ \ \ \ \ \ \ \ \ \ \ \ \ \ \ \ \ \ \ \ \ \ \ \ \
\ \ \ \ \ \ \ \ \ \ \ \ \ \ \ \ \ \ \ \ \ \ \ \ \ \ \ \ \ \ \ \ \ \ \ \ \ \
\ \ \ \ \ \ \ \ \ \ \ \ \ \ \ \ \ \ \ \ \ \ \ \ \ \ \ \ \ \ \ \ \ \ \ \ \ \
\ \ \ \ \ \ \ \ \ \ \ \ \ \ \ \ \ \ \ \ \ \ \ \ \ \ \ \ \ \ \ \ \ \ \ \ \ \
\ \ \ \ \ \ \ \ \ \ \ \ \ \ \ \ \ \ \ \ \ \ \ \ \ \ \ \ \ \ \ \ \ \ \ \ \ \
\ \ \ \ \ \ \ \ \ \ \ \ \ \ \ \ \ \ \ \ \ \ \ \ \ \ \ \ \ \ \ \ \ \ \ \ \ \
\ \ \ \ \ \ \ \ \ \ \ \ \ \ \ \ \ \ \ \ \ \ \ \ \ \ \ \ \ \ \ \ \ \ \ \ \ \
\ \ \ \ \ \ \ \ \ \ \ \ \ \ \ \ \ \ \ \ \ \ \ \ \ \ \ \ \ \ \ \ \ \ \ \ \ \
\ \ \ \ \ \ \ \ \ \ \ \ \ \ \ \ \ \ \ \ \ \ \ \ \ \ \ \ \ \ \ \ \ \ \ \ \ \
\ \ \ \ \ \ \ \ \ \ \ \ \ \ \ \ \ \ \ \ \ \ \ \ \ \ \ \ \ \ \ \ \ \ \ \ \ \
\ \ \ \ \ \ \ \ \ \ \ \ \ \ \ \ \ \ \ \ \ \ \ \ \ \ \ \ \ \ \ \ \ \ \ \ \ \
\ \ \ \ \ \ \ \ \ \ \ \ \ \ \ \ \ \ \ \ \ \ \ \ \ \ \ \ \ \ \ \ \ \ \ \ \ \
\ \ \ \ \ \ \ \ \ \ \ \ \ \ \ \ \ \ \ \ \ \ \ \ \ \ \ \ \ \ \ \ \ \ \ \ \ \
\ \ \ \ \ \ \ \ \ \ \ \ \ \ \ \ \ \ \ \ \ \ \ \ \ \ \ \ \ \ \ \ \ \ \ \ \ \
\ \ \ \ \ \ \ \ \ \ \ \ \ \ \ \ \ \ \ \ \ \ \ \ \ \ \ \ \ \ \ \ \ \ \ \ \ \
\ \ \ \ \ \ \ \ \ \ \ \ \ \ \ \ \ \ \ \ \ \ \ \ \ \ \ \ \ \ \ \ \ \ \ \ \ \
\ \ \ \ \ \ \ \ \ \ \ \ \ \ \ \ \ \ \ \ \ \ \ \ \ \ \ \ \ \ \ \ \ \ \ \ \ \
\ \ \ \ \ \ \ \ \ \ \ \ \ \ \ \ \ \ \ \ \ \ \ \ \ \ \ \ \ \ \ \ \ \ \ \ \ \
\ \ \ \ \ \ \ \ \ \ \ \ \ \ \ \ \ \ \ 

To recapulate then, after including the x-dependence from (2.3), we have the
complete expression for $U$ given by$\ $\ \ \ \ \ \ \ \ \ \ \ \ \ \ \ \ \ \
\ \ \ \ \ \ \ \ \ \ \ \ \ \ \ \ \ \ \ \ \ \ \ \ \ \ \ \ \ \ \ \ \ \ \ \ \ \
\ \ \ \ \ \ \ \ \ \ \ \ \ \ \ \ \ \ \ \ \ \ \ \ \ \ \ \ \ \ \ \ \ \ \ \ \ \
\ \ \ \ \ \ \ \ \ \ \ \ \ \ \ \ \ \ \ \ \ \ \ \ \ \ \ \ \ \ \ \ \ \ \ \ \ \
\ \ \ \ \ \ \ \ \ \ \ \ \ \ \ \ \ \ \ \ \ \ \ \ 

\begin{equation}
U=%
\begin{bmatrix}
s_{2}^{2} & s_{1}s_{2} & s_{2} \\ 
s_{1}s_{2} & s_{1}^{2} & s_{1} \\ 
s_{2} & s_{1} & 1%
\end{bmatrix}%
\lambda _{0t}\left[ 1+x\lambda _{0t}^{2}\right] ^{-\frac{1}{2}}  \tag{4.14}
\label{4.14}
\end{equation}%
$\ \ \ \ \ \ \ \ \ \ \ \ \ \ \ \ \ \ \ \ \ \ \ \ \ \ \ \ \ \ \ \ \ \ \ \ \ \
\ \ \ \ \ \ \ \ \ \ \ \ \ \ \ \ \ \ \ \ \ \ \ \ \ \ \ \ \ \ \ \ \ \ \ \ \ \
\ \ \ \ \ \ \ \ \ \ \ \ \ \ \ \ \ \ \ \ \ \ \ \ \ \ \ \ \ \ \ \ \ \ \ \ \ \
\ \ \ \ \ \ \ \ \ \ \ \ \ \ \ \ \ \ \ \ \ \ \ \ \ \ \ \ \ \ \ \ \ \ \ \ \ \
\ \ \ \ \ \ \ \ \ \ \ \ \ \ \ \ \ \ \ \ \ \ \ \ \ \ \ \ \ \ \ \ \ \ \ \ \ \
\ \ \ \ \ \ \ \ \ \ \ \ \ \ \ \ \ \ \ \ \ \ \ \ \ \ \ \ \ \ \ \ \ \ \ \ \ \
\ \ \ \ \ \ \ \ \ \ \ \ \ \ \ \ \ \ \ \ \ \ \ \ \ \ \ \ \ \ \ \ \ \ \ \ \ \
\ \ \ \ \ \ \ \ \ \ \ \ \ \ \ \ \ \ \ \ \ \ \ \ \ \ \ \ \ \ \ \ \ \ \ \ \ \
\ \ \ \ \ \ \ \ \ \ \ \ \ \ \ \ \ \ \ \ \ \ \ \ \ \ \ \ \ \ \ \ \ \ $where
where $s_{i}$ are the "input" CKM parameters. \ \ 

The eigenvalues of $U_{0}$ are, of course, the same as $M_{diag}^{\left(
1\right) }$ i.e $(0,0,1)$. \ To obtain the correct values, as we stated in
the Introduction, one must go beyond the standard model which we propose to
do in Section V by incorporating texture zeros.

\textbf{B. D-matrix}

We assume here that $D$ is different from $U$ by the fact that the solution
of the "uncoupled" part

\begin{equation}
-\dfrac{dD}{dx}=\frac{1}{4}\left[ DD^{+}+Tr(DD^{+})\mathbf{1}\right] D 
\tag{4.15}  \label{4.15}
\end{equation}%
is now given by $M_{0}^{\left( 2\right) },$the "semi-hierarchical" solution
(3.11) and (3.12)\textbf{\ }with $V_{1}=V_{2}=V$\ \ \ \ \ \ \ \ \ \ \ \ \ \
\ \ \ \ \ \ \ \ \ \ \ \ \ \ \ \ \ \ \ \ \ \ \ \ \ \ \ \ \ \ \ \ \ \ \ \ \ \
\ \ \ \ \ \ \ \ \ \ \ \ \ \ \ \ \ \ \ \ \ \ \ \ \ \ \ \ \ \ \ \ \ \ \ \ \ \
\ \ \ \ \ \ \ \ \ \ \ \ \ \ \ \ \ \ \ \ \ \ \ \ \ \ \ \ \ \ \ \ \ \ \ \ \ \
\ \ \ \ \ \ \ \ \ \ \ \ \ \ \ \ \ \ \ \ \ \ \ \ \ \ \ \ \ \ \ \ \ \ \ \ \ \
\ \ \ \ \ \ \ \ \ \ \ \ \ \ \ \ \ \ \ \ \ \ \ \ \ \ \ \ \ \ \ \ \ \ \ \ \ \
\ \ \ \ \ \ \ \ \ \ \ \ \ \ \ \ \ \ \ \ \ \ \ \ \ \ \ \ \ \ \ \ \ \ \ \ \ \
\ 

\begin{equation*}
M_{diag}^{\left( 2\right) }=\sqrt{\frac{2}{3}}%
\begin{bmatrix}
0 & 0 & 0 \\ 
0 & 1 & 0 \\ 
0 & 0 & 1%
\end{bmatrix}%
\end{equation*}%
\ \ \ \ \ \ \ \ \ \ \ \ \ \ \ \ \ \ \ \ \ \ \ \ \ \ \ \ \ \ \ \ \ \ \ \ \ \
\ \ \ \ \ \ \ \ \ \ \ \ \ \ \ \ \ \ \ \ \ \ \ \ \ \ \ \ \ \ \ \ \ \ \ \ \ \
\ \ \ \ \ \ \ \ \ \ \ \ \ \ \ \ \ \ \ \ \ \ \ \ \ \ \ \ \ \ \ \ \ \ \ \ \ \
\ \ \ \ \ \ \ \ \ \ \ \ \ \ \ \ \ \ \ \ \ \ \ \ \ \ \ \ \ \ \ \ \ \ \ \ \ \
\ \ \ \ \ \ \ \ \ \ \ \ \ \ \ \ \ \ \ \ \ \ \ \ \ \ \ \ \ \ \ \ \ \ \ \ \ \
\ \ \ \ \ \ \ \ \ \ \ \ \ \ \ \ \ \ \ \ 

\begin{equation*}
M_{0}^{\left( 2\right) }=VM_{diag}^{\left( 2\right) }V^{\dagger }
\end{equation*}%
\ However, this can not be a complete solution for the $D$-equation given by
(2.5)

\begin{equation*}
-\dfrac{dD}{dx}=\frac{1}{4}\left[ DD^{+}+Tr(DD^{+})\mathbf{1}\right] D+\frac{%
1}{12}UU^{+}D
\end{equation*}%
\ \ \ \ \ \ \ \ \ \ \ \ \ \ \ \ \ \ \ \ \ \ \ \ \ \ \ \ \ \ \ \ \ \ \ \ \ \
\ \ \ \ \ \ \ \ $\ \ \ \ \ \ \ \ \ \ \ \ \ \ \ \ \ \ \ \ \ \ \ \ \ \ \ \ \ \
\ \ \ \ \ \ \ \ \ \ \ \ \ \ \ \ \ \ \ \ \ \ \ \ $ \ \ \ \ \ \ \ \ \ \ \ \ \
\ \ \ \ \ \ \ \ \ \ \ \ \ \ \ \ \ \ \ \ \ \ \ \ \ \ \ \ \ \ \ since the
coupling term neglected in (4.15)

\begin{equation}
\frac{1}{12}UU^{+}D  \tag{4.16}  \label{4.16}
\end{equation}%
can be very large because $\lambda _{t}>>\lambda _{b}$. This is in stark
contrast to $U$ where the coupling term was negligible.\ \ \ \ \ \ \ \ \ \ \
\ \ \ \ \ \ \ \ \ \ \ \ \ \ \ \ \ \ \ \ \ \ \ \ \ \ \ \ \ \ \ \ \ \ \ \ \ \
\ \ \ \ \ \ \ \ \ \ \ \ \ \ \ \ \ \ \ \ \ \ \ \ \ \ \ \ \ \ \ \ \ \ \ \ \ \
\ \ \ \ \ \ \ \ \ \ \ \ \ \ \ \ \ \ \ \ \ \ \ \ \ \ \ \ \ \ \ \ \ \ \ \ \ \
\ \ \ \ \ \ \ \ \ \ \ \ \ \ \ \ \ \ \ \ \ \ \ \ \ \ \ \ \ \ \ \ \ \ \ \ \ \
\ \ \ \ \ \ \ \ \ \ \ \ \ \ \ \ \ \ \ \ \ \ \ \ \ \ \ \ \ \ \ \ \ \ \ \ \ \
\ \ \ \ \ \ \ \ \ \ \ \ \ \ \ \ \ \ \ \ \ \ \ \ \ \ \ \ \ \ \ \ \ \ \ \ \ \
\ \ \ \ \ \ \ \ \ \ \ \ \ \ \ \ \ \ \ \ \ \ \ \ \ \ \ \ \ \ \ \ \ \ \ \ \ \
\ \ \ \ \ \ \ \ \ \ \ \ \ \ \ \ \ \ \ \ \ \ \ \ \ \ \ \ \ \ \ \ \ \ \ \ 

If we consider the $x$-independent (scale-independent) coupling term in
(4.16), then we have

\begin{equation}
\frac{1}{12}U_{0}U_{0}^{+}D_{0}=\ \ \frac{1}{12}M_{0}^{\left( 1\right)
}M_{0}^{\left( 1\right) }M_{0}^{\left( 2\right) }  \tag{4.17}  \label{4.17}
\end{equation}

\begin{equation}
\ =\ \frac{1}{12}\sqrt{\frac{2}{3}}\left[ V%
\begin{bmatrix}
0 & 0 & 0 \\ 
0 & 0 & 0 \\ 
0 & 0 & 1%
\end{bmatrix}%
V^{\dagger }\right] \left[ V%
\begin{bmatrix}
0 & 0 & 0 \\ 
0 & 0 & 0 \\ 
0 & 0 & 1%
\end{bmatrix}%
V^{\dagger }\right] \left[ V%
\begin{bmatrix}
0 & 0 & 0 \\ 
0 & 1 & 0 \\ 
0 & 0 & 1%
\end{bmatrix}%
V^{\dagger }\ \right]  \tag{4.18}  \label{4.18}
\end{equation}%
Therefore,

\begin{equation}
\frac{1}{12}U_{0}U_{0}^{+}D_{0}=\ \ \frac{1}{12}\sqrt{\frac{2}{3}}\left[ V%
\begin{bmatrix}
0 & 0 & 0 \\ 
0 & 0 & 0 \\ 
0 & 0 & 1%
\end{bmatrix}%
V^{\dagger }\right] =\frac{1}{12}\sqrt{\frac{2}{3}}M_{0}^{\left( 1\right) } 
\tag{4.19}  \label{4.19}
\end{equation}%
\ \ \ \ \ \ \ \ \ \ \ \ \ \ \ \ \ \ \ \ \ \ \ \ \ \ \ \ \ \ \ \ \ \ \ \ \ \
\ \ \ \ \ \ \ \ \ \ \ \ \ \ \ \ \ \ \ \ \ \ \ \ \ \ \ \ \ \ \ \ \ \ \ \ \ \
\ \ \ \ \ \ \ \ \ \ \ \ \ \ \ \ \ \ \ \ \ \ \ \ \ \ \ \ \ \ \ \ \ \ \ \ \ \
\ \ \ \ \ \ \ \ \ \ \ \ \ \ \ \ \ \ \ \ \ \ \ \ \ \ \ \ \ \ \ \ \ \ \ \ \ 

So there is a contribution from $M_{0}^{\left( 1\right) }$ in the $D$-matrix
that needs to be added to $M_{0}^{\left( 2\right) }$. One can, to a good
approximation, write the differential equation for $D$ in (2.5) as

\begin{equation}
-\dfrac{dD}{dx}=V\left[ \frac{1}{2}\lambda _{b}^{3}\sqrt{\frac{2}{3}}%
\begin{bmatrix}
0 & 0 & 0 \\ 
0 & 1 & 0 \\ 
0 & 0 & 1%
\end{bmatrix}%
+\frac{1}{12}\sqrt{\frac{2}{3}}\lambda _{t}^{2}\lambda _{b}%
\begin{bmatrix}
0 & 0 & 0 \\ 
0 & 0 & 0 \\ 
0 & 0 & 1%
\end{bmatrix}%
\right] V^{\dagger }  \tag{4.20}  \label{4.20}
\end{equation}

\begin{equation}
=\frac{1}{2}\lambda _{b}^{3}M_{0}^{\left( 2\right) }+\frac{1}{12}\sqrt{\frac{%
2}{3}}\lambda _{t}^{2}\lambda _{b}M_{0}^{\left( 1\right) }  \tag{4.21}
\label{4.21}
\end{equation}%
\ \ \ \ \ \ \ \ \ \ \ \ \ \ \ \ \ \ \ \ \ \ \ \ \ \ \ \ \ \ \ \ \ \ \ \ \ \
\ \ \ \ \ \ \ \ \ \ \ \ \ \ \ \ \ \ \ \ \ \ \ \ \ \ \ \ \ \ \ 

We can solve for the 33-elements on both sides of the equation (4.21),
assumed dominated by $\lambda _{b}$ and $\lambda _{t}$, and obtain

\begin{equation}
-\dfrac{d\lambda _{b}}{dx}=\frac{1}{2}\lambda _{b}^{3}+\ \frac{1}{12}\lambda
_{t}^{2}\lambda _{b}\   \tag{4.22}  \label{4.22}
\end{equation}%
\ \ \ \ \ \ \ \ \ \ \ \ \ \ \ \ \ \ \ \ \ \ \ \ \ \ \ \ \ \ \ \ \ \ \ \ \ \
\ \ \ \ \ \ \ \ \ \ \ \ \ \ \ \ \ \ \ \ \ \ \ \ \ \ \ \ \ \ \ \ \ \ \ \ \ \
\ \ \ \ \ \ \ \ \ \ \ \ \ \ \ \ \ \ \ \ \ \ \ \ \ \ \ \ \ \ \ \ \ \ \ \ \ \
\ \ \ \ \ \ \ \ \ \ \ \ \ \ \ \ \ \ \ \ \ \ \ \ \ \ \ \ \ \ \ \ \ \ \ \ \ \
\ \ \ \ \ \ \ \ \ \ \ \ \ \ \ \ \ \ \ \ \ \ \ \ \ \ \ \ \ \ \ \ \ \ \ \ \ \
\ \ \ \ \ \ \ \ \ \ \ \ \ \ \ \ \ \ \ \ \ \ \ \ \ \ \ \ \ \ \ \ \ \ \ \ \ \
\ \ \ \ \ \ \ \ \ \ \ \ \ \ \ \ \ \ \ \ \ \ \ \ \ \ \ \ \ \ \ \ \ \ \ \ \ \
\ \ 

An exact analytical solution of (4.22) is given in Appendix II

It is important to note that, as far as the 22-elements of the two diagonal
matrices in (4.20) are concerned, the second matrix is a "small
perturbation" on the first because it is, in fact, negligible (zero)
compared to the first . In contrast, with respect to the 33-matrix elements,
it is the first matrix\ which is a "small perturbation" since $\lambda
_{b}<<\lambda _{t}.$Furthermore, we note that the x-dependence of the
22-diagonal element will not be the same as that of the 33-element, which is
inconsistent with our assumption that the entire $D$ matix is represented by
a single function of $x$.

However, since the first matrix is extremely small we can write an
approximate solution for $D$ as follows

\begin{equation}
D_{diag}=\left[ 
\begin{array}{ccc}
0 & 0 & 0 \\ 
0 & \epsilon & 0 \\ 
0 & 0 & 1%
\end{array}%
\right]  \tag{4.23}  \label{4.23}
\end{equation}%
\ \ \ \ \ \ \ \ \ \ \ \ \ \ \ \ \ \ \ \ \ \ \ \ \ \ \ \ \ \ \ \ \ \ \ \ \ \
\ \ \ \ \ \ \ \ \ \ \ \ \ \ \ \ \ \ \ \ \ \ \ \ \ \ \ \ \ \ \ \ \ \ \ \ \ \
\ \ \ \ \ \ \ \ \ \ \ \ \ \ \ \ \ \ \ \ \ \ \ \ \ \ \ \ \ \ \ \ \ \ \ \ \ \
\ \ \ \ \ \ \ \ \ \ \ \ \ \ \ \ \ 

\begin{equation}
D_{0}=V\left[ 
\begin{array}{ccc}
0 & 0 & 0 \\ 
0 & \epsilon & 0 \\ 
0 & 0 & 1%
\end{array}%
\right] V^{\dagger }  \tag{4.24}  \label{4.24}
\end{equation}%
\ \ \ \ \ \ \ \ \ \ \ \ \ \ \ \ \ \ \ \ \ \ \ \ \ \ \ \ \ \ \ \ \ \ \ \ \ \
\ \ \ \ \ \ \ \ \ \ \ \ \ \ \ \ \ \ \ \ \ \ \ \ \ \ \ \ \ \ \ \ \ \ \ \ \ \
\ \ \ \ \ \ \ \ \ \ \ \ \ \ \ \ \ \ \ \ \ \ \ \ \ \ \ \ \ \ \ \ \ \ \ \ \ \
\ \ \ \ \ \ \ \ \ \ \ \ \ \ \ \ \ \ \ \ \ \ \ \ \ \ \ \ \ \ \ \ \ \ \ \ \ \
\ \ \ \ \ \ \ \ \ \ \ \ \ \ \ \ \ \ \ \ \ \ \ \ \ \ \ \ \ \ \ \ \ \ \ \ \ \
\ \ \ \ \ \ \ \ \ \ \ \ \ \ \ \ \ \ \ \ \ \ \ \ \ \ \ \ \ \ \ \ \ \ \ \ \ \
\ \ \ \ \ \ \ \ \ \ \ \ \ \ \ \ \ \ \ \ \ \ \ \ \ \ \ \ \ \ \ \ \ \ \ \ \ \
\ \ \ \ \ \ \ \ \ \ \ \ \ \ \ \ \ \ \ \ \ \ \ \ \ \ \ \ \ \ \ \ \ \ \ \ \ \
\ \ \ \ \ \ \ \ \ \ \ \ \ \ \ \ \ \ \ \ \ \ \ \ \ \ \ \ \ \ \ \ \ \ \ \ \ \
\ \ \ \ \ \ \ \ \ \ \ \ \ \ \ \ \ \ \ \ \ \ \ \ \ \ \ \ \ \ \ \ \ \ \ \ \ \
\ \ \ \ \ \ \ \ where $\epsilon $ is assumed to be a constant, independent
of $x,$ and where in the 33-element of $D_{diag}$ we have ignored the
contribution of $\epsilon $ compared to 1.

To determine $\epsilon $ one can (numerically) integrate the first term in
(4.22), $\frac{1}{2}\lambda _{b}^{3}$, from the knowledge of the analytic
expression for $\lambda _{b}$ given in Appendix II, and compare it to the
integral of the second term, $\frac{1}{12}\lambda _{t}^{2}\lambda _{b},$
with $\lambda _{t}$ given in (2.3). The ratio of the two will estimate the
value of $\epsilon $. \ If we take the ratio of the Yukawa couplings $%
\lambda _{0b}$ and $\lambda _{0t}$ to be the same as the ratio of the masses
then we obtain $\epsilon \thickapprox .01$ which is found to be essentially
independent of $x.$

Another, rough order of magnitude, estimate of $\epsilon $ can also be
obtained on the basis of the ratio of first to the second term in (2.5) above

\begin{equation}
\epsilon \thickapprox \left( \dfrac{\lambda _{b}^{3}}{\lambda
_{t}^{2}\lambda _{b}}\right) ^{\frac{1}{2}}=\dfrac{m_{b}}{m_{t}}\thickapprox
.025  \tag{4.25}  \label{4.25}
\end{equation}%
\ \ \ \ \ \ \ \ \ \ \ \ \ \ \ \ \ \ \ \ \ \ \ \ \ \ \ \ \ \ \ \ \ \ \ \ \ \
\ \ \ \ \ \ \ \ \ \ \ \ \ \ \ \ \ \ \ \ \ \ \ \ \ \ \ \ \ \ \ \ \ \ \ \ \ \
\ \ \ \ \ \ \ \ \ \ \ \ \ \ \ \ \ \ \ \ \ \ \ \ \ \ \ \ \ \ \ \ \ \ \ \ \ \
\ \ \ \ \ \ \ \ \ \ \ \ \ \ \ \ \ \ \ \ \ \ 

The square root is taken because, effectively, the first term in the
equation (2.5) for $\dfrac{dD}{dx}$ is proprtional to $D^{3}$ while the
second is proprtional to $U^{2}D,$ so the ratio is $\thickapprox D^{2}$. \
Therefore, the contribution to $D$ will involve the square root. The
numerical value of $\epsilon $ from (4.25) \textbf{is} obtained by taking
the ratio of the Yukawa couplings to be the same as the ratio of the
corresponding masses.\ \ \ \ \ \ \ \ \ \ \ \ \ \ \ \ \ \ \ \ \ \ \ \ \ \ \ \
\ \ \ \ \ \ \ \ \ \ \ \ \ \ \ \ \ \ \ \ \ \ \ \ \ \ \ \ \ \ \ \ \ \ \ \ \ \
\ \ \ \ \ \ \ \ \ \ \ \ \ \ \ \ \ \ \ \ \ \ \ \ \ \ \ \ \ \ \ \ \ \ \ \ \ \
\ \ \ \ \ \ \ \ \ \ \ \ \ \ \ \ \ \ \ \ \ \ \ \ \ \ \ \ \ \ \ \ \ \ \ \ \ \
\ \ \ \ \ \ \ \ \ \ \ \ \ \ \ \ \ \ \ \ \ \ \ \ \ \ \ \ \ \ \ \ \ \ \ \ \ \
\ \ \ \ \ \ \ \ \ \ \ \ \ \ \ \ \ \ \ \ \ \ \ \ \ \ \ \ \ \ \ \ \ \ \ \ \ \
\ \ \ \ \ \ \ \ \ \ \ \ \ \ \ \ \ \ \ \ \ \ \ \ \ \ \ \ \ \ \ \ \ \ \ \ \ \
\ \ \ \ \ \ \ \ \ \ \ \ \ \ \ 

In our calculations to follow we will take $\epsilon $ as an arbitrary
parameter keeping in mind, however, that its order of magnitude is $%
\thickapprox 10^{-2}$ .\ \ \ \ \ \ \ \ \ \ \ \ \ \ \ \ \ \ \ \ \ \ \ \ \ \ \
\ \ \ \ \ \ \ \ \ \ \ \ \ \ \ \ \ \ \ \ \ \ \ \ \ \ \ \ \ \ \ \ \ \ \ \ \ \
\ \ \ \ \ \ \ \ \ \ \ \ \ \ \ \ \ \ \ \ \ \ \ \ \ \ \ \ \ \ \ \ \ \ \ \ \ \
\ \ \ \ \ \ \ \ \ \ \ \ \ \ \ \ \ \ \ \ \ \ \ \ \ \ \ \ \ \ 

To obtain an approximate solution for $\lambda _{b},$we can ignore the first
term in (4.22)\textbf{, }which is very small\textbf{,} to write, after
substituting expression (2.3) for $\lambda _{t}$ ,

\begin{equation}
-\dfrac{d\lambda _{b}}{dx}\thickapprox \frac{1}{12}\lambda _{t}^{2}\lambda
_{b}\ =\frac{1}{12}\lambda _{0t}^{2}\left[ 1+x\lambda _{0t}^{2}\right]
^{-1}\lambda  \tag{4.26}  \label{4.26}
\end{equation}%
the solution to which is\ \ \ \ \ \ \ \ \ \ \ \ \ \ \ \ \ \ \ \ \ \ \ \ \ \
\ \ \ \ \ \ \ \ \ \ \ \ \ \ \ \ \ \ \ \ \ \ \ \ \ \ \ \ \ \ \ \ \ \ \ \ \ \
\ \ \ \ \ \ \ \ \ \ \ \ \ \ \ \ \ \ \ \ \ \ \ \ \ \ \ \ \ \ \ \ \ \ \ \ \ \
\ \ \ \ \ \ \ \ \ \ \ \ \ \ \ \ \ \ \ \ \ \ \ \ \ \ \ \ \ \ \ \ \ \ \ \ \ \
\ \ \ \ \ \ \ \ \ \ \ \ \ \ \ \ \ \ \ \ \ \ \ \ \ \ \ \ \ \ \ \ \ \ \ \ \ \
\ \ \ \ \ \ \ 

\begin{equation}
\lambda _{b}\left( x\right) =\lambda _{0b}\left[ 1+x\lambda _{0t}^{2}\right]
^{-\frac{1}{12}}  \tag{4.27}  \label{4.27}
\end{equation}%
\ \ \ \ \ \ \ \ \ \ \ \ \ \ \ \ \ \ \ \ \ \ \ \ \ \ \ \ \ \ \ \ \ \ \ \ \ \
\ \ \ \ \ \ \ \ \ \ \ \ \ \ \ \ \ \ \ \ \ \ \ \ \ \ \ \ \ \ \ \ \ \ \ \ \ \
\ \ \ \ \ \ \ \ \ \ \ \ \ \ \ \ \ \ \ \ \ \ \ \ \ \ \ \ \ \ \ \ \ \ \ \ \ \
\ \ \ \ \ \ \ \ \ \ \ \ \ \ \ \ \ \ \ \ \ \ \ \ \ \ \ \ \ \ \ \ \ \ \ \ \ \
\ \ \ \ \ \ \ \ \ \ \ \ \ \ \ \ \ \ \ \ \ \ \ \ \ \ \ \ \ \ \ \ \ \ \ \ \ \
\ \ \ \ \ \ \ \ \ \ \ \ \ \ \ \ \ \ \ \ \ \ \ \ \ \ \ \ \ \ \ \ \ \ \ \ \ \
\ \ \ \ \ \ \ \ \ \ \ \ \ \ \ \ \ \ \ \ \ \ \ \ \ \ \ \ \ \ \ \ \ \ \ \ \ \
\ \ \ \ \ \ \ \ \ \ \ \ \ \ \ \ \ \ \ \ \ \ \ \ \ \ \ \ \ \ \ \ \ \ \ \ \ \
\ \ \ \ \ \ \ \ \ \ \ \ \ \ \ \ \ \ \ \ \ \ \ \ \ \ \ \ \ \ \ \ \ \ \ \ \ \
\ \ \ \ \ \ \ \ \ \ \ \ \ \ \ \ \ \ \ \ \ \ \ \ \ \ \ \ \ \ \ \ \ \ \ \ \
where \ \ \ $\lambda _{0b}=$\ $\lambda _{b}\left( t_{0}\right) =$\ $\lambda
_{b}\left( x=0\right) $\ \ \ \ \ \ \ \ \ \ \ \ \ \ \ \ \ \ \ \ \ \ \ \ \ \ \
\ \ \ \ \ \ \ \ \ \ \ \ \ \ \ \ \ \ \ \ \ \ \ \ \ \ \ \ \ \ \ \ \ \ \ \ \ \
\ \ \ \ \ \ \ \ \ \ \ \ \ \ \ \ \ \ \ \ \ \ \ \ \ \ \ \ \ \ \ \ \ \ \ \ \ \
\ \ \ \ \ \ \ \ \ \ \ \ \ \ \ \ \ \ \ \ \ 

\bigskip From (4.24) we can write down the expression for $D$

\begin{equation}
D=%
\begin{bmatrix}
\epsilon s_{3}^{2} & \epsilon s_{3} & s_{2}-\epsilon s_{1}s_{3} \\ 
\epsilon s_{3} & s_{1}^{2}+\epsilon & s_{1} \\ 
s_{2}-\epsilon s_{1}s_{3} & s_{1} & 1%
\end{bmatrix}%
\lambda _{0b}\left[ 1+x\lambda _{0t}^{2}\right] ^{-\frac{1}{12}}  \tag{4.28}
\label{4.28}
\end{equation}

where $s_{i}$'s are the "input" CKM parameters. We can express $s_{i}$ in
terms of $\lambda $, as we did previously when we considered $U_{0}$, to
obtain\ \ \ \ \ \ \ \ \ \ \ \ \ \ \ \ \ \ \ \ \ \ \ \ \ \ \ \ \ \ \ \ \ \ \
\ \ \ \ \ \ \ \ \ \ \ \ \ \ \ \ \ \ \ \ \ \ \ \ \ \ \ \ \ \ \ \ \ \ \ \ \ \
\ \ \ \ \ \ \ \ \ \ \ \ \ \ \ \ \ \ \ \ \ \ \ \ \ \ \ \ \ \ \ \ \ \ \ \ \ \
\ 

\begin{equation}
D_{0}\approx 
\begin{bmatrix}
\epsilon \lambda ^{2} & \epsilon \lambda & \lambda ^{4}-\epsilon \lambda ^{3}
\\ 
\epsilon \lambda & \lambda ^{4}+\epsilon & \lambda ^{2} \\ 
\lambda ^{4}-\epsilon \lambda ^{3} & \lambda ^{2} & 1%
\end{bmatrix}
\tag{4.29}  \label{4.29}
\end{equation}

The estimate for $\epsilon $ given by (4.25) is, numerically $\thickapprox
\lambda ^{2}$, and, correspondingly we have

\begin{equation}
D_{0}\approx 
\begin{bmatrix}
\lambda ^{4} & \lambda ^{3} & \lambda ^{4} \\ 
\lambda ^{3} & \lambda ^{2} & \lambda ^{2} \\ 
\lambda ^{4} & \lambda ^{2} & 1%
\end{bmatrix}
\tag{4.30}  \label{4.30}
\end{equation}

This expression shows that the hierarchy in $D_{0}$ is not as pronounced as
it was found in $U,$ which is consistent with generally accepted form for $%
D_{0}$ in [1]

\begin{equation}
D_{0}\approx 
\begin{bmatrix}
\lambda ^{4} & \lambda ^{3} & \lambda ^{3} \\ 
\lambda ^{3} & \lambda ^{2} & \lambda ^{2} \\ 
\lambda ^{3} & \lambda ^{2} & 1%
\end{bmatrix}
\tag{4.31}  \label{4.31}
\end{equation}

Our results are in excellent agreement with the above if we ignore the
slight discrepancy in the 13- and 31- components which, in fact, are quite
inconsequential as we will see below when we discuss texture zeros.

The eigenvalues of $D_{0}$ \ are given by the eigenvalues of $D_{diag}$,
i.e. $(0,\epsilon ,1)$. As pointed out in the Introduction, to obtain the
correct values one must go beyond the standard model which we propose to do
in Section V by incorporating texture zeros.

\pagebreak \medskip \medskip

\textbf{V. Mass eigenvalues and CKM matrix in the quark sector with }

\ \ \ \textbf{texture zeros}

As stated in the Introduction the most acceptable structures with texture
zeros are the following

\begin{equation*}
U_{0}=\left[ 
\begin{array}{ccc}
0 & 0 & X \\ 
0 & X & 0 \\ 
X & 0 & X%
\end{array}%
\right] ,\ \ \ \ D_{0}=\left[ 
\begin{array}{ccc}
0 & X & 0 \\ 
X & X & X \\ 
0 & X & X%
\end{array}%
\right]
\end{equation*}%
In terms of our results in section IV we then have\ \ \ \ \ \ \ \ \ \ \ \ \
\ \ \ \ \ \ \ \ \ \ \ \ \ \ \ \ \ \ \ \ \ \ \ \ \ \ \ \ \ \ \ \ \ \ \ \ \ \
\ \ \ \ \ \ \ \ \ \ \ \ \ \ \ \ \ \ \ \ \ \ \ \ \ \ \ \ \ \ \ \ \ \ \ \ \ \
\ \ \ \ \ \ \ \ \ \ \ \ \ \ \ \ \ \ \ \ \ \ \ \ \ \ \ \ \ \ \ \ \ \ \ \ \ \
\ \ \ \ \ \ \ \ \ \ \ \ \ \ \ \ \ \ \ \ \ \ \ \ \ \ \ \ \ \ \ \ \ \ \ \ \ \
\ \ \ \ \ \ \ \ \ \ \ \ \ \ \ \ \ \ \ \ \ \ \ \ \ \ \ \ \ \ \ \ \ \ \ \ \ \
\ \ \ \ \ \ \ \ \ \ \ \ \ \ \ \ \ \ \ \ \ \ \ \ \ \ \ \ \ \ \ \ \ \ \ \ \ \
\ \ \ \ \ \ \ \ \ \ \ \ \ \ \ \ \ \ \ \ \ \ \ \ \ \ \ \ \ \ \ \ \ \ \ \ \ \
\ \ \ \ \ \ \ \ \ \ 

\begin{equation}
U_{0}=%
\begin{bmatrix}
0 & 0 & s_{2} \\ 
0 & s_{1}^{2} & 0 \\ 
s_{2} & 0 & 1%
\end{bmatrix}%
,\ \ D_{0}=%
\begin{bmatrix}
0 & \epsilon s_{3} & 0 \\ 
\epsilon s_{3} & s_{1}^{2}+\epsilon & s_{1} \\ 
0 & s_{1} & 1%
\end{bmatrix}
\tag{5.1}  \label{5.1}
\end{equation}%
\ \ \ \ \ \ \ \ \ \ \ \ \ \ \ \ \ \ \ \ \ \ \ \ \ \ \ \ \ \ \ \ \ \ \ \ \ \
\ \ \ \ \ \ \ \ \ \ \ \ \ \ \ \ \ \ \ \ \ \ \ \ \ \ \ \ \ \ \ \ \ \ \ \ \ \
\ \ \ \ \ 

\textbf{A. CKM matrix}

The matrix elements above involve the "input" CKM parameters $s_{i}$ already
defined (4.8)

\begin{equation}
V_{CKM}^{(in)}=%
\begin{bmatrix}
1 & s_{3} & s_{2} \\ 
-s_{3} & 1 & s_{1} \\ 
s_{1}s_{3}-s_{2} & -s_{1} & 1%
\end{bmatrix}
\tag{5.2}  \label{5.2}
\end{equation}

The numerical values of the parameters are known from experiments [22]and
are given by

\begin{equation}
s_{1}=\left( .038-.044\right) ,s_{2}=\left( .0025-.0048\right) ,s_{3}=\left(
.219-.226\ \right)  \tag{5.3}  \label{5.3}
\end{equation}%
\ \ \ \ \ \ \ \ \ \ \ \ \ \ \ \ \ \ \ \ \ \ \ \ \ \ \ \ \ \ \ \ \ \ \ \ \ \
\ \ \ \ \ \ \ \ \ \ \ \ \ \ \ \ \ \ \ \ \ \ \ \ \ \ \ \ \ \ \ \ \ \ \ \ \ \
\ \ \ \ \ \ \ \ \ \ \ \ \ \ \ \ \ \ \ \ \ \ \ \ \ \ \ \ \ \ \ \ \ \ \ \ \ \
\ \ \ \ \ \ \ \ \ \ \ \ \ \ \ \ \ \ \ \ \ \ \ \ \ \ \ \ \ \ \ \ \ \ \ \ \ \
\ \ \ \ \ \ \ \ \ \ \ \ \ \ \ \ \ \ \ \ \ \ \ \ \ \ \ \ \ \ \ \ \ \ \ \ \ \
\ \ \ \ \ \ \ \ \ \ \ \ \ \ \ \ \ \ \ \ \ \ \ \ \ \ \ \ \ \ \ \ \ \ \ \ \ \
\ \ \ \ \ \ \ \ \ \ \ \ \ \ \ \ \ \ \ \ \ \ \ \ \ \ \ \ \ \ \ \ \ \ \ \ \ \
\ \ \ \ \ \ \ \ \ \ \ \ \ \ \ \ \ \ \ \ \ \ \ \ \ \ \ \ \ \ \ \ \ \ \ 

The "output" CKM is given by the standard definition $\ \ \ \ \ \ \ \ \ \ \
\ \ \ \ \ \ \ \ \ \ \ \ \ \ \ \ \ \ \ \ \ \ \ \ \ \ \ \ \ \ \ \ \ \ \ \ \ \
\ \ \ \ \ \ \ \ \ \ \ \ \ \ \ \ \ \ \ \ \ \ \ \ $

\begin{equation}
V_{CKM}^{(out)}=V_{u}^{\dagger }V_{d}  \tag{5.4}  \label{5.4}
\end{equation}%
\ \ \ \ \ \ \ \ \ \ \ \ \ \ \ \ \ \ \ \ \ \ \ \ \ \ \ \ \ \ \ \ \ \ \ \ \ \
\ \ \ \ \ \ \ \ \ \ \ \ \ \ \ \ \ \ \ \ \ \ \ \ \ \ \ \ \ \ \ \ \ \ \ \ \ \
\ \ \ \ \ \ \ \ \ \ \ \ \ \ \ \ \ \ \ \ \ \ \ \ \ \ \ \ \ \ \ \ \ \ \ \ \ \
\ \ \ \ \ \ \ \ \ \ \ \ \ \ \ \ \ \ \ \ \ \ \ \ \ \ \ \ \ \ \ \ \ \ \ \ \ \
\ \ \ \ \ \ \ \ \ \ \ \ \ \ \ \ \ \ \ \ \ \ \ \ \ \ \ \ \ \ \ \ \ \ \ \ \ \
\ \ \ \ \ \ \ \ \ \ \ \ \ \ \ \ \ \ \ \ \ \ \ \ \ \ \ \ \ \ \ \ \ \ \ \ \ \
\ \ \ \ \ \ \ \ \ \ \ \ \ \ \ \ \ \ \ \ \ \ \ \ \ \ \ \ \ \ \ \ \ \ \ \ \ \
\ \ \ \ \ \ \ \ \ \ \ \ \ \ \ \ \ \ \ \ \ \ \ \ \ \ \ \ \ \ \ \ \ \ \ \ \ \
\ \ \ \ \ \ \ \ \ \ \ \ \ \ where $V_{d}$ is the unitary matrix which
diagonalizes $D_{0};$ and $V_{u}$ is the unitary matrix which diagonalizes $%
U_{0}$. \ In terms of the rotation angles it is given by (assuming $%
c_{i}^{\prime }\thickapprox 1)$

\begin{equation}
V_{CKM}^{(out)}=%
\begin{bmatrix}
1 & s_{3}^{\prime } & s_{2}^{\prime } \\ 
-s_{3}^{\prime } & 1 & s_{1}^{\prime } \\ 
s_{1}^{\prime }s_{3}^{\prime }-s_{2}^{\prime } & -s_{1}^{\prime } & 1%
\end{bmatrix}
\tag{5.5}  \label{5.5}
\end{equation}%
\ \ \ \ \ \ \ \ \ \ \ \ \ \ \ \ \ \ \ \ \ \ \ \ \ \ \ \ \ \ \ \ \ \ \ \ \ \
\ \ \ \ \ \ \ \ \ \ \ \ \ \ \ \ \ \ \ \ \ \ \ \ \ \ \ \ \ \ \ \ \ \ \ \ \ \
\ \ \ \ \ \ \ \ \ \ \ \ \ \ \ \ \ \ \ \ \ \ \ \ \ \ \ \ \ \ \ \ \ \ \ \ \ \
\ \ \ \ \ \ \ \ \ \ \ \ \ \ \ \ \ \ \ \ \ \ \ \ \ \ \ \ \ \ \ \ \ \ \ \ \ \
\ \ \ \ \ \ \ \ \ \ \ \ \ \ \ \ \ \ \ \ \ \ \ \ \ \ \ \ \ \ \ \ \ \ \ \ \ \
\ \ \ \ \ \ \ \ \ \ \ \ \ \ \ \ \ \ \ \ \ \ \ \ \ \ \ \ \ \ \ \ \ \ \ \ \ \
\ \ \ \ \ \ \ \ \ \ \ \ \ \ \ \ \ \ \ \ \ \ \ \ \ \ \ \ \ \ \ \ \ \ \ \ \ \
\ \ \ \ \ \ \ \ \ \ \ \ \ \ \ \ \ \ \ \ \ \ \ \ \ \ \ where primes are used
to distinguish from the "input" case.

It is now easy to observe from the expressions in (5.1)\textbf{\ }that the
rotation for diagonalizing the 23-sub matrix involves only $D_{0}$, and for
the 13-sub matrix it involves only $U_{0}$. \ Therefore, by the nature of
the location of the texture zeros a (complete) mismatch is achieved between
the rotation angles for $U_{0}$ and $D_{0}$ for these submatrices. We then
have

\begin{equation}
s_{1}^{\prime }=s_{1},\ s_{2}^{\prime }=s_{2}  \tag{5.6}  \label{5.6}
\end{equation}%
\ \ \ \ \ \ \ \ \ \ \ \ \ \ \ \ \ \ \ \ \ \ \ \ \ \ \ \ \ \ \ \ \ \ \ \ \ \
\ \ \ \ \ \ \ \ \ \ \ \ \ \ \ \ \ \ \ \ \ \ \ \ \ \ \ \ \ \ \ \ \ \ \ \ \ \
\ \ \ \ \ \ \ \ \ \ \ \ \ \ \ \ \ \ \ \ \ \ \ \ \ \ \ \ \ \ \ \ \ \ \ \ \ \
\ \ \ \ \ \ \ \ \ \ \ \ \ \ \ \ \ \ \ \ \ \ \ \ \ \ \ \ \ \ \ \ \ \ \ \ \ \
\ \ \ \ \ \ \ \ \ \ \ \ \ \ \ \ \ \ \ \ \ \ \ \ \ \ \ \ \ \ \ \ \ \ \ \ \ \
\ \ \ \ \ \ \ \ \ \ \ \ \ \ \ \ \ \ \ \ \ \ \ \ \ \ \ \ \ \ \ \ \ \ \ \ \ \
\ \ \ \ \ \ \ \ \ \ \ \ \ \ \ \ \ \ \ \ \ \ \ \ \ \ \ \ \ \ \ \ \ \ \ \ \ \
\ \ \ \ \ \ \ \ \ \ \ \ \ \ \ \ \ \ \ \ \ \ \ \ \ \ \ \ \ \ \ \ \ \ \ \ \ \
\ \ \ \ \ \ \ \ \ \ \ \ \ This mismatch continues for the 12-submatrix,
where the rotation only involves $D_{0}$ and we have

\begin{equation}
s_{3}^{\prime }=\dfrac{\epsilon s_{3}}{s_{1}^{2}+\epsilon }  \tag{5.7}
\label{5.7}
\end{equation}%
\ \ \ \ \ \ \ \ \ \ \ \ \ \ \ \ \ \ \ \ \ \ \ \ \ \ \ \ \ \ \ \ \ \ \ \ \ \
\ \ \ \ \ \ \ \ \ \ \ \ \ \ \ \ \ \ \ \ \ \ \ \ \ \ \ \ \ \ \ \ \ \ \ \ \ \
\ \ \ \ \ \ \ \ \ \ \ \ \ \ \ \ \ \ \ \ \ \ \ \ \ \ \ \ \ \ \ \ \ \ \ \ \ \
\ \ \ \ \ \ \ \ \ \ \ \ \ \ \ \ \ \ \ \ \ \ \ \ \ \ \ \ \ \ \ \ \ \ \ \ \ \
\ \ \ \ \ \ \ \ \ \ \ \ \ \ \ \ \ \ \ \ \ \ \ \ \ \ \ \ \ \ \ \ \ \ \ \ \ \
\ \ \ \ \ \ \ \ \ \ \ \ \ \ \ \ \ \ \ \ \ \ \ \ \ \ \ \ \ \ \ \ \ \ \ \ \ \
\ \ \ \ \ \ \ \ \ \ \ \ \ \ \ \ \ \ \ \ \ \ \ \ \ \ \ \ \ \ \ \ \ \ \ \ \ \
\ \ \ \ \ \ \ \ \ \ \ \ \ \ \ \ \ \ \ \ \ \ \ \ \ \ \ \ \ \ \ \ \ \ \ \ \ \
\ \ \ \ \ \ \ \ \ \ \ \ \ \ \ \ \ \ \ \ \ \ \ \ \ \ But as we discussed in
section IV the magnitude of $\epsilon $ was estimated to be $\thickapprox
10^{-2}$ which is much larger than $s_{1}^{2}\left( \thickapprox
.0016\right) .$Therefore,

\begin{equation}
s_{3}^{\prime }\thickapprox s_{3}  \tag{5.8}  \label{5.8}
\end{equation}%
\ \ \ \ \ \ \ \ \ \ \ \ \ \ \ \ \ \ \ \ \ \ \ \ \ \ \ \ \ \ \ \ \ \ \ \ \ \
\ \ \ \ \ \ \ \ \ \ \ \ \ \ \ \ \ \ \ \ \ \ \ \ \ \ \ \ \ \ \ \ \ \ \ \ \ \
\ \ \ \ \ \ \ \ \ \ \ \ \ \ \ \ \ \ \ \ \ \ \ \ \ \ \ \ \ \ \ \ \ \ \ \ \ \
\ \ \ \ \ \ \ \ \ \ \ \ \ \ \ \ \ \ \ \ \ \ \ \ \ \ \ \ \ \ \ \ \ \ \ \ \ \
\ \ \ \ \ \ \ \ \ \ \ \ \ \ \ \ \ \ \ \ \ \ \ \ \ \ \ \ \ \ \ \ \ \ \ \ \ \
\ \ \ \ \ \ \ \ \ \ \ \ \ \ \ \ \ \ \ \ \ \ \ \ \ \ \ \ \ \ \ \ \ \ \ \ \ \
\ \ \ \ \ \ \ \ \ \ \ \ \ \ \ \ \ \ \ \ \ \ \ \ \ \ \ \ \ \ \ \ \ \ \ \ \ \
\ \ \ \ \ \ \ \ \ \ \ \ \ \ \ \ \ \ \ \ \ \ \ \ \ \ \ \ \ \ \ \ \ \ \ \ \ \
\ \ \ \ \ \ \ \ \ \ \ \ \ \ \ \ \ \ \ \ Hence we conclude that

\begin{equation}
V_{CKM}^{(out)}=V_{CKM}^{(in)}  \tag{5.9}  \label{5.9}
\end{equation}%
\ \ \ \ \ \ \ \ \ \ \ \ \ \ \ \ \ \ \ \ \ \ \ \ \ \ \ \ \ \ \ \ \ \ \ \ \ \
\ \ \ \ \ \ \ \ \ \ \ \ \ \ \ \ \ \ \ \ \ \ \ \ \ \ \ \ \ \ \ \ \ \ \ \ \ \
\ \ \ \ \ \ \ \ \ \ \ \ \ \ \ \ \ \ \ \ \ \ \ \ \ \ \ \ \ \ \ \ \ \ \ \ \ \
\ \ \ \ \ \ \ \ \ \ \ \ \ \ \ \ \ \ \ \ \ \ \ \ \ \ \ \ \ \ \ \ \ 

\textbf{B. Mass eigenvlues}

Plugging in the following "input" values$\ \ $

\begin{equation}
s_{1}=.04,s_{2}=.004,s_{3}=.22\ \   \tag{5.10}  \label{5.10}
\end{equation}%
$\ \ \ \ \ \ \ \ \ \ \ \ \ \ \ \ \ \ \ \ \ \ \ \ \ \ \ \ \ \ \ \ \ \ \ \ \ \
\ \ \ \ \ \ \ \ \ \ \ \ \ \ \ \ \ \ \ \ \ \ \ \ \ \ \ \ \ \ \ \ \ \ \ \ \ \
\ \ \ \ \ \ \ \ \ \ \ \ \ \ \ \ \ \ \ \ \ \ \ \ \ \ \ \ \ \ \ \ \ \ \ \ \ \
\ \ \ \ \ \ \ \ \ \ \ \ \ \ \ \ \ \ \ \ \ \ \ \ \ \ \ \ \ \ \ \ \ \ \ \ \ \
\ \ \ \ \ \ \ \ \ \ \ \ \ \ \ \ \ \ \ \ \ \ \ \ \ \ \ \ \ \ \ \ \ \ \ \ \ \
\ \ \ \ \ \ \ \ \ \ \ \ \ \ \ \ \ \ \ \ \ \ \ \ \ \ \ \ \ \ \ \ \ \ \ \ \ \
\ \ \ \ \ \ \ \ \ \ \ \ \ \ \ \ \ \ \ \ \ \ \ \ \ \ \ \ \ \ \ \ \ \ \ \ \ \
\ \ \ \ \ \ \ \ \ \ \ $whichwhich which are within the range of the
experimental values given by (5.3) and taking $\ \epsilon $ $=.05$, we
obtain the following mass ratios, keeping in mind that the 33-matrix
elements $\lambda _{t}$ and $\lambda _{b}$ are normalized to their observed
mass values ("exp" means experimental values).

\begin{equation*}
\dfrac{m_{u}}{m_{t}}=4.\allowbreak 8\times 10^{-5}\ \left( \text{exp}%
\thickapprox 3\times 10^{-5}\right)
\end{equation*}

\begin{equation*}
\dfrac{m_{c}}{m_{t}}=\allowbreak 1.\,\allowbreak 9\times 10^{-3}\ \left( 
\text{exp}\thickapprox 8\times 10^{-3}\right)
\end{equation*}%
\ \ \ \ \ \ \ \ \ \ \ \ \ \ \ \ \ \ \ \ \ \ \ \ \ \ \ \ \ \ \ \ \ \ \ \ \ \
\ \ \ \ \ \ \ \ \ \ \ \ \ \ \ \ \ \ \ \ \ \ \ \ \ \ \ \ \ \ \ \ \ \ \ \ \ \
\ \ \ \ \ \ \ \ \ \ \ \ \ \ \ \ \ \ \ \ \ \ \ \ \ \ \ \ \ \ 

\begin{equation*}
\dfrac{m_{d}}{m_{b}}=2.4\times 10^{-3}\ \left( \text{exp}\thickapprox
2\times 10^{-3}\right)
\end{equation*}

\begin{equation}
\dfrac{m_{s}}{m_{b}}=5.\,\allowbreak 5\times 10^{-2}\ \left( \text{exp}%
\thickapprox 3\times 10^{-2}\right)  \tag{5.11}  \label{5.11}
\end{equation}%
\ \ \ \ \ \ \ \ \ \ \ \ \ \ \ \ \ \ \ \ \ \ \ \ \ \ \ \ \ \ \ \ \ \ \ \ \ \
\ \ \ \ \ \ \ \ \ \ \ \ \ \ \ \ \ \ \ \ \ \ \ \ \ \ \ \ \ \ \ \ \ \ \ \ \ \
\ \ \ \ \ \ \ \ \ \ \ \ \ \ \ \ \ \ \ \ \ \ \ \ \ \ \ \ \ \ \ \ \ \ \ \ \ \
\ \ \ \ \ \ \ \ \ \ \ \ \ \ \ \ \ \ \ \ \ \ \ \ \ \ \ \ \ \ \ \ \ \ \ \ \ \
\ \ \ \ \ \ \ \ \ \ \ \ \ \ \ \ \ \ \ \ \ \ \ \ \ \ \ \ \ \ \ \ \ \ \ \ \ \
\ \ \ \ \ \ \ \ \ \ \ \ \ \ \ \ \ \ \ \ \ \ \ \ \ \ \ \ \ \ \ \ \ \ \ \ \ \
\ \ \ \ \ \ \ \ \ \ \ \ \ \ \ \ \ \ \ \ \ \ \ \ \ \ \ \ \ \ \ \ \ \ \ \ \ \
\ \ \ \ \ \ \ \ \ \ \ \ \ \ \ \ \ \ \ \ \ \ \ \ \ \ \ 

The agreement with experiments is very good considering the fact that no
attempt was made to do a detailed numerical analysis to fit all the data.

\bigskip \medskip \medskip \medskip \medskip \medskip \medskip \medskip

\textbf{VI. Solutions to the RGEs in the lepton sector}

The RG equations we wish to solve for $N$ and $E$ are given by (2.6) and
(2.7). We notice that the last term in each equation involves a trace which
multiplies equally all the matrix elements of $N$ and $E$. Therefore, it
will effectively change the scale, though we realize that this is not
completely correct since there are also non-linear terms present in the
equation. We will ignore the last terms in (2.6) and (2.7), nevertheless, as
a simplifying assumption. We then have the following

\begin{equation}
-\dfrac{dE}{dx}=\frac{1}{12}\left[ 3EE^{+}+Tr(EE^{+})\mathbf{1}\right] E\ +%
\frac{1}{12}NN^{+}E\   \tag{6.1}  \label{6.1}
\end{equation}%
\ \ \ \ \ \ \ \ \ \ \ \ \ \ \ \ \ \ \ \ \ \ \ \ \ \ \ \ \ \ \ \ \ \ \ \ \ \
\ \ \ \ \ \ \ \ \ \ \ \ \ \ \ \ \ \ \ \ \ \ \ \ \ \ \ \ \ \ \ \ \ \ \ \ \ \
\ \ \ \ \ \ \ \ \ \ \ \ \ \ \ \ \ \ \ \ \ \ \ \ 

\begin{equation}
-\dfrac{dN}{dx}=\frac{1}{12}\left[ 3NN^{+}+Tr(NN^{+})\mathbf{1}\right] N\ +%
\frac{1}{12}EE^{+}N\   \tag{6.2}  \label{6.2}
\end{equation}%
\ \ \ \ \ \ \ \ \ \ \ \ \ \ \ \ \ \ \ \ \ \ \ \ \ \ \ \ \ \ \ \ \ \ \ \ \ \
\ \ \ \ \ \ \ \ \ \ \ \ \ \ \ \ \ \ \ \ \ \ \ \ \ \ \ \ \ \ \ \ \ \ \ \ \ \
\ \ \ \ \ \ \ \ \ \ \ \ \ \ \ \ \ \ \ \ \ \ \ \ \ \ \ \ \ \ \ \ \ \ \ \ \ \
\ \ \ \ \ \ \ \ \ \ \ \ \ \ \ \ \ \ \ \ \ \ \ \ \ \ \ \ \ \ \ \ \ \ \ \ \ \
\ \ \ \ \ \ \ \ \ \ \ \ \ \ \ \ \ \ \ \ \ \ \ \ \ \ \ \ \ \ \ \ \ \ \ \ \ \
\ \ \ \ \ \ \ \ \ \ \ \ \ \ \ \ \ \ \ \ \ \ \ \ \ \ \ \ \ \ \ \ \ \ \ \ \ \
\ \ \ \ \ \ \ \ \ \ \ \ \ \ \ \ \ \ \ \ \ \ \ \ \ \ \ \ \ \ \ \ \ \ \ \ \ \
\ \ \ \ \ \ \ \ \ \ \ \ \ \ \ \ \ In solving these equations we want to\
follow, as closely as possible, the analogy to $U$ and $D$, in particular to
identify $N$ with $U,$ and $E$ with $D$.

Considering first, as we did for $U$ and $D$, the uncoupled terms in the
square brackets. We find that they satisfy an equation very similar to
(3.4), except for the numerical factors.

\begin{equation}
-\dfrac{dL}{dx}=\frac{1}{12}\left[ 3LL^{\dagger }+Tr\left( LL^{\dagger
}\right) \mathbf{1}\right] L  \tag{6.3}  \label{6.3}
\end{equation}%
Following the same procedure as before, we write\ \ \ \ \ \ \ \ \ \ \ \ \ \
\ \ \ \ \ \ \ \ \ \ \ \ \ \ \ \ \ \ \ \ \ \ \ \ \ \ \ \ \ \ \ \ \ \ \ \ \ \
\ \ \ \ \ \ \ \ \ \ \ \ \ \ \ \ \ \ \ \ \ \ \ \ \ \ \ \ \ \ \ \ \ \ \ \ \ \
\ \ \ \ \ \ \ \ \ \ \ \ \ \ \ \ \ \ \ \ \ \ \ \ \ \ \ \ \ \ \ \ \ \ \ \ \ \
\ \ \ \ \ \ \ \ \ \ \ \ \ \ \ \ \ \ \ \ \ \ \ \ \ \ \ \ \ \ \ \ \ \ \ \ \ \
\ \ \ \ \ \ \ \ \ \ \ \ \ \ \ \ \ \ \ \ \ \ \ \ \ \ \ \ \ \ \ \ \ \ \ \ \ \
\ \ \ \ \ \ \ \ \ \ \ \ \ \ \ \ \ \ \ \ \ \ \ \ \ \ \ \ \ \ \ \ \ \ \ \ \ \
\ \ \ \ \ \ \ \ \ \ \ \ \ \ \ \ \ \ \ \ \ \ \ \ \ \ \ \ \ \ \ \ \ \ \ \ \ \
\ \ \ \ \ \ \ \ \ \ \ \ \ \ \ \ \ \ \ \ \ \ \ \ 

\begin{equation}
L=L_{0}\lambda _{l}\left( x\right)  \tag{6.4}  \label{6.4}
\end{equation}%
\ \ \ \ \ \ \ \ \ \ \ \ \ \ \ \ \ \ \ \ \ \ \ \ \ \ \ \ \ \ \ \ \ \ \ \ \ \
\ \ \ \ \ \ \ \ \ \ \ \ \ \ \ \ \ \ \ \ \ \ \ \ \ \ \ \ \ \ \ \ \ \ \ \ \ \
\ \ \ \ \ \ \ \ \ \ \ \ \ \ \ \ \ \ \ \ \ \ \ \ \ \ \ \ \ \ \ \ \ \ \ \ \ \
\ \ \ \ \ \ \ \ \ \ \ \ \ \ \ \ \ \ \ \ \ \ \ \ \ \ \ \ \ \ \ \ \ \ \ \ \ \
\ \ \ \ \ \ \ \ \ \ \ \ \ \ \ \ \ \ \ \ \ \ \ \ \ \ \ \ \ \ \ \ \ \ \ \ \ \
\ \ \ \ \ \ \ \ \ \ \ \ \ \ \ \ \ \ \ \ \ \ \ \ \ \ \ \ \ \ \ \ \ \ \ \ \ \
\ \ \ \ \ \ \ \ \ \ \ \ \ \ \ \ \ \ \ \ \ \ \ \ \ \ \ \ \ \ \ \ \ \ \ \ \ \
\ \ \ \ \ \ \ \ \ \ \ \ \ \ \ \ \ \ \ \ \ \ \ \ \ \ \ \ \ \ \ \ \ \ \ \ \ \
\ \ \ \ \ \ \ \ \ \ \ \ \ \ \ \ \ \ \ \ \ \ \ where $\lambda _{l}\left(
x\right) $ represents the dominant 33-matrix element, and,

\begin{equation}
L_{0}=\frac{1}{4}\left[ 3L_{0}L_{0}^{\dagger }+trL_{0}L_{0}^{\dagger }%
\mathbf{1}\right] L_{0}  \tag{6.5}  \label{6.5}
\end{equation}%
\ \ \ \ \ \ \ \ \ \ \ \ \ \ \ \ \ \ \ \ \ \ \ \ \ \ \ \ \ \ \ \ \ \ \ \ \ \
\ \ \ \ \ \ \ \ \ \ \ \ \ \ \ \ \ \ \ \ \ \ \ \ \ \ \ \ \ \ \ \ \ \ \ \ \ \
\ \ \ 

\begin{equation}
L_{diag}=V_{1}^{\dagger }L_{0}V_{2}  \tag{6.6}  \label{6.6}
\end{equation}%
\ \ \ \ \ \ \ \ \ \ \ \ \ \ \ \ \ \ \ \ \ \ \ \ \ \ \ \ \ \ \ \ \ \ \ \ \ \
\ \ \ \ \ \ \ \ \ \ \ \ \ \ \ \ \ \ \ \ \ \ \ \ \ \ \ \ \ \ \ \ \ \ \ \ \ \
\ \ \ \ \ \ \ \ \ \ \ \ \ \ \ \ \ \ \ \ \ \ \ \ \ \ \ \ \ \ \ \ \ \ \ \ \ \
\ \ \ \ \ \ \ \ \ \ \ \ \ \ \ \ \ \ \ \ \ \ \ \ \ \ \ \ \ \ \ \ \ \ \ \ \ \
\ \ \ \ \ \ \ \ \ \ \ \ \ \ \ \ \ \ \ \ \ \ \ \ \ \ \ \ \ \ \ \ \ \ \ \ \ \
\ \ \ \ \ \ \ \ \ \ \ \ \ \ \ \ \ \ \ \ \ \ \ \ \ \ \ \ \ \ \ \ \ \ \ \ \ \
\ \ \ \ \ \ \ \ \ \ \ \ \ \ \ \ \ \ \ \ \ \ \ \ \ \ \ \ \ \ \ \ \ \ \ \ \ \
\ \ \ \ \ \ \ \ \ \ \ 

\begin{equation}
L_{diag}=\frac{1}{4}\left[ 3L_{diag}^{2}+trL_{diag}^{2}\mathbf{1}\right]
L_{diag}  \tag{6.7}  \label{6.7}
\end{equation}%
\ \ \ \ \ \ \ \ \ \ \ \ \ \ \ \ \ \ \ \ \ \ \ \ \ \ \ \ \ \ \ \ \ \ \ \ \ \
\ \ \ \ \ \ \ \ \ \ \ \ \ \ \ \ \ \ \ \ \ \ \ \ \ \ \ \ \ \ \ \ \ \ \ \ \ \
\ \ \ \ \ \ \ \ \ \ \ \ \ \ \ \ \ \ \ \ \ \ \ \ \ \ \ \ \ \ \ \ \ \ \ \ \ \
\ \ \ \ \ \ \ \ \ \ \ \ \ \ \ \ \ \ \ \ \ \ \ \ \ \ \ \ \ \ \ \ \ \ \ \ \ \
\ \ \ \ \ \ \ \ \ \ \ \ \ \ \ \ \ \ \ \ \ \ \ \ \ \ \ \ \ \ \ \ \ \ \ \ \ \
\ \ \ \ \ \ \ \ \ \ \ \ \ \ \ \ \ \ \ \ \ \ \ \ \ \ \ \ \ \ \ \ \ \ \ \ \ \
\ \ \ \ \ \ \ \ \ \ \ \ \ \ \ \ \ \ \ \ \ \ \ \ \ \ \ \ \ \ \ \ \ \ \ \ \ \
\ \ \ \ \ \ \ \ \ \ \ \ \ \ \ \ \ \ \ \ \ \ \ \ \ \ \ \ \ \ \ \ \ \ \ \ \ \
\ \ \ \ \ \ \ \ \ \ \ \ \ \ \ \ \ \ \ \ \ \ \ \ \ \ \ \ \ \ \ \ \ \ \ \ \ \
\ \ \ \ \ \ \ \ \ \ \ \ \ \ \ \ \ \ \ \ \ \ \ \ \ \ \ \ \ \ \ \ \ \ \ \ \ \
\ \ \ \ \ \ \ \ \ \ \ \ \ \ \ \ \ \ \ \ \ \ \ \ \ \ \ \ \ \ \ \ \ \ \ \ \ \
\ \ \ \ \ \ \ \ \ \ \ \ \ \ \ \ \ \ \ \ \ \ \ \ \ \ \ \ \ \ \ \ \ \ \ \ \ \
\ \ \ \ \ \ \ \ \ \ \ \ \ \ \ \ \ \ \ \ \ \ \ \ \ \ \ \ \ \ \ \ \ \ \ \ \ \
\ \ \ \ \ \ \ \ \ \ \ \ \ \ \ \ \ \ \ \ \ \ \ \ \ \ \ \ \ \ \ \ \ \ \ \ \ \
\ \ \ \ \ \ \ \ \ \ \ \ \ \ \ \ \ \ \ \ \ \ \ \ \ \ \ \ \ \ \ \ \ \ $\ \ \ $%
\ 

We express the above relation in terms of the eigenvalues

\begin{equation}
L_{diag}=%
\begin{bmatrix}
\lambda _{1} & 0 & 0 \\ 
0 & \lambda _{2} & 0 \\ 
0 & 0 & \lambda _{3}%
\end{bmatrix}
\tag{6.8}  \label{6.8}
\end{equation}%
\ \ \ \ \ \ \ \ \ \ \ \ \ \ \ \ \ \ \ \ \ \ \ \ \ \ \ \ \ \ \ \ \ \ \ \ \ \
\ \ \ \ \ \ \ \ \ \ \ \ \ \ \ \ \ \ \ \ \ \ \ \ \ \ \ \ \ \ \ \ \ \ \ \ \ \ $%
\ \ \ \ \ \ \ \ \ \ \ \ \ \ \ \ \ \ \ \ \ \ \ \ \ \ \ \ \ \ \ \ \ \ \ \ \ \
\ \ \ \ \ $\ \ \ \ \ \ \ \ \ \ \ \ \ \ \ \ \ \ \ \ \ \ \ \ \ \ \ \ \ \ \ \ \
\ \ \ \ \ \ \ \ \ \ \ \ \ \ \ \ \ \ \ \ \ \ \ \ \ \ \ \ \ \ \ \ \ \ \ \ \ \
\ \ \ \ \ \ \ \ \ \ \ \ \ \ \ \ \ \ \ \ \ \ \ \ \ \ \ \ \ \ \ \ \ \ \ \ \ \
\ \ \ \ \ \ \ \ \ \ \ \ \ \ \ \ \ \ \ \ \ \ \ \ \ \ \ \ \ \ \ \ \ \ \ \ \ \
\ \ \ \ \ \ \ \ \ \ \ \ \ \ \ \ \ \ \ \ \ \ \ \ \ \ \ \ \ \ \ \ \ \ \ \ \ \
\ \ \ \ \ \ \ \ \ \ \ \ \ \ \ \ then we obtain the following relations

\begin{equation}
\lambda _{1}=\frac{1}{4}\left[ 4\lambda _{1}^{3}+\lambda _{1}\left( \lambda
_{2}^{2}+\lambda _{3}^{2}\right) \right]  \tag{6.9a}  \label{6,9a}
\end{equation}

\begin{equation}
\lambda _{2}=\frac{1}{4}\left[ 4\lambda _{2}^{3}+\lambda _{2}\left( \lambda
_{3}^{2}+\lambda _{1}^{2}\right) \right]  \tag{6.9b}  \label{6.9b}
\end{equation}

\begin{equation}
\lambda _{3}=\frac{1}{4}\left[ 4\lambda _{3}^{3}+\lambda _{3}\left( \lambda
_{1}^{2}+\lambda _{2}^{2}\right) \right]  \tag{6.9c}  \label{6.9c}
\end{equation}%
\ \ \ \ \ \ \ \ \ \ \ \ \ \ \ \ \ \ \ \ \ \ \ \ \ \ \ \ \ \ \ \ \ \ \ \ \ \
\ \ \ \ \ \ \ \ \ \ \ \ \ \ \ \ \ \ \ \ \ \ \ \ \ \ \ \ \ \ \ \ \ \ \ \ \ \
\ \ \ \ \ \ \ \ \ \ \ \ \ \ \ \ \ \ \ \ \ \ \ \ \ \ \ \ \ \ \ \ \ \ \ \ \ \
\ \ \ \ \ \ \ \ \ 

Once again we have only two non-trivial solutions.\ \ \ \ \ \ \ \ \ \ \ \ \
\ \ \ \ \ \ \ \ \ \ \ \ \ \ \ \ \ \ \ \ \ \ \ \ \ \ \ \ \ \ \ \ \ \ \ \ \ \
\ \ \ \ \ \ \ \ \ \ \ \ \ \ \ \ \ \ $\ $\ $\ \ \ \ \ \ \ \ \ \ \ \ \ \ \ \ \
\ \ \ \ \ \ \ \ \ \ \ \ \ \ \ \ \ \ \ \ $\ \ \ \ \ \ \ $\ \ \ \ $\ \ \ \ \ \
\ \ \ \ \ \ \ \ \ \ \ \ \ \ \ \ \ \ \ \ \ \ \ \ \ \ \ \ \ \ \ \ \ \ \ \ \ \
\ \ \ \ \ \ \ \ \ \ \ \ \ \ \ \ \ \ \ \ \ \ \ \ \ \ \ \ \ \ \ \ \ \ \ \ \ \
\ \ \ \ \ \ \ \ \ \ \ \ \ \ \ \ \ \ \ 

(i) "Hierarchical" solution with $\lambda _{1}=\lambda _{2}=0$, $\lambda
_{3}=1$

\begin{equation}
L_{diag}^{\left( 1\right) }=%
\begin{bmatrix}
0 & 0 & 0 \\ 
0 & 0 & 0 \\ 
0 & 0 & 1%
\end{bmatrix}
\tag{6.10}  \label{6.10}
\end{equation}%
\ \ \ \ \ \ \ \ \ \ \ \ \ \ \ \ \ \ \ \ \ \ \ \ \ \ \ \ \ \ \ \ \ \ \ \ \ \
\ \ \ \ \ \ \ \ \ \ \ \ \ \ \ \ \ \ \ \ \ \ \ \ \ \ \ \ \ \ \ \ \ \ \ \ \ \
\ \ \ \ \ \ \ \ \ \ \ \ \ \ \ \ \ \ \ \ \ \ \ \ \ \ \ \ \ \ \ \ \ \ \ \ \ \
\ \ \ \ \ \ \ \ \ \ \ \ \ \ \ \ \ \ \ \ $\ \ \ \ \ \ \ \ \ \ \ \ \ \ \ \ \ \
\ \ \ \ \ \ \ \ \ \ \ \ \ \ \ \ \ $

\begin{equation}
L_{0}^{\left( 1\right) }=V_{1}L_{diag}^{\left( 1\right) }V_{2}^{\dagger } 
\tag{6.11}  \label{6.11}
\end{equation}%
\ 

(ii) "Semi-hierarchical" solution with $\lambda _{1}=0,\lambda _{2}=\lambda
_{3}=\sqrt{\frac{4}{5}}$

\begin{equation}
L_{diag}^{\left( 2\right) }=\sqrt{\frac{4}{5}}%
\begin{bmatrix}
0 & 0 & 0 \\ 
0 & 1 & 0 \\ 
0 & 0 & 1%
\end{bmatrix}
\tag{6.12}  \label{6.12}
\end{equation}%
\ \ \ \ \ \ \ \ \ \ \ \ \ \ \ \ \ \ \ \ \ \ \ \ \ \ \ \ \ \ \ \ \ \ \ \ \ \
\ \ \ \ \ \ \ \ \ \ \ \ \ \ \ \ \ \ \ \ \ \ \ \ \ \ \ \ \ \ \ \ \ \ \ \ \ \
\ \ \ \ \ \ \ \ \ \ \ \ \ \ $\ \ \ \ \ \ \ \ \ \ \ \ \ \ \ \ \ \ \ \ \ \ \ \
\ \ \ \ \ \ \ \ \ \ \ \ \ \ \ \ \ \ $

\begin{equation}
L_{0}^{\left( 2\right) }=V_{1}L_{diag}^{\left( 2\right) }V_{2}^{\dagger } 
\tag{6.13}  \label{6.13}
\end{equation}%
\ 

We now make the following important identifications:

(i) As we mentioned earlier, we identify $N$ with $U$ so that we take it to
be of the hierchical type i.e.

\begin{equation}
N_{diag}=L_{diag}^{\left( 1\right) }=%
\begin{bmatrix}
0 & 0 & 0 \\ 
0 & 0 & 0 \\ 
0 & 0 & 1%
\end{bmatrix}
\tag{6.14}  \label{6.14}
\end{equation}%
\ \ \ \ \ \ \ \ \ \ \ \ \ \ \ \ \ \ \ \ \ \ \ \ \ \ \ \ \ \ \ \ \ \ \ \ \ \
\ \ \ \ \ \ \ \ \ \ \ \ \ \ \ \ \ \ \ \ \ \ \ \ \ \ \ \ \ \ \ \ \ \ \ \ \ \
\ \ \ \ \ \ \ \ \ \ \ \ \ \ \ \ \ \ \ \ \ \ \ \ \ \ \ \ \ \ \ \ \ \ \ \ \ \
\ \ \ \ \ \ \ \ \ \ \ \ \ \ \ \ \ \ \ \ \ \ \ \ \ \ \ \ \ \ \ \ \ \ \ \ \ \
\ \ \ \ \ \ \ \ \ \ \ \ \ \ \ \ \ \ \ \ \ \ \ \ \ \ \ \ \ \ \ \ \ \ \ \ \ \
\ \ \ \ \ \ \ \ \ \ \ \ \ \ \ \ \ \ \ \ \ \ \ \ \ \ \ \ \ \ \ \ \ \ \ \ \ \
\ \ \ \ \ \ \ \ \ 

(ii) We, therefore, take the 33-element of $N$ much larger than that of $E,$
which implies that

\begin{equation}
\lambda _{3v}>>\lambda _{\tau }  \tag{6.15}  \label{6.15}
\end{equation}%
\ \ \ \ \ \ \ \ \ \ \ \ \ \ \ \ \ \ \ \ \ \ \ \ \ \ \ \ \ \ \ \ \ \ \ \ \ \
\ \ \ \ \ \ \ \ \ \ \ \ \ \ \ \ \ \ \ \ \ \ \ \ \ \ \ \ \ \ \ \ \ \ \ \ \ \
\ \ \ \ \ \ \ \ \ \ \ \ \ \ \ \ \ \ \ \ \ \ \ \ \ \ \ \ \ \ \ \ \ \ \ \ \ \
\ \ \ \ \ \ \ \ \ \ \ \ \ \ \ \ \ \ \ \ \ \ \ \ \ \ \ \ \ \ \ \ \ \ \ \ \ \
\ \ \ \ \ \ 

(iii) We observe that unlike $U,D,E,$ which are entirely standard model
constituents, $N$ couples a standard model object, the left-handed neutrino, 
$\nu _{L},$ to a particle not in the standard model framework. This leads us
to identify $V_{1}$ as the small angle rotation matrix of the standard model
i.e. $V_{CKM},$as we did for $U$, $D$ and $E$ and we assume that $V_{2}$
represents the large-angle mixing matrix of the neutrino data i.e. the
analog of the CKM matrix \textbf{[10]}, we have called it $V_{large}.$
Therefore,

\begin{equation}
V_{1}=V_{CKM}  \tag{6.16}  \label{6.16}
\end{equation}

\begin{equation}
V_{2}=V_{large}  \tag{6.17}  \label{6.17}
\end{equation}%
\ \ \ \ \ \ \ \ \ \ \ \ \ \ \ \ \ \ \ \ \ \ \ \ \ \ \ \ \ \ \ \ \ \ \ \ \ \
\ \ \ \ \ \ \ \ \ \ \ \ \ \ \ \ \ \ \ \ \ \ \ \ \ \ \ \ \ \ \ \ \ \ \ \ \ \
\ \ \ \ \ \ \ \ \ \ \ \ \ \ \ \ \ \ \ \ \ \ \ \ \ \ \ \ \ \ \ \ \ \ \ \ \ \
\ \ \ \ \ \ \ \ \ \ \ \ \ \ \ \ \ \ \ \ \ \ \ \ \ \ \ \ \ \ \ \ \ \ \ \ \ \
\ \ \ \ \ \ \ \ \ \ \ \ \ \ \ \ \ \ \ \ \ \ \ \ \ \ \ \ \ \ \ \ \ \ \ \ \ \
\ \ \ \ \ \ \ \ \ \ \ \ \ \ \ \ \ \ \ \ \ \ \ \ \ \ \ \ \ \ \ \ \ \ \ \ \ \
\ \ \ \ \ \ \ \ \ \ \ \ \ \ \ \ \ \ \ \ \ \ \ \ \ \ \ \ \ \ \ \ \ \ \ \ \ \
\ \ \ \ \ \ \ \ \ \ \ \ \ \ \ \ \ \ \ \ \ \ \ \ \ \ \ \ \ \ \ \ \ \ \ \ \ \
\ \ \ \ Thus, from (6.11)

\begin{equation}
N_{0}=V_{CKM}L_{diag}^{\left( 1\right) }V_{large}^{\dagger }=V_{CKM}%
\begin{bmatrix}
0 & 0 & 0 \\ 
0 & 0 & 0 \\ 
0 & 0 & 1%
\end{bmatrix}%
V_{large}^{\dagger }  \tag{6.18}  \label{6.18}
\end{equation}%
\ \ \ \ \ \ \ \ \ \ \ \ \ \ \ \ \ \ \ \ \ \ \ \ \ \ \ \ \ \ \ \ \ \ \ \ \ \
\ \ \ \ \ \ \ \ \ \ \ \ \ \ \ \ \ \ \ \ \ \ \ \ \ \ \ \ \ \ \ \ \ \ \ \ \ \
\ \ \ \ \ \ \ 

\begin{equation}
N=N_{0}\lambda _{3v}(x)  \tag{6.19a}  \label{6.19a}
\end{equation}

We can write down the expression for $\lambda _{3v}$ in analogy to $U$
taking account of the differences in the numerical factors in the equation
for $U$ in (2.4) and for $N$ in (6.2)\textbf{,}

\begin{equation}
\lambda _{3v}(x)=\lambda _{03v}\left[ 1+x\lambda _{03v}^{2}\right] ^{-\frac{1%
}{3}}  \tag{6.19b}  \label{6.19b}
\end{equation}%
where, $\lambda _{03v}=$ $\lambda _{3v}(0).$

\bigskip

(iv) We identify $E$ with $D$, and write

\begin{equation}
E_{diag}=%
\begin{bmatrix}
0 & 0 & 0 \\ 
0 & \epsilon ^{\prime } & 0 \\ 
0 & 0 & 1%
\end{bmatrix}
\tag{6.20}  \label{6.20}
\end{equation}%
\ \ \ \ \ \ \ \ \ \ \ \ \ \ \ \ \ \ \ \ \ \ \ \ \ \ \ \ \ \ \ \ \ \ \ \ \ \
\ \ \ \ \ \ \ \ \ \ \ \ \ \ \ \ \ \ \ \ \ \ \ \ \ \ \ \ \ \ \ \ \ \ \ \ \ \
\ \ \ \ \ \ \ \ \ \ \ \ \ \ \ \ \ \ \ \ \ \ \ \ \ \ \ \ \ \ \ \ \ \ \ \ \ \
\ \ \ \ \ \ \ \ \ \ \ \ \ \ \ \ \ \ \ \ \ \ \ \ \ \ \ \ \ \ \ \ \ \ \ \ \ \
\ \ \ \ \ \ \ \ \ \ \ \ \ \ \ \ \ \ \ \ \ \ \ \ \ \ \ \ \ \ \ \ \ \ \ \ \ \
\ \ \ \ \ \ \ \ \ \ \ \ \ \ \ \ \ \ \ \ \ \ \ \ \ \ \ \ \ \ \ \ \ \ \ \ \ \
\ \ \ \ \ \ \ \ \ \ \ \ \ \ \ \ \ \ \ \ \ \ \ \ \ \ \ \ \ \ \ \ \ \ \ \ \ \
\ \ \ \ \ \ where $\epsilon ^{\prime }$,as in the case of $D$, is the mixing
parameter between the hierarchical and semi-hierarchical matrices. And, of
course, since it is a standard model particle, we take, as we did for $U_{0}$
and $D_{0},$

\begin{equation}
V_{1}=V_{2}=V_{CKM}\   \tag{6.21}  \label{6.21}
\end{equation}%
Therfore,\ \ \ \ \ \ \ \ \ \ \ \ \ \ \ \ \ \ \ \ \ \ \ \ \ \ \ \ \ \ \ \ \ \
\ \ \ \ \ \ \ \ \ \ \ \ \ \ \ \ \ \ \ \ \ \ \ \ \ \ \ \ \ \ \ \ \ \ \ \ \ \
\ \ \ \ \ \ \ \ \ \ \ \ \ \ \ \ \ \ \ \ \ \ \ \ \ \ \ \ \ \ \ \ \ \ \ \ \ \
\ \ \ \ \ \ \ \ \ \ \ \ \ \ \ \ \ \ \ \ \ \ \ \ \ \ \ \ \ \ \ \ \ \ \ \ \ \
\ \ \ \ \ \ \ \ \ \ \ \ \ \ \ \ \ \ \ \ \ \ \ \ \ \ \ \ \ \ \ \ \ \ \ \ \ \
\ \ \ \ \ \ \ \ \ \ \ \ \ \ \ \ \ \ \ \ \ \ \ \ \ \ \ \ \ \ \ \ \ \ \ \ \ \
\ \ \ \ \ \ \ \ \ \ \ \ \ \ \ \ \ \ \ \ \ \ \ \ \ \ \ \ \ \ \ \ \ \ \ \ \ \
\ \ \ \ \ \ \ \ \ \ \ \ \ \ \ \ \ \ \ \ \ \ \ \ \ \ \ \ \ \ \ \ 

\begin{equation}
E_{0}=V_{CKM}%
\begin{bmatrix}
0 & 0 & 0 \\ 
0 & \epsilon ^{\prime } & 0 \\ 
0 & 0 & 1%
\end{bmatrix}%
V_{CKM}^{\dagger }  \tag{6.22a}  \label{6.22a}
\end{equation}

\begin{equation}
E\ =E_{0}\lambda _{\tau }(x)  \tag{6.22b}  \label{6.22b}
\end{equation}%
\ \ \ \ \ \ \ \ \ \ \ \ \ \ \ \ \ \ \ \ \ \ \ \ \ \ \ \ \ \ \ \ \ \ \ \ \ \
\ \ \ \ \ \ \ \ \ \ \ \ \ \ \ \ \ \ \ \ \ \ \ \ \ \ \ \ \ \ \ \ \ \ \ \ \ \
\ \ \ \ \ \ \ \ \ \ \ \ \ \ \ \ \ \ \ \ \ \ \ \ \ \ \ \ \ \ \ \ \ \ \ \ 

Finally, for the Majorana neutrino mass matrix $\kappa $, described by the
seesaw mechanism, as given by (1.5) we have

\begin{equation}
\kappa =N^{T}\left[ M_{R}^{-1}\right] N  \tag{6.23}  \label{6.23}
\end{equation}%
\ \ \ \ \ \ \ \ \ \ \ \ \ \ \ \ \ \ \ \ \ \ \ \ \ \ \ \ \ \ \ \ \ \ \ \ \ \
\ \ \ \ \ \ \ \ \ \ \ \ \ \ \ \ \ \ \ \ \ \ \ \ \ \ \ \ \ \ \ \ \ \ \ \ \ \
\ \ \ \ \ \ \ \ \ \ \ \ \ \ \ \ \ \ \ \ \ \ \ \ \ \ \ \ \ \ \ \ \ \ \ \ \ \
\ \ \ \ \ \ \ \ \ \ \ \ \ \ \ \ \ \ \ \ \ \ \ \ \ \ \ \ \ \ \ \ \ \ \ \ \ \
\ \ \ \ \ \ \ \ \ \ \ \ \ \ \ \ \ \ \ \ \ \ \ \ \ \ \ \ \ \ \ \ \ \ \ \ \ \
\ \ \ \ \ \ \ \ \ \ \ \ \ \ \ \ \ \ \ \ \ \ \ \ \ \ \ \ \ \ \ \ \ \ \ \ \ \
\ \ \ \ \ \ \ \ \ \ \ \ \ \ \ \ \ \ \ \ \ \ \ \ \ \ \ \ \ \ \ \ \ \ \ \ \ \
\ \ \ \ \ \ \ \ \ \ \ \ \ \ \ \ \ \ \ \ \ \ \ \ \ \ \ \ \ \ \ \ \ \ \ \ \ \
\ \ \ \ \ where as mentioned in the Introduction, $\left[ M_{R}^{-1}\right] $
is the distribution of the reciprocal of the seesaw neutrino masses. We then
have from (1.5) and (6.18) and (6.19a) the following\ \ \ \ \ \ \ \ \ \ \ \
\ \ \ \ \ \ \ \ \ \ \ \ \ \ \ \ \ \ \ \ \ \ \ \ \ \ \ \ \ \ \ \ \ \ \ \ \ \
\ \ \ \ \ \ \ \ \ \ \ \ \ \ \ \ \ \ \ \ \ \ \ \ \ \ \ \ \ \ \ \ \ \ \ \ \ \
\ \ \ \ \ \ \ \ \ \ \ \ \ \ \ \ \ \ \ \ \ \ \ \ \ \ \ \ \ \ \ \ \ \ \ \ \ \
\ 

\begin{equation}
\kappa =\left[ V_{large}%
\begin{bmatrix}
0 & 0 & 0 \\ 
0 & 0 & 0 \\ 
0 & 0 & 1%
\end{bmatrix}%
V_{CKM}^{\dagger }\right] \left[ M_{R}^{-1}\right] \left[ V_{CKM}%
\begin{bmatrix}
0 & 0 & 0 \\ 
0 & 0 & 0 \\ 
0 & 0 & 1%
\end{bmatrix}%
V_{large}^{\dagger }\right] \lambda _{3\nu }^{2}  \tag{6.24}  \label{6.24}
\end{equation}%
We then obtain

\begin{equation}
\kappa \thickapprox \kappa _{0}\dfrac{\lambda _{3\nu }^{2}}{M_{33}} 
\tag{6.25}  \label{6.25}
\end{equation}%
\ \ \ \ \ \ \ \ \ \ \ \ \ \ \ \ \ \ \ \ \ \ \ \ \ \ \ \ \ \ \ \ \ \ \ \ \ \
\ \ \ \ \ \ \ \ \ \ \ \ \ \ \ \ \ \ \ \ \ \ \ \ \ \ \ \ \ \ \ \ \ \ \ \ \ \
\ \ \ \ \ \ \ \ \ \ \ \ \ \ \ \ \ \ \ \ \ \ \ \ \ \ \ \ \ \ \ \ \ \ \ \ \ \
\ \ \ \ \ \ \ \ \ \ \ \ \ \ \ \ \ \ \ \ \ \ \ \ \ \ \ \ \ \ \ \ \ 

\begin{equation}
\kappa _{0}=V_{large}%
\begin{bmatrix}
0 & 0 & 0 \\ 
0 & 0 & 0 \\ 
0 & 0 & 1%
\end{bmatrix}%
V_{large}^{\dagger }  \tag{6.26}  \label{6.26}
\end{equation}%
where $\left( \dfrac{1}{M_{33}}\right) $ is the 33-matrix element of $\left[
M_{R}^{-1}\right] ,$ and where we have used the CKM properties in the
product in (6.24), namely small\ \ \ \ \ \ \ \ \ \ \ \ \ \ \ \ \ \ \ \ \ \ \
\ \ \ \ \ \ \ \ \ \ \ \ \ \ \ \ \ \ \ \ \ \ \ \ \ \ \ \ \ \ \ \ \ \ \ \ \ \
\ \ \ \ \ \ \ \ \ \ \ \ \ \ \ \ \ \ \ \ \ \ \ \ \ \ \ \ \ \ \ \ \ \ \ \ \ \
\ \ \ \ \ \ \ \ \ \ \ \ \ \ \ \ \ \ \ \ \ \ \ \ \ \ \ \ \ \ \ \ \ \ \ \ \ \
\ \ \ \ \ \ \ \ \ \ \ \ \ \ \ \ \ \ \ \ \ \ \ \ \ \ \ \ \ \ \ \ \ \ \ \ \ \
\ \ \ \ \ \ \ \ \ \ \ \ \ \ \ \ \ \ \ \ \ \ \ \ \ \ \ \ \ \ \ \ \ \ \ \ \ \
\ \ \ \ \ \ \ \ \ \ \ \ \ \ \ \ \ \ \ \ \ \ \ \ \ \ \ \ \ \ \ \ \ \ \ \ \ \
\ \ \ \ \ \ \ \ \ \ \ \ \ \ angles, implying $c_{i}\thickapprox 1$. \ The
rotation matrix $V_{large}$ is the CKM-analog [10] given by

\begin{equation}
V_{large}=%
\begin{bmatrix}
C_{2}C_{3} & C_{2}S_{3} & S_{2} \\ 
-C_{1}S_{3}-S_{1}S_{2}C_{3} & C_{1}C_{3}-S_{1}S_{2}S_{3} & S_{1}C_{2} \\ 
S_{1}S_{3}-C_{1}S_{2}C_{3} & -S_{1}C_{3}-C_{1}S_{2}S_{3} & C_{1}C_{2}%
\end{bmatrix}
\tag{6.27}  \label{6.27}
\end{equation}%
where, in order to distinguish it from the CKM matrix, we have used capital
letters.\ \ \ \ \ \ \ \ \ \ \ \ \ \ \ \ \ \ \ \ \ \ \ \ \ \ \ \ \ \ \ \ \ \
\ \ \ \ \ \ \ \ \ \ \ \ \ \ \ \ \ \ \ \ \ \ \ \ \ \ \ \ \ \ \ \ \ \ \ \ \ \
\ \ \ \ \ \ \ \ \ \ \ \ \ \ \ \ \ \ \ \ \ \ \ \ \ \ \ \ \ \ \ \ \ \ \ \ \ \
\ \ \ \ \ \ \ \ \ \ \ \ \ \ \ \ \ \ \ \ \ \ \ \ \ \ \ \ \ \ \ \ \ \ \ \ \ \
\ \ \ \ \ \ \ \ \ \ \ \ \ \ \ \ \ \ \ \ \ \ \ 

Therefore, the neutrino mass matrix, $\kappa _{0}$ is given by

\begin{equation}
\kappa _{0}=%
\begin{bmatrix}
S_{2}^{2} & C_{2}S_{1}S_{2} & C_{1}C_{2}S_{2} \\ 
C_{2}S_{1}S_{2} & C_{2}^{2}S_{1}^{2} & C_{1}C_{2}^{2}S_{1} \\ 
C_{1}C_{2}S_{2} & C_{1}C_{2}^{2}S_{1} & C_{1}^{2}C_{2}^{2}%
\end{bmatrix}
\tag{6.28}  \label{6.28}
\end{equation}%
which is similar to $U$ except for the presence of large angles$\allowbreak $%
, and the seesaw\ contribution.\ \ \ \ \ \ \ \ \ \ \ \ \ \ \ \ \ \ \ \ \ \ \
\ \ \ \ \ \ \ \ \ \ \ \ \ \ \ \ \ \ \ \ \ \ \ \ \ \ \ \ \ \ \ \ \ \ \ \ \ \
\ \ \ \ \ \ \ \ \ \ \ \ \ \ \ \ \ \ \ \ \ \ \ \ \ \ \ \ \ \ \ \ \ \ \ \ \ \
\ \ \ \ \ \ \ \ \ \ \ \ \ \ \ \ \ \ \ \ \ \ \ \ \ \ \ \ \ \ \ \ \ \ \ \ \ \
\ \ \ \ \ \ \ \ \ \ \ \ \ \ \ \ \ \ \ \ \ \ \ \ \ \ \ \ \ \ \ \ \ \ \ \ \ \
\ \ \ \ \ \ \ \ \ \ \ \ \ \ \ \ \ \ \ \ \ \ \ \ \ \ \ \ \ 

Needless to say, the parameters of $V_{large}$ in (6.27) and (6.28) are what
we have defined as "input" parameters.\ \ \ \medskip \medskip \medskip\ \ \
\ \ \ \ \ \ \ \ \ \ \ \ \ \ \ \ \ \ \ \ \ \ \ \ \ \ \ \ \ \ \ \ \ \ \ \ \ \
\ \ \ \ \ \ \ \ 

\textbf{VII. Mass eigenvalues and mixing angles in the lepton sector with }

\ \ \ \ \ \ \textbf{texture zeros}

For the charged leptons, as we mentioned earlier, the structure of $E_{0}$
is assumed to be the same as that of $D_{0}$

\begin{equation}
E_{0}=\left[ 
\begin{array}{ccc}
0 & X & 0 \\ 
X & X & X \\ 
0 & X & X%
\end{array}%
\right]  \tag{7.1}  \label{7.1}
\end{equation}%
\ \ \ \ \ \ \ \ \ \ \ \ \ \ \ \ \ \ \ \ \ \ \ \ \ \ \ \ \ \ \ \ \ \ \ \ \ \
\ \ \ \ \ \ \ \ \ \ \ \ \ \ \ \ \ \ \ \ \ \ \ \ \ \ \ \ \ \ \ \ \ \ \ \ \ \
\ \ \ \ \ \ \ \ \ \ \ \ \ \ \ \ \ \ \ \ \ \ \ \ \ \ \ \ \ \ \ \ \ \ \ \ \ \
\ \ \ \ \ \ \ \ \ \ \ \ \ \ \ \ \ \ \ \ \ \ \ \ \ \ \ \ \ \ \ \ \ \ \ \ \ \
\ \ \ \ \ \ \ \ \ \ \ \ \ \ \ \ \ \ \ \ \ \ \ \ \ \ \ \ \ \ \ \ \ \ \ \ \ \
\ \ \ \ \ \ \ \ \ \ \ \ \ \ \ \ \ \ \ \ \ \ \ \ \ \ \ \ \ \ \ \ \ \ \ \ \ \
\ \ \ \ \ \ \ \ \ \ \ \ \ \ \ \ \ \ \ \ \ \ \ \ \ \ \ \ \ \ \ \ \ \ \ \ \ \
\ \ \ \ \ \ \ \ \ \ \ \ \ \ \ \ \ \ \ \ \ \ \ \ \ \ \ \ \ \ \ \ \ \ \ \ \ \
\ \ \ \ and for $\kappa _{0}$, as stated in the Introduction, we have two
types of structures[18].

\bigskip (a) A-type structure [18]

\begin{equation}
\kappa _{0}=\left[ 
\begin{array}{ccc}
0 & 0 & X \\ 
0 & X & X \\ 
X & X & X%
\end{array}%
\right] \ \ or\ \kappa _{0}=\left[ 
\begin{array}{ccc}
0 & X & 0 \\ 
X & X & X \\ 
0 & X & X%
\end{array}%
\right]  \tag{7.2}  \label{7.2}
\end{equation}%
\ \ \ \ \ \ \ \ \ \ \ \ \ \ \ \ \ \ \ \ \ \ \ \ \ \ \ \ \ \ \ \ \ \ \ \ \ \
\ \ \ \ \ \ \ \ \ \ \ \ \ \ \ \ \ \ \ \ \ \ \ \ \ \ \ \ \ \ \ \ \ \ \ \ \ \
\ \ \ \ \ \ \ \ \ \ \ \ \ \ \ \ \ \ \ \ \ \ \ \ \ \ \ \ \ \ \ \ \ \ \ \ \ \
\ \ \ \ \ \ \ \ \ \ \ \ \ \ \ \ \ \ \ \ \ \ \ \ \ \ \ \ \ \ \ \ \ \ \ \ \ \
\ \ \ \ \ \ \ \ \ \ \ \ \ \ \ \ \ \ \ \ \ \ \ \ \ \ \ \ \ \ \ \ \ \ \ \ \ \
\ \ \ \ \ \ \ \ \ \ \ \ \ \ \ \ \ \ \ \ \ \ \ \ \ \ \ \ \ \ \ \ \ \ \ \ \ \
\ \ \ \ \ \ \ \ \ \ \ \ \ \ \ \ \ \ \ \ \ \ \ \ \ \ \ \ \ \ \ \ \ \ \ \ \ \
\ \ \ \ \ \ \ \ \ \ These matrices give hierarchical neutrino mass values.
We will only consider the first matrix above since the second matrix gives
the same results [18,19].

(b) C-type structure [18]

\begin{equation}
\ \kappa _{0}=\left[ 
\begin{array}{ccc}
X & X & X \\ 
X & 0 & X \\ 
X & X & 0%
\end{array}%
\right]  \tag{7.3}  \label{7.3}
\end{equation}%
\ \ \ \ \ \ \ \ \ \ \ \ \ \ \ \ \ \ \ \ \ \ \ \ \ \ \ \ \ \ \ \ \ \ \ \ \ \
\ \ \ \ \ \ \ \ \ \ \ \ \ \ \ \ \ \ \ \ \ \ \ \ \ \ \ \ \ \ \ \ \ \ \ \ \ \
\ \ \ \ \ \ \ \ \ \ \ \ \ \ \ \ \ \ \ \ \ \ \ \ \ \ \ \ \ \ \ \ \ \ \ \ \ \
\ \ \ \ \ \ \ \ \ \ \ \ \ \ \ \ \ \ \ \ \ \ \ \ \ \ \ \ \ \ \ \ \ \ \ \ \ \
\ \ \ \ \ \ \ \ \ \ \ \ \ \ \ \ \ \ \ \ \ \ \ \ \ \ \ \ \ \ \ \ \ \ \ \ \ \
\ \ \ \ \ \ \ \ \ \ \ \ \ \ \ \ \ \ \ \ \ \ \ \ \ \ \ \ \ \ \ \ \ \ \ \ \ \
\ \ \ \ \ \ \ \ \ \ \ \ \ \ \ \ \ \ \ \ \ \ \ \ \ \ \ \ \ \ \ \ \ \ \ \ \ \
\ \ \ \ \ \ \ \ \ \ \ \ \ \ \ \ \ \ \ \ \ \ \ \ \ \ \ \ \ \ \ \ This gives
an inverted hierarchy [18,19].

In the following we will not consider the C-type structure for a simple
reason. Our entire construction of the Yukawa matrices has had at its basis
the "hierarchical" and "semi-hierarchical" primordial systems which clearly
can not be compatible with inverted hierarchy. A quick calculation confirms
this conclusion.

We will revisit the C-type structure in section VIII.

\textbf{A. Mass eigenvalues for charged leptons}

We simply follow the results from the D-matrix, and write

\begin{equation}
E_{0}=%
\begin{bmatrix}
0 & \epsilon ^{\prime }s_{3} & 0 \\ 
\epsilon ^{\prime }s_{3} & s_{1}^{2}+\epsilon ^{\prime } & s_{1} \\ 
0 & s_{1} & 1%
\end{bmatrix}
\tag{7.4}  \label{7.4}
\end{equation}%
\ \ \ \ \ \ \ \ \ \ \ \ \ \ \ \ \ \ \ \ \ \ \ \ \ \ \ \ \ \ \ \ \ \ \ \ \ \
\ \ \ \ \ \ \ \ \ \ \ \ \ \ \ \ \ \ \ \ \ \ \ \ \ \ \ \ \ \ \ \ \ \ \ \ \ \
\ \ \ \ \ \ \ \ \ \ \ \ \ \ \ \ \ \ \ \ \ \ \ \ \ \ \ \ \ \ \ \ \ \ \ \ \ \
\ \ \ \ \ \ \ \ \ \ \ \ \ \ \ \ \ \ \ \ \ \ \ \ \ \ \ \ \ \ \ \ \ \ \ \ \ \
\ \ \ \ \ \ \ \ \ \ \ \ \ \ \ \ \ \ \ \ \ \ \ \ \ \ \ \ \ \ \ \ \ \ \ \ \ \
\ \ \ \ \ \ \ \ \ \ \ \ \ \ \ \ \ \ \ \ \ \ \ \ \ \ \ \ \ \ \ \ \ \ \ \ \ \
\ \ \ \ \ \ \ \ \ \ \ \ \ \ \ \ \ \ \ \ \ \ \ \ \ \ \ \ \ \ \ \ \ \ \ \ \ \
\ \ \ \ \ \ \ \ \ \ \ \ \ \ \ \ \ \ \ \ \ \ \ \ \ \ \ \ \ \ \ \ \ \ \ \
where the angles $s_{i}$ are the usual "input" CKM angles, and $\epsilon
^{\prime }$ is the analog of $\epsilon $ in $D_{0}$ which is the mixing
parameter between the "hierarchical" and "semi-hierarchical"
representations. For the "input" CKM values given in (5.10) and \ $\epsilon
^{\prime }=.02,$ we find\ \ \ \ \ \ \ \ \ \ \ \ \ \ \ \ \ \ \ \ \ \ \ \ \ \
\ \ \ \ \ \ \ \ \ \ \ \ \ \ \ \ \ \ \ \ \ \ \ \ \ \ \ \ \ \ \ \ \ \ \ \ \ \
\ \ \ \ \ \ \ \ \ \ \ \ \ \ \ \ \ \ \ \ \ \ \ \ \ \ \ \ \ \ \ \ \ \ \ \ \ \
\ \ \ \ \ \ \ \ \ \ \ \ \ \ \ \ \ \ \ \ \ \ \ \ \ \ \ \ \ \ \ \ \ \ \ \ \ \
\ \ \ \ \ 

\begin{equation}
\dfrac{m_{\mu }}{m_{\tau }}=2.\,\allowbreak 1\times 10^{-2}\allowbreak \
\left( \text{exp}\thickapprox 6\times 10^{-2}\right)  \tag{7.5a}
\label{7.5a}
\end{equation}

\begin{equation}
\dfrac{m_{e}}{m_{\tau }}=9.\,\allowbreak 2\times 10^{-4}\ \ \left( \text{exp}%
\allowbreak \thickapprox 3\times 10^{-4}\right)  \tag{7.5b}  \label{7.5b}
\end{equation}%
\ \ \ \ \ \ \ \ \ \ \ \ \ \ \ \ \ \ \ \ \ \ \ \ \ \ \ \ \ \ \ \ \ \ \ \ \ \
\ \ \ \ \ \ \ \ \ \ \ \ \ \ \ \ \ \ \ \ \ \ \ \ \ \ \ \ \ \ \ \ \ \ \ \ \ \
\ \ \ \ \ \ \ \ \ \ \ \ \ \ \ \ \ \ \ \ \ \ \ \ \ \ \ \ \ \ \ \ \ \ \ \ \ \
\ \ \ \ \ \ \ \ \ \ \ \ \ \ \ \ \ \ \ \ \ \ \ \ \ \ \ \ \ \ \ \ \ \ \ \ \ \
\ \ \ \ \ \ \ \ \ \ \ \ \ \ \ \ \ \ \ \ \ \ \ \ \ \ \ \ \ \ \ \ \ \ \ \ \ \
\ \ \ \ \ \ \ \ \ \ \ \ \ \ \ \ \ \ \ \ \ \ \ \ \ \ \ \ \ \ \ \ \ \ \ \ \ \
\ \ \ \ \ \ \ \ \ \ \ \ \ \ \ \ \ \ \ \ \ \ \ \ \ \ \ \ \ \ \ \ \ \ \ \ \ \
\ \ \ \ \ \ \ \ \ \ \ \ \ \ \ \ \ \ \ \ \ \ \ \ \ where "exp" means
experimental values [22]. \ Once again, there is a good agreement with
experiments.

\textbf{B. Mixing angles and mass eigenvalues for Majorana neutrinos}

The A-type Majorana matrix, $\kappa _{0}$, has texture zeros as given by
[18,19]$.$

\begin{equation}
\kappa _{0}=%
\begin{bmatrix}
0 & 0 & C_{1}C_{2}S_{2} \\ 
0 & C_{2}^{2}S_{1}^{2} & C_{1}C_{2}^{2}S_{1} \\ 
C_{1}C_{2}S_{2} & C_{1}C_{2}^{2}S_{1} & C_{1}^{2}C_{2}^{2}%
\end{bmatrix}
\tag{7.6}  \label{7.6}
\end{equation}%
where $C_{i}$ and $S_{i}$ are the parameters of the "input" CKM-analog,
already defined in (6.27)\ \ \ \ \ \ \ \ \ \ \ \ \ \ \ \ \ \ \ \ \ \ \ \ \ \
\ \ \ \ \ \ \ \ \ \ \ \ \ \ \ \ \ \ \ \ \ \ \ \ \ \ \ \ \ \ \ \ \ \ \ \ \ \
\ \ \ \ \ \ \ \ \ \ \ \ \ \ \ \ \ \ \ \ \ \ \ \ \ \ \ \ \ \ \ \ \ \ \ \ \ \
\ \ \ \ \ \ \ \ \ \ \ \ \ \ \ \ \ \ \ \ \ \ \ \ \ \ \ \ \ \ \ \ \ \ \ \ \ \
\ \ \ \ \ \ \ \ \ \ \ \ \ \ \ \ \ \ \ \ \ \ \ \ \ \ \ \ \ \ \ \ \ \ \ \ \ \
\ \ \ \ \ \ \ \ \ \ \ \ \ \ \ \ \ \ \ \ \ \ \ \ \ \ \ \ \ \ \ \ \ \ \ \ \ \
\ \ \ \ \ \ \ \ \ \ \ \ \ \ \ \ \ \ \ \ \ \ \ \ \ \ \ \ \ \ \ \ \ \ \ \ \ \
\ \ \ \ \ \ \ \ \ \ 

\begin{equation}
V_{large}^{(in)}=%
\begin{bmatrix}
C_{2}C_{3} & C_{2}S_{3} & S_{2} \\ 
-C_{1}S_{3}-S_{1}S_{2}C_{3} & C_{1}C_{3}-S_{1}S_{2}S_{3} & S_{1}C_{2} \\ 
S_{1}S_{3}-C_{1}S_{2}C_{3} & -S_{1}C_{3}-C_{1}S_{2}S_{3} & C_{1}C_{2}%
\end{bmatrix}
\tag{7.7}  \label{7.7}
\end{equation}%
This matrix is defined in the basis when $E_{0}$ is diagonal. The
experimental values of the parameters are [3,4,5,6,7,8]\ \ \ \ \ \ \ \ \ \ \
\ \ \ \ \ \ \ \ \ \ \ \ \ \ \ \ \ \ \ \ \ \ \ \ \ \ \ \ \ \ \ \ \ \ \ \ \ \
\ \ \ \ \ \ \ \ \ \ \ \ \ \ \ \ \ \ \ \ \ \ \ \ \ \ \ \ \ \ \ \ \ \ \ \ \ \
\ \ \ \ \ \ \ \ \ \ \ \ \ \ \ \ \ \ \ \ \ \ \ \ \ \ \ \ \ \ \ \ \ \ \ \ \ \
\ \ \ \ \ \ \ \ \ \ \ \ \ \ \ \ \ \ \ \ \ \ \ \ \ \ \ \ \ \ \ \ \ \ \ \ \ \
\ \ \ \ \ \ \ \ \ \ \ \ \ \ \ \ \ \ \ \ \ \ \ \ \ \ \ \ \ \ \ \ \ \ \ \ \ \
\ \ \ \ \ \ \ \ \ \ \ \ \ \ \ \ \ \ \ \ \ \ \ \ \ \ \ \ \ \ \ \ \ \ \ \ \ \
\ \ \ \ \ \ \ \ \ \ \ \ \ \ \ \ \ \ \ \ \ \ 

\begin{equation}
S_{1}=\left( .54-.83\right) ;S_{3}=\left( .40-.70\right) ;S_{2}\leq .16 
\tag{7.8}  \label{7.8}
\end{equation}%
\ \ \ \ \ \ \ \ \ \ \ \ \ \ \ \ \ \ \ \ \ \ \ \ \ \ \ \ \ \ \ \ \ \ \ \ \ \
\ \ \ \ \ \ \ \ \ \ \ \ \ \ \ \ \ \ \ \ \ \ \ \ \ \ \ \ \ \ \ \ \ \ \ \ \ \
\ \ \ \ \ \ \ \ \ \ \ \ \ \ \ \ \ \ \ 

The matrix that will diagonalize $\kappa _{0}$ in (6.28) is

\begin{equation}
\begin{bmatrix}
C_{2}^{\prime }C_{3}^{\prime } & C_{2}^{\prime }S_{3}^{\prime } & 
S_{2}^{\prime } \\ 
-C_{1}^{\prime }S_{3}^{\prime }-S_{1}^{\prime }S_{2}^{\prime }C_{3}^{\prime }
& C_{1}^{\prime }C_{3}^{\prime }-S_{1}^{\prime }S_{2}^{\prime }S_{3}^{\prime
} & S_{1}^{\prime }C_{2}^{\prime } \\ 
S_{1}^{\prime }S_{3}^{\prime }-C_{1}^{\prime }S_{2}^{\prime }C_{3}^{\prime }
& -S_{1}^{\prime }C_{3}^{\prime }-C_{1}^{\prime }S_{2}^{\prime
}S_{3}^{\prime } & C_{1}^{\prime }C_{2}^{\prime }%
\end{bmatrix}
\tag{7.9}  \label{7.9}
\end{equation}

However, this matrix is not exactly the CKM-analog [10], $V_{large}^{(out)}.$%
We need to include rotations that diagonalize $E_{0}$ as well, since the
CKM-analog is defined in the mass basis of $E_{0}$. \ We note, however, that
the rotation angles to diagonalize the 2-3 and 1-3 sub matrices of $E_{0}$
are very small and can be ignored. We will only consider the 1-2 submatrix
in $E_{0}$ since, here, for $\epsilon ^{\prime }$ of the order $.02$, needed
to give the correct charged lepton masses, as discussed earlier, the $s_{3}$
parameter turns out to be $\thickapprox .22$ [22]. \ Thus we define

\begin{equation}
V_{large}^{(out)}=\ 
\begin{bmatrix}
C_{2}^{\prime }C_{3}^{\prime } & C_{2}^{\prime }S_{3}^{\prime } & 
S_{2}^{\prime } \\ 
-C_{1}^{\prime }S_{3}^{\prime }-S_{1}^{\prime }S_{2}^{\prime }C_{3}^{\prime }
& C_{1}^{\prime }C_{3}^{\prime }-S_{1}^{\prime }S_{2}^{\prime }S_{3}^{\prime
} & S_{1}^{\prime }C_{2}^{\prime } \\ 
S_{1}^{\prime }S_{3}^{\prime }-C_{1}^{\prime }S_{2}^{\prime }C_{3}^{\prime }
& -S_{1}^{\prime }C_{3}^{\prime }-C_{1}^{\prime }S_{2}^{\prime
}S_{3}^{\prime } & C_{1}^{\prime }C_{2}^{\prime }%
\end{bmatrix}%
\begin{bmatrix}
c_{3} & -s_{3} & 0 \\ 
s_{3} & c_{3} & 0 \\ 
0 & 0 & 1%
\end{bmatrix}
\tag{7.10}  \label{7.10}
\end{equation}

The effect of the second bracket above is to change only the angle $\theta
_{3}^{\prime },$ by $\theta _{3E},$where $sin\theta _{3E}=s_{3}$ $\left(
\thickapprox .22\right) $. \ Thus

\begin{equation}
\theta _{1}^{(out)}=\theta _{1}^{\prime },\text{ }\theta _{2}^{(out)}=\theta
_{2}^{\prime },\text{ }\theta _{3}^{(out)}=\theta _{3}^{\prime }-\theta _{3E}
\tag{7.11}  \label{7.11}
\end{equation}%
\ \ \ \ \ \ \ \ \ \ \ \ \ \ \ \ \ \ \ \ \ \ \ \ \ \ \ \ \ \ \ \ \ \ \ \ \ \
\ \ \ \ \ \ \ \ \ \ \ \ \ \ \ \ \ \ \ \ \ \ \ \ \ \ \ \ \ \ \ \ \ \ \ \ \ \
\ \ \ \ \ \ \ \ \ \ \ \ \ \ \ \ \ \ \ \ \ \ 

First of all a simple examination of the structure of the 23-sub matrix of $%
\kappa _{0}$ in (7.6) shows that the angle of rotation, $\theta _{1}^{\prime
},$ that would diagonalize $\kappa _{0}$ is precisely $\theta _{1}$itself,

\begin{equation}
S_{1}^{\prime }=S_{1}  \tag{7.12}  \label{7.12}
\end{equation}%
\ \ \ \ \ \ \ \ \ \ \ \ \ \ \ \ \ \ \ \ \ \ \ \ \ \ \ \ \ \ \ \ \ \ \ \ \ \
\ \ \ \ \ \ \ \ \ \ \ \ \ \ \ \ \ \ \ \ \ \ \ \ \ \ \ \ \ \ \ \ \ \ \ \ \ \
\ \ \ \ \ \ \ \ \ \ \ \ \ \ \ \ \ \ \ \ \ \ \ \ \ \ \ \ \ \ \ \ \ \ \ \ \ \
\ \ \ \ \ \ \ \ \ \ \ \ \ \ \ \ \ \ \ \ \ \ \ \ \ \ \ \ \ \ \ \ \ \ \ \ \ \
\ \ \ \ \ \ \ \ \ \ \ \ \ \ \ \ \ \ \ \ \ \ \ \ \ \ \ \ \ \ \ \ \ \ \ \ \ \
\ \ \ \ \ \ \ \ \ \ \ \ \ \ \ \ \ \ \ \ \ \ \ \ \ \ \ \ \ \ \ \ \ \ \ \ \ \
\ \ \ \ \ \ \ \ \ \ \ \ \ \ \ \ \ \ \ \ \ \ \ \ \ \ \ \ \ \ \ \ \ \ \ \ \ \
\ \ \ \ \ \ \ \ \ \ \ \ \ \ \ \ \ \ \ \ \ \ \ \ \ \ \ \ \ \ \ \ \ \ \ \ \ \
\ \ \ \ \ \ \ \ \ \ \ \ \ \ \ \ \ \ \ \ \ \ \ \ \ \ \ \ \ \ \ \ \ \ \ \ \ \
\ \ \ \ \ \ \ \ \ \ \ \ \ \ \ \ \ \ \ \ \ \ \ \ \ \ \ \ \ \ \ \ \ \ \ \ \ \
\ \ \ \ \ \ \ \ \ \ \ \ \ \ \ \ \ \ \ \ \ \ \ \ \ \ \ \ \ \ \ \ \ \ \ \ \ \
\ \ \ \ \ \ \ \ \ \ \ \ \ \ \ \ \ \ \ \ \ \ \ \ \ \ \ \ \ \ \ \ \ \ \ \ \ \
The actual value we take for $S_{1}$ will be determined when we calculate
the neutrino masses.

For the other two angles we find

\begin{equation}
S_{2}^{\prime }\thickapprox C_{1}^{2}S_{2}  \tag{7.13}  \label{7.13}
\end{equation}%
where we take $C_{2}=C_{2}^{\prime }=1$ since $S_{2}$ is small.\ And for $%
S_{3}^{\prime },$we obtain the following equation\ \ \ \ \ \ \ \ \ \ \ \ \ \
\ \ \ \ \ \ \ \ \ \ \ \ \ \ \ \ \ \ \ \ \ \ \ \ \ \ \ \ \ \ \ \ \ \ \ \ \ \
\ \ \ \ \ \ \ \ \ \ \ \ \ \ \ \ \ \ \ \ \ \ \ \ \ \ \ \ \ \ \ \ \ \ \ \ \ \
\ \ \ \ \ \ \ \ \ \ \ \ \ \ \ \ \ \ \ \ \ \ \ \ \ \ \ \ \ \ \ \ \ \ \ \ \ \
\ \ \ \ \ \ \ \ \ \ \ \ \ \ \ \ \ \ \ \ \ \ \ \ \ \ \ \ \ \ \ \ \ \ \ \ \ \
\ \ \ \ \ \ \ \ \ \ \ \ \ \ \ \ \ \ \ \ \ \ \ \ \ \ \ \ \ 

\begin{equation}
T_{3}^{\prime 2}-S_{2}\left( \frac{C_{1}^{3}}{S_{1}}\right) T_{3}^{\prime
}-1=0  \tag{7.14}  \label{7.14}
\end{equation}%
\ \ \ \ \ \ \ \ \ \ \ \ \ \ \ \ \ \ \ \ \ \ \ \ \ \ \ \ \ \ \ \ \ \ \ \ \ \
\ \ \ \ \ \ \ \ \ \ \ \ \ \ \ \ \ \ \ \ \ \ \ \ \ \ \ \ \ \ \ \ \ \ \ \ \ \
\ \ \ \ \ \ \ \ \ \ \ \ \ \ \ \ \ \ \ \ \ \ \ \ \ \ \ \ \ \ \ \ \ \ \ \ \ \
\ \ \ \ \ \ \ \ \ \ \ \ \ \ \ \ \ \ \ \ \ \ \ \ \ \ \ \ \ \ \ \ \ \ \ \ \ \
\ \ \ \ \ \ \ \ \ \ \ \ \ \ \ \ \ \ \ \ \ \ \ \ \ \ \ \ \ \ \ \ \ \ \ \ \ \
\ \ \ \ \ \ \ \ \ \ \ \ \ \ \ \ \ \ \ \ \ \ \ \ \ \ \ \ \ \ \ \ \ \ \ \ \ \
\ \ \ \ \ \ \ \ \ \ \ \ \ \ \ \ \ \ \ \ \ \ \ \ \ \ \ \ \ \ \ \ \ \ \ \ \ \
\ \ \ \ \ \ \ \ \ \ \ \ \ \ \ \ \ \ \ \ \ \ \ \ \ \ \ \ \ \ \ \ \ \ \ \
where $T_{3}^{\prime }=$ $tan\theta _{3}^{\prime }$ $=\tfrac{S_{3}^{\prime }%
}{C_{3}^{\prime }}.$\ \ 

The diagonalized form for the full mass matrix $\kappa $, defined in terms
of $\kappa _{0}$ in (6.25) and (7.6)\textbf{,} is found to be the following
(again we take $C_{2}=C_{2}^{\prime }=1$)

\begin{equation}
\dfrac{\lambda _{03\nu }^{2}}{M_{33}}\left[ 
\begin{array}{ccc}
\left( \dfrac{C_{1}S_{1}S_{2}}{T_{3}^{\prime }}\right) & 0 & 0 \\ 
0 & -T_{3}^{\prime }C_{1}S_{1}S_{2} & 0 \\ 
0 & 0 & 1%
\end{array}%
\right]  \tag{7.15}  \label{7.15}
\end{equation}%
\ \ \ \ \ \ \ \ \ \ \ \ \ \ \ \ \ \ \ \ \ \ \ \ \ \ \ \ \ \ \ \ \ \ \ \ \ \
\ \ \ \ \ \ \ \ \ \ \ \ \ \ \ \ \ \ \ \ \ \ \ \ \ \ \ \ \ \ \ \ \ \ \ \ \ \
\ \ \ \ \ \ \ \ \ \ \ \ \ \ \ \ \ \ \ \ \ \ \ \ \ \ \ \ \ \ \ \ \ \ \ \ \ \
\ \ \ \ \ \ \ \ \ \ \ \ \ \ \ \ \ \ \ \ \ \ \ \ \ \ \ \ \ \ \ \ \ \ \ \ \ \
\ \ \ \ \ \ \ \ \ \ \ \ \ \ \ \ \ \ \ \ \ \ \ \ \ \ \ \ \ \ \ \ \ \ \ \ \ \
\ \ \ \ \ \ \ \ \ \ \ \ \ \ \ \ \ \ \ \ \ \ \ \ \ \ \ \ \ \ \ \ \ \ \ \ \ \
\ \ \ \ \ \ \ \ \ \ \ \ \ \ \ \ \ \ \ \ \ \ \ \ \ \ \ \ \ \ \ \ \ \ \ \ \ \
\ \ \ \ \ \ \ \ \ \ \ \ \ \ \ \ \ \ \ \ \ \ \ \ \ \ \ \ \ \ \ \ \ \ \ \ \ \
\ \ \ \ \ \ \ \ \ \ \ \ \ \ \ \ \ \ \ \ \ \ \ \ \ \ \ \ \ \ \ \ \ \ \ \ \ \
\ \ \ \ \ \ \ \ \ \ \ \ \ \ \ \ \ \ \ \ \ \ \ \ \ \ \ \ \ \ \ \ \ \ \ \ \ \
\ \ \ \ \ \ \ \ \ \ \ \ \ \ \ \ \ \ \ \ \ \ \ \ \ \ \ \ \ \ \ \ \ \ \ \ \ \
\ \ \ \ \ \ \ \ \ \ \ \ \ \ \ \ \ \ \ \ \ \ \ \ \ \ \ \ \ \ \ \ \ \ \ \ \ \
\ \ \ \ \ \ \ \ \ \ \ \ \ \ \ \ \ \ \ \ \ \ \ \ \ \ \ \ \ \ \ \ \ \ \ \ \ \
\ \ \ \ \ \ \ \ \ \ \ \ \ \ \ \ \ \ \ \ \ \ \ \ \ \ \ \ \ \ \ \ \ \ \ \ \ \
\ \ \ \ \ \ \ \ \ \ \ \ \ \ \ \ \ \ \ \ \ \ \ \ \ \ \ \ \ \ \ \ \ \ \ \ \ \
\ \ \ \ \ \ \ \ \ \ \ \ \ \ \ \ \ \ \ \ \ \ \ \ \ \ \ \ \ \ \ \ \ \ \ \ \ \
\ \ \ \ \ \ \ \ \ \ \ \ \ \ \ \ \ 

We now compare the above mass values with the neutrino oscillation data
which give the following mass-squared differences[3,4,5,6,7,8]

\begin{equation}
\Delta m_{12}^{2}=\left( 2-50\right) \times 10^{-5}eV^{2},\ \Delta
m_{23}^{2}=\left( 1.2-5\right) \times 10^{-3}eV^{2}  \tag{7.16}  \label{7.16}
\end{equation}%
\ \ \ \ \ \ \ \ \ \ \ \ \ \ \ \ \ \ \ \ \ \ \ \ \ \ \ \ \ \ \ \ \ \ \ \ \ \
\ \ \ \ \ \ \ \ \ \ \ \ \ \ \ \ \ \ \ \ \ \ \ \ \ \ \ \ \ \ \ \ \ 

The mixing angle values that are consistent with the experimental values
given in (7.8) and (7.16) are found to be

\begin{equation}
S_{2}=.16,\ S_{1}=.54  \tag{7.17}  \label{7.17}
\end{equation}%
With these parameters and equation (7.14) we obtain\ \ \ \ \ \ \ \ \ \ \ \ \
\ \ \ \ \ \ \ \ \ \ \ \ \ \ \ \ \ \ \ \ \ \ \ \ \ \ \ \ \ \ \ \ \ \ \ \ \ \
\ \ \ \ \ \ \ \ \ \ \ \ \ \ \ \ \ \ \ \ \ \ \ \ \ \ \ \ \ \ \ \ \ \ \ \ \ \
\ \ \ \ \ \ \ \ \ \ \ \ \ \ \ \ \ \ \ \ \ \ \ \ \ \ \ \ \ \ \ \ \ \ \ \ \ \
\ \ \ \ \ \ \ \ \ \ \ \ \ \ \ \ \ \ \ \ \ \ \ \ \ \ \ \ 

\begin{equation}
S_{3}^{\prime }=0.74  \tag{7.18}  \label{7.18}
\end{equation}%
and from the relation between $\theta _{3}^{(out)}$, $\theta _{3}^{\prime }$
and $\theta _{3E}$ given in (7.11) we obtain\ \ \ \ \ \ \ \ \ \ \ \ \ \ \ \
\ \ \ \ \ \ \ \ \ \ \ \ \ \ \ \ \ \ \ \ \ \ \ \ \ \ \ \ \ \ \ \ \ \ \ \ \ \
\ \ \ \ \ \ \ \ \ \ \ \ \ \ \ \ \ \ \ \ \ \ \ \ \ \ \ \ \ \ \ \ \ \ \ \ \ \
\ \ \ \ \ \ \ \ \ \ \ \ \ \ \ \ \ \ \ \ \ \ \ \ \ \ \ \ \ \ \ \ \ \ \ \ \ \
\ \ \ \ \ \ \ \ \ \ \ \ \ \ \ \ \ \ \ \ \ \ \ \ \ \ \ \ \ \ \ \ \ \ 

\begin{equation}
S_{3}^{(out)}=.57  \tag{7.19}  \label{7.19}
\end{equation}

The mass values for $\kappa _{0}$ are determined from the above parameters
and by taking

\begin{equation}
\dfrac{m_{3\nu }^{2}}{M_{33}}=7.1\times 10^{-2}eV  \tag{7.20}  \label{7.20}
\end{equation}%
where $m_{3\nu }$ is the dominant 33-component of the (Dirac) neutrino
Yukawa matrix, $N$. We then obtain\ \ \ \ \ \ \ \ \ \ \ \ \ \ \ \ \ \ \ \ \
\ \ \ \ \ \ \ \ \ \ \ \ \ \ \ \ \ \ \ \ \ \ \ \ \ \ \ \ \ \ \ \ \ \ \ \ \ \
\ \ \ \ \ \ \ \ \ \ \ \ \ \ \ \ \ \ \ \ \ \ \ \ \ \ \ \ \ \ \ \ \ \ \ \ \ \
\ \ \ \ \ \ \ \ \ \ \ \ \ \ \ \ \ \ \ \ \ \ \ \ \ \ \ \ \ \ \ \ \ \ \ \ \ \
\ \ \ \ \ 

\begin{equation}
m_{1}=\allowbreak 4.\,\allowbreak 8\times 10^{-3}eV  \tag{7.21a}
\label{7.21a}
\end{equation}

\begin{equation}
m_{2}=\allowbreak 5.\,\allowbreak 8\times 10^{-3}eV  \tag{7.21b}
\label{7.21b}
\end{equation}

\begin{equation}
m_{3}=7.1\times 10^{-2}eV  \tag{7.21c}  \label{7.21c}
\end{equation}%
These values are consisitent with the experimental results (7.8).\ \ \ \ \ \
\ \ \ \ \ \ \ \ \ \ \ \ \ \ \ \ \ \ \ \ \ \ \ \ \ \ \ \ \ \ \ \ \ \ \ \ \ \
\ \ \ \ \ \ \ \ \ \ \ \ \ \ \ \ \ \ \ \ \ \ \ \ \ \ \ \ \ \ \ \ \ \ \ \ \ \
\ \ \ \ \ \ \ \ \ \ \ \ \ \ \ \ \ \ \ \ \ \ \ \ \ \ \ \ \ \ \ \ \ \ \ \ \ \
\ \ \ \ \ \ \ \ \ \ \ \ \ \ \ \ \ \ \ \ \ \ \ \ \ \ \ \ \ 

As for comparing the "input" and "output" values of the mixing matrices, we
have

\begin{equation}
S_{1}^{(out)}=S_{1}^{(in)}=.54  \tag{7.22a}  \label{7.22a}
\end{equation}

\begin{equation}
S_{2}^{(out)}=\allowbreak .11,\text{ }S_{2}^{(in)}=.16  \tag{7.22b}
\label{7.22b}
\end{equation}

\begin{equation}
S_{3}^{(out)}=S_{3}^{(in)}=.57  \tag{7.22c}  \label{7.22c}
\end{equation}%
We point out that through (7.9) and (7.10)\textbf{\ }we first obtained $%
S_{3}^{(out)}$ and then simply assumed $S_{3}^{(in)}$ to have the same value
since $S_{3}\left( =S_{3}^{(in)}\right) $ was not involved\ \ \ \ \ \ \ \ \
\ \ \ \ \ \ \ \ \ \ \ \ \ \ \ \ \ \ \ \ \ \ \ \ \ \ \ \ \ \ \ \ \ \ \ \ \ \
\ \ \ \ \ \ \ \ \ \ \ \ \ \ \ \ \ \ \ \ \ \ \ \ \ \ \ \ \ \ \ \ \ \ \ \ \ \
\ \ \ \ \ \ \ \ \ \ \ \ \ \ \ \ \ \ \ \ \ \ \ \ \ \ \ \ \ \ \ \ \ \ \ \ \ \
\ \ \ \ \ \ \ \ \ \ \ \ \ \ \ \ \ \ \ \ \ \ \ \ \ \ \ \ \ \ \ \ \ \ \ \ \ \
\ \ \ \ \ \ \ in constructing the mass matrix (7.6) for $\kappa _{0}$ and,
therefore, there were no constraints on its value.

The above predictions are consistent with experiments. Moreover, apart from
the differences in $S_{2},$which are minor, the "input" and "output" mixing
parameters are the same.

We also point out that, in order to get the correct neutrino masses, the
value of $S_{2}$ needs to be large, as already noted for the case of A-type
structure in ref.[18] and [19]. As for the possibility of having a maximal
coupling in the 23 sector of $\kappa _{0}$ $($i.e. $S_{1}=\tfrac{1}{\sqrt{2}}%
=.71)$ we opted instead to reconcile the "input" and "output" values of $%
S_{2}$ which implied, through relation (7.13), that we have as large a $%
C_{1} $ as possible, within experimental bounds, which led us to the value
of $S_{1}$ given by (7.17)$.$

We can estimate the mass, $M_{33}$, for the seesaw neutrino in (7.20) by
following our assumption that $N$ and $E$ have mass properties that are
similar to $U$ and $D$ respectively, in which case

\begin{equation}
\dfrac{m_{3\nu }}{m_{\tau }}\thickapprox \dfrac{m_{t}}{m_{b}}  \tag{7.23}
\label{7.23}
\end{equation}%
and, therefore, putting in the other known mass values, we get $m_{3\nu
}^{2}\thickapprox 10^{3}GeV^{2}.$ \ From (7.20) we then find

\begin{equation}
M_{33}\thickapprox 10^{13}GeV  \tag{7.24}  \label{7.24}
\end{equation}

\bigskip

\textbf{VIII. Inverted Hierarchy in Neutrinos, revisited}

One might ask if in the eigenvalue equation for the leptons (6.9a,b,c), we
could have chosen the following "primordial" solutions to generate an
inverted hierarchy for the neutrinos, leaving the charged leptons in the
hierarchical pattern

\begin{equation}
\lambda _{1}=1,\lambda _{2}=0,\lambda _{3}=0(neutrinos)  \tag{8.1a}
\label{8.1a}
\end{equation}

\begin{equation}
\lambda _{1}=0,\text{ }\lambda _{2}=0,\text{ }\lambda _{3}=1\ \ \ \ or\ \ \
\ \ \ \ \lambda _{1}=0,\text{ }\lambda _{2}=1,\text{ }\lambda _{3}=1\text{ }%
(charged\text{ }leptons)  \tag{8.1b}  \label{8.1b}
\end{equation}%
\ \ \ \ \ \ \ \ \ \ \ \ \ \ \ \ \ \ \ \ \ \ \ \ \ \ \ \ \ \ \ \ \ \ \ \ \ \
\ \ \ \ \ \ \ \ \ 

First of all, we notice immediately that for the charged lepton RGE (6.1)
the coupling term

\begin{equation}
\frac{1}{12}NN^{+}E  \tag{8.2}  \label{8.2}
\end{equation}%
will vanish as it involves the product\ \ \ \ \ \ \ \ \ \ \ \ \ \ \ \ \ \ \
\ \ \ \ \ \ \ \ \ \ \ \ \ \ \ \ \ \ \ \ \ \ \ \ \ \ \ \ \ \ \ \ \ \ \ \ \ \
\ \ \ \ \ \ \ \ \ \ \ \ \ \ \ \ \ \ \ \ \ \ \ \ \ \ \ \ \ \ \ \ \ \ \ \ \ \
\ \ \ \ \ \ \ \ \ \ \ \ \ \ \ \ \ \ \ \ \ \ \ \ \ \ \ \ \ \ \ \ \ \ \ \ \ \
\ \ \ \ \ \ \ \ \ \ \ \ \ \ \ \ \ \ \ \ \ \ \ \ \ \ \ \ \ \ \ \ \ \ \ \ \ \
\ \ \ \ \ \ \ 

\begin{equation}
\begin{bmatrix}
1 & 0 & 0 \\ 
0 & 0 & 0 \\ 
0 & 0 & 0%
\end{bmatrix}%
\begin{bmatrix}
0 & 0 & 0 \\ 
0 & 0 & 0 \\ 
0 & 0 & 1%
\end{bmatrix}
\tag{8.3}  \label{8.3}
\end{equation}%
In the absence of the coupling contribution, the charged lepton matrix will
have two choices, instead of the previous situation when it was, like $D_{0}$%
, a mixture of "hierarchical" and "semi-hierarchical" bases. Either it has
to be strictly "semi-hierarchical", which will be inconsistent with
experiments since it would mean two mass eigenvalues very close to each
other. Or it has to be "hierarchical" which would mean it would be like the
up-quark matrix, $U_{0}$, which would, however, give mass ratios an order of
magnitude smaller than observed. \ Thus this model will not work for the
charged leptons. We will proceed anyway, with the hope that there may be
another, totally different, scenario possible for it, and take the charged
lepton matrix to be diagonal so its parameters do not enter into the
calculation for neutrinos.

The neutrino mass matrix $\kappa _{0}$ will have the following expression
(before incorporating texture zeros)\ \ \ \ \ \ \ \ \ \ \ \ \ \ \ \ \ \ \ \
\ \ \ \ \ \ \ \ \ \ \ \ \ \ \ \ \ \ \ \ \ \ \ \ \ \ \ \ \ \ \ \ \ \ \ \ \ \
\ \ \ \ \ \ \ \ \ \ \ \ \ \ \ \ \ \ \ \ \ \ \ \ \ \ \ \ \ \ \ \ \ \ \ \ \ \
\ \ \ \ \ \ \ \ \ \ \ \ \ \ \ \ \ \ \ \ \ \ \ \ \ \ \ \ \ \ \ \ \ \ \ \ \ \
\ \ \ \ \ \ \ \ \ \ \ \ \ \ \ \ \ \ \ \ \ \ \ \ \ \ \ \ \ \ \ \ \ \ \ \ \ \
\ \ \ \ \ \ \ \ \ \ \ \ \ \ \ \ \ \ \ \ \ 

\begin{equation}
\kappa _{0}=V_{large}%
\begin{bmatrix}
1 & 0 & 0 \\ 
0 & 0 & 0 \\ 
0 & 0 & 0%
\end{bmatrix}%
V_{large}^{\dagger }  \tag{8.4}  \label{8.4}
\end{equation}%
\ \ \ \ 

\begin{equation}
\kappa \thickapprox \kappa _{0}\dfrac{\lambda _{1\nu }^{2}}{M_{11}} 
\tag{8.5}  \label{8.5}
\end{equation}%
with $V_{large}$ as the "input" matrix as before, and \ $\left( \dfrac{1}{%
M_{11}}\right) $ the 11-element of $\left[ M_{R}^{-1}\right] $\ .\ \ \ \ \ \
\ \ \ \ \ \ \ \ \ \ \ \ \ \ \ \ \ \ \ \ \ \ \ \ \ \ \ \ \ \ \ \ \ \ \ \ \ \
\ \ \ \ \ \ \ \ \ \ \ \ \ \ \ \ \ \ \ \ \ \ \ \ \ \ \ \ \ \ \ \ \ \ \ \ \ \
\ \ \ \ \ \ \ \ \ \ \ \ \ \ \ \ \ \ \ \ \ \ \ \ \ \ \ \ \ \ \ \ \ \ \ \ \ \
\ \ \ \ \ \ \ \ \ \ \ \ \ \ \ \ \ \ \ \ \ \ \ \ \ \ \ \ \ \ \ \ \ \ \ \ \ \
\ \ \ \ \ \ \ \ \ \ \ \ \ \ \ \ \ \ \ \ \ \ \ \ \ \ \ \ \ \ \ \ \ \ \ \ \ \
\ \ \ \ \ \ \ \ \ \ \ \ \ \ \ \ \ \ \ \ \ \ \ \ \ \ \ \ \ \ \ \ \ \ \ \ \ \
\ \ \ \ \ \ \ \ \ \ \ \ \ \ \ \ \ \ \ \ \ \ \ \ \ \ \ \ \ \ \ \ \ \ \ \ \ \
\ \ \ \ \ \ \ \ 

Instead of writing a long and complicated matrix that will result from the
above product we write it after the texture zeros, of the C-variety, are
already imposed. At the same time, we take account of the fact, evident from 
$\kappa _{0}$'s structure given in (7.3), that with vanishing 22 and 33
diagonal elements, the rotation angle for the matrix diagonalizing the
23-sub matrix will be $45^{0}$. Thus, the "output" $S_{1}$ will
automatically be

\begin{equation}
S_{1}^{(out)}=\dfrac{1}{\sqrt{2}}  \tag{8.6}  \label{8.6}
\end{equation}%
\ \ \ \ \ \ \ \ \ \ \ \ \ \ \ \ \ \ \ \ \ \ \ \ \ \ \ \ \ \ \ \ \ \ \ \ \ \
\ \ \ \ \ \ \ \ \ \ \ \ \ \ \ \ \ \ \ \ \ \ \ \ \ \ \ \ \ \ \ \ \ \ \ \ \ \
\ \ \ \ \ \ \ \ \ \ \ \ \ \ \ \ \ \ \ \ \ \ \ \ \ \ \ \ \ \ \ \ \ \ \ \ \ \
\ \ \ \ \ \ \ \ \ \ \ \ \ \ \ \ \ \ \ \ \ \ \ \ \ \ \ \ \ \ \ \ \ \ \ \ \ \
\ \ \ \ \ \ \ \ \ \ \ \ \ \ \ \ \ \ \ \ \ \ \ \ \ \ \ \ \ \ \ \ \ \ \ \ \ \
\ \ \ \ \ \ \ \ \ \ \ \ \ \ \ \ \ \ \ \ \ \ \ \ \ \ \ \ \ \ \ \ \ \ \ \ \ \
\ \ \ \ \ \ \ \ \ \ \ \ \ \ \ \ \ \ \ \ \ \ \ \ \ \ \ \ \ \ \ \ \ \ \ \ \ \
\ \ \ \ \ \ \ \ \ \ \ \ \ \ \ \ Since the above value is, indeed, allowed by
the experiments we also take

\begin{equation}
S_{1}^{(in)}=\dfrac{1}{\sqrt{2}}  \tag{8.7}  \label{8.7}
\end{equation}%
\ \ \ \ \ \ \ \ \ \ \ \ \ \ \ \ \ \ \ \ \ \ \ \ \ \ \ \ \ \ \ \ \ \ \ \ \ \
\ \ \ \ \ \ \ \ \ \ \ \ \ \ \ \ \ \ \ \ \ \ \ \ \ \ \ \ \ \ \ \ \ \ \ \ \ \
\ \ \ \ \ \ \ \ \ \ \ \ \ \ \ \ \ \ \ \ \ \ \ \ \ \ \ \ \ \ \ \ \ \ \ \ \ \
\ \ \ \ \ \ \ \ \ \ \ \ \ \ \ \ \ \ \ \ \ \ \ \ \ \ \ \ \ \ \ \ \ \ \ \ \ \
\ \ \ \ \ \ \ \ \ \ \ \ \ \ \ \ \ \ \ \ \ \ \ \ \ \ \ \ \ \ \ \ \ \ \ \ \ \
\ \ \ \ \ \ \ \ \ \ \ \ \ \ \ \ \ \ \ \ \ \ \ \ \ \ \ \ \ \ \ \ \ \ \ \ \ \
\ \ \ \ \ \ \ \ \ \ \ \ \ \ \ \ \ \ \ \ \ \ \ \ \ \ \ \ \ \ \ \ \ \ \ \ \ \
\ \ \ \ \ \ \ \ \ \ \ \ \ \ \ \ \ \ \ \ \ \ \ \ A further simplication
occurs because of (8.6), namely [19]

\begin{equation}
S_{2}^{(out)}=0  \tag{8.8}  \label{8.8}
\end{equation}%
and, here again, we take the same value for the "input" parameter\ \ \ \ \ \
\ \ \ \ \ \ \ \ \ \ \ \ \ \ \ \ \ \ \ \ \ \ \ \ \ \ \ \ \ \ \ \ \ \ \ \ \ \
\ \ \ \ \ \ \ \ \ \ \ \ \ \ \ \ \ \ \ \ \ \ \ \ \ \ \ \ \ \ \ \ \ \ \ \ \ \
\ \ \ \ \ \ \ \ \ \ \ \ \ \ \ \ \ \ \ \ \ \ \ \ \ \ \ \ \ \ \ \ \ \ \ \ \ \
\ \ \ \ \ \ \ \ \ \ \ \ \ \ \ \ \ \ \ \ \ \ \ \ \ \ \ \ \ \ \ \ \ \ \ \ \ \
\ \ \ \ \ \ \ \ \ \ \ \ \ \ \ \ 

\begin{equation}
S_{2}^{(in)}=0  \tag{8.9}  \label{8.9}
\end{equation}%
since it is allowed by the experiments.\ \ \ \ \ \ \ \ \ \ \ \ \ \ \ \ \ \ \
\ \ \ \ \ \ \ \ \ \ \ \ \ \ \ \ \ \ \ \ \ \ \ \ \ \ \ \ \ \ \ \ \ \ \ \ \ \
\ \ \ \ \ \ \ \ \ \ \ \ \ \ \ \ \ \ \ \ \ \ \ \ \ \ \ \ \ \ \ \ \ \ \ \ \ \
\ \ \ \ \ \ \ \ \ \ \ \ \ \ \ \ \ \ \ \ \ \ \ \ \ \ \ \ \ \ \ \ \ \ \ \ \ \
\ \ \ \ \ \ \ \ \ \ \ \ \ \ \ \ \ \ \ \ \ \ \ \ \ \ \ \ \ \ \ \ \ \ \ \ \ \
\ \ 

Only $S_{3}$ now remains to be determined. With the above values of the
"input" parameters we obtain the following, simplified expression,

\begin{equation}
\kappa _{0}=%
\begin{bmatrix}
C_{3}^{2} & -\dfrac{C_{3}S_{3}}{\sqrt{2}} & \dfrac{C_{3}S_{3}}{\sqrt{2}} \\ 
-\dfrac{C_{3}S_{3}}{\sqrt{2}} & 0 & -\dfrac{S_{3}^{2}}{2} \\ 
\dfrac{C_{3}S_{3}}{\sqrt{2}} & -\dfrac{S_{3}^{2}}{2} & 0%
\end{bmatrix}
\tag{8.10}  \label{8.10}
\end{equation}%
Diagonalizing it we obtain $\left( \text{with }T_{3}^{\prime }=\frac{%
S_{3}^{\prime }}{C_{3}^{\prime }}\right) $\ \ \ \ \ \ \ \ \ \ \ \ \ \ \ \ \
\ \ \ \ \ \ \ \ \ \ \ \ \ \ \ \ \ \ \ \ \ \ \ \ \ \ \ \ \ \ \ \ \ \ \ \ \ \
\ \ \ \ \ \ \ \ \ \ \ \ \ \ \ \ \ \ \ \ \ \ \ \ \ \ \ \ \ \ \ \ \ \ \ \ \ \
\ \ \ \ \ \ \ \ \ \ \ \ \ \ \ \ \ \ \ \ \ \ \ \ \ \ \ \ \ \ \ \ \ \ \ \ \ \
\ \ \ \ \ \ \ \ \ \ \ \ \ \ \ \ \ \ \ \ \ \ \ \ \ \ \ \ \ \ \ \ \ \ \ \ \ \
\ \ \ \ \ \ \ \ \ \ \ \ \ \ \ \ \ \ \ \ \ \ \ \ \ \ \ \ \ \ \ \ \ \ \ \ \ \
\ \ \ \ \ \ \ \ \ \ \ \ \ \ \ \ \ \ \ \ \ \ \ \ \ \ \ \ \ \ \ \ \ \ \ \ \ \
\ \ \ \ \ \ \ \ \ \ \ \ \ \ \ \ \ \ \ \ \ \ \ \ \ \ \ \ \ \ \ \ \ \ \ \ \ \
\ \ \ \ \ \ \ \ \ \ \ \ \ \ 

\begin{equation}
\left[ 
\begin{array}{ccc}
\dfrac{\left( C_{3}S_{3}+\frac{1}{2}T_{3}^{\prime }S_{3}^{2}\right) }{%
T_{3}^{\prime }} & 0 & 0 \\ 
0 & \left( -T_{3}^{\prime }C_{3}S_{3}+\frac{1}{2}S_{3}^{2}\right) & 0 \\ 
0 & 0 & -\left( \frac{1}{2}\right) S_{3}^{2}%
\end{array}%
\right]  \tag{8.11}  \label{8.11}
\end{equation}%
where $T_{3}^{\prime }$ satisfies\ \ \ \ \ \ \ \ \ \ \ \ \ \ \ \ \ \ \ \ \ \
\ \ \ \ \ \ \ \ \ \ \ \ \ \ \ \ \ \ \ \ \ \ \ \ \ \ \ \ \ \ \ \ \ \ \ \ \ \
\ \ \ \ \ \ \ \ \ \ \ \ \ \ \ \ \ \ \ \ \ \ \ \ \ \ \ \ \ \ \ \ \ \ \ \ \ \
\ \ \ \ \ \ \ \ \ \ \ \ \ \ \ \ \ \ \ \ \ \ \ \ \ \ \ \ \ \ \ \ \ \ \ \ \ \
\ \ \ \ \ \ \ \ \ \ \ \ \ \ \ \ \ \ \ \ \ \ \ \ \ \ \ \ \ \ \ \ \ \ \ \ \ \
\ \ \ \ \ \ \ \ \ \ \ \ \ \ \ \ \ \ \ \ \ \ \ \ \ \ \ \ \ \ \ \ \ \ \ \ \ \
\ \ \ \ \ \ \ \ \ \ \ \ \ \ \ \ \ \ \ \ \ \ \ \ \ \ \ \ \ \ \ \ \ \ \ \ \ \
\ \ \ \ \ \ \ 

\begin{equation}
T_{3}^{\prime 2}+\left( \dfrac{C_{3}^{2}-\frac{1}{2}S_{3}^{2}}{C_{3}S_{3}}%
\right) T_{3}^{\prime }-1=0  \tag{8.12}  \label{8.12}
\end{equation}%
\ \ \ \ \ \ \ \ \ \ \ \ \ \ \ \ \ \ \ \ \ \ \ \ \ \ \ \ \ \ \ \ \ \ \ \ \ \
\ \ \ \ \ \ \ \ \ \ \ \ \ \ \ \ \ \ \ \ \ \ \ \ \ \ \ \ \ \ \ \ \ \ \ \ \ \
\ \ \ \ \ \ \ \ \ \ \ \ \ \ \ \ \ \ \ \ \ \ \ \ \ \ \ \ \ \ \ \ \ \ \ \ \ \
\ \ \ \ 

We choose $S_{3}\left( =S_{3}^{\prime }\right) =.58,$ which is within the
experimental range (7.8). From equation (8.12) we obtain, $S_{3}^{\prime
}=.51$ which is also the "output" value, since we have taken $\theta _{3E}=0$%
, as explained above. \ Thus we have

\begin{equation}
S_{1}^{(out)}=S_{1}^{(in)}=\frac{1}{\sqrt{2}}  \tag{8.13a}  \label{8.13a}
\end{equation}

\begin{equation}
S_{2}^{(out)}=S_{2}^{(in)}=0  \tag{8.13b}  \label{8.13b}
\end{equation}%
\ 

\begin{equation}
S_{3}^{(out)}=\ .51,\ \ \ \ \ S_{3}^{(in)}=.58  \tag{8.13c}  \label{8.13c}
\end{equation}%
\ \ \ \ \ \ \ \ \ \ \ \ \ \ \ \ \ \ \ \ \ \ \ \ \ \ \ \ \ \ \ \ \ \ \ \ \ \
\ \ \ \ \ \ \ \ \ \ \ \ \ \ \ \ \ \ \ \ \ \ \ \ \ \ \ \ \ \ \ \ \ \ \ \ \ \
\ \ \ \ \ \ \ \ \ \ \ \ \ \ \ \ \ \ \ \ \ \ \ \ \ \ \ \ \ \ \ \ \ \ \ \ \ \
\ \ \ \ \ \ \ \ \ \ \ \ \ \ \ \ \ \ \ \ \ \ \ \ \ \ \ \ \ \ \ \ \ \ \ \ \ \
\ \ \ \ \ \ \ \ \ \ \ \ \ \ \ \ \ \ \ \ \ \ \ \ \ \ \ \ \ \ \ \ \ \ \ \ \ \
\ \ \ \ \ \ \ \ \ \ \ \ \ \ \ \ \ \ \ \ \ \ \ \ \ \ \ \ \ \ \ \ \ \ \ \ \ \
\ \ \ \ \ \ \ \ \ \ \ \ \ \ \ \ \ \ \ \ \ \ \ \ \ \ \ \ \ \ \ \ \ \ \ \ \ \
\ \ \ \ \ \ \ 

The "output" and "input" values are, therefore, identical for two angles and
very close for the third.

With the above parameters and with

\begin{equation}
\dfrac{m_{1\nu }^{2}}{M_{11}}\ =7.3\times 10^{-2}eV  \tag{8.14}  \label{8.14}
\end{equation}%
\ \ \ \ \ \ \ \ \ \ \ \ \ \ \ \ \ \ \ \ \ \ \ \ \ \ \ \ \ \ \ \ \ \ \ \ \ \
\ \ \ \ \ \ \ \ \ \ \ \ \ \ \ \ \ \ \ \ \ \ \ \ \ \ \ \ \ \ \ \ \ \ \ \ \ \
\ \ \ \ \ \ \ \ \ \ \ \ \ \ \ \ \ \ \ \ \ \ \ \ \ \ \ \ \ \ \ \ \ \ \ \ \ \
\ \ \ \ \ \ \ \ \ \ \ \ \ \ \ \ \ \ \ \ \ \ \ \ \ \ \ \ \ \ \ \ \ \ \ \ \ \
\ \ \ \ \ \ \ \ \ \ \ \ \ \ \ \ \ \ \ \ \ \ \ \ \ \ \ \ \ \ \ \ \ \ \ \ \ \
\ \ \ \ \ \ \ \ \ \ \ \ \ \ \ \ \ \ \ \ \ \ \ \ \ \ \ \ \ \ \ \ \ \ \ \ \ \
\ \ \ \ \ \ \ \ \ \ \ \ \ \ \ \ \ \ \ \ \ \ \ \ \ \ \ \ \ \ \ \ \ \ \ \ \ \
\ \ \ \ \ \ \ \ \ \ \ \ \ \ \ \ \ \ \ \ \ \ \ \ \ \ \ \ \ \ \ \ \ \ \ \ \ \
\ \ \ \ \ \ \ \ \ \ \ \ \ \ (as before, $m_{1\nu }^{2}\thickapprox
10^{3}GeV^{2}$ and $M_{11}\thickapprox $ $10^{13}$ $GeV$) we have the
following mass eigenvalues

\begin{equation}
m_{1}=\allowbreak 7.\,\allowbreak 1\times 10^{-2}eV  \tag{8.15a}
\label{8.15a}
\end{equation}

\begin{equation}
m_{2}=\allowbreak 1.\,2\times 10^{-2}eV  \tag{8.15b}  \label{8.15b}
\end{equation}

\begin{equation}
m_{3}=\allowbreak 8.\,\allowbreak 3\times 10^{-3}eV  \tag{8.15c}
\label{8.15c}
\end{equation}%
\ \ \ \ \ \ \ \ \ \ \ \ \ \ \ \ \ \ \ \ \ \ \ \ \ \ \ \ \ \ \ \ \ \ \ \ \ \
\ \ \ \ \ \ \ \ \ \ \ \ \ \ \ \ \ \ \ \ \ \ \ \ \ \ \ \ \ \ \ \ \ \ \ \ \ \
\ \ \ \ \ \ \ \ \ \ \ \ \ \ \ \ \ \ \ \ \ \ \ \ \ \ \ \ \ \ \ \ \ \ \ \ \ \
\ \ \ \ \ \ \ \ \ \ \ \ \ \ \ \ \ \ \ \ \ \ \ \ \ \ \ \ \ \ \ \ \ \ \ \ \ \
\ \ \ \ \ \ \ \ \ \ \ \ \ \ \ \ \ \ \ \ \ \ \ \ \ \ \ \ \ \ \ \ \ \ \ \ \ \
\ \ \ \ \ \ \ \ \ \ \ \ \ \ \ \ \ \ \ \ \ \ \ \ \ \ \ \ \ \ \ \ \ \ \ \ \ \
\ \ \ \ \ \ \ \ \ \ \ \ \ \ \ \ \ \ \ \ \ \ \ \ \ \ \ \ \ \ \ \ \ \ \ \ \ \
\ \ \ \ \ \ \ \ \ \ \ \ \ \ \ \ \ \ \ \ \ \ \ \ \ \ \ \ \ \ \ \ \ \ \ \ \ \
\ \ \ \ \ \ \ \ These values are consistent with the experimental values
given by (7.16)[3,4,5,6,7,8]

\bigskip \medskip \bigskip \medskip \bigskip \bigskip

\textbf{IX. Conclusion}

One of our central assumptions was that the scale dependence of a Yukawa
matrix is dictated entirely by the dominant 33-matrix element which can be
factored out leaving behind a matrix which is independent of the scale. As a
conseqence, the renormalization group equations for the Yukawa matrix can be
expressed as two separate equations: one a differential equation for the
33-element and another an algebraic equation for the scale-independent 3x3
matrix.

It is the properties of the scale-independent matrices that has concerned us
primarily. After constructing the solutions in terms of the mixing angles
and incorporating texture zeros, consistent with hierarchical behavior, we
made the following identifications for $U,$ $D$, $E$, and (A-type) $\kappa $,

\begin{equation*}
U=%
\begin{bmatrix}
0 & 0 & s_{2} \\ 
0 & s_{1}^{2} & 0 \\ 
s_{2} & 0 & 1%
\end{bmatrix}%
\lambda _{t},\ \ \ \ \ \ \ \ \ D=%
\begin{bmatrix}
0 & \epsilon s_{3} & 0 \\ 
\epsilon s_{3} & s_{1}^{2}+\epsilon & s_{1} \\ 
0 & s_{1} & 1%
\end{bmatrix}%
\lambda _{b}
\end{equation*}

\begin{equation*}
E=%
\begin{bmatrix}
0 & \epsilon ^{\prime }s_{3} & 0 \\ 
\epsilon ^{\prime }s_{3} & s_{1}^{2}+\epsilon ^{\prime } & s_{1} \\ 
0 & s_{1} & 1%
\end{bmatrix}%
\lambda _{\tau },\ \ \ \ \ \kappa =%
\begin{bmatrix}
0 & 0 & C_{1}C_{2}S_{2} \\ 
0 & C_{2}^{2}S_{1}^{2} & C_{1}C_{2}^{2}S_{1} \\ 
C_{1}C_{2}S_{2} & C_{1}C_{2}^{2}S_{1} & C_{1}^{2}C_{2}^{2}%
\end{bmatrix}%
\dfrac{\lambda _{3\nu }^{2}}{M_{33}}
\end{equation*}

For the C-type neutrinos we have, as discussed in section VII, an especially
simple mass matrix because of the location of the texture zeros which,
automatically, result in $S_{1}=\frac{1}{\sqrt{2}}$ and $S_{2}=0$, and give

\begin{equation*}
\kappa =%
\begin{bmatrix}
C_{3}^{2} & -\dfrac{C_{3}S_{3}}{\sqrt{2}} & \dfrac{C_{3}S_{3}}{\sqrt{2}} \\ 
-\dfrac{C_{3}S_{3}}{\sqrt{2}} & 0 & -\dfrac{S_{3}^{2}}{2} \\ 
\dfrac{C_{3}S_{3}}{\sqrt{2}} & -\dfrac{S_{3}^{2}}{2} & 0%
\end{bmatrix}%
\dfrac{\lambda _{1\nu }^{2}}{M_{11}}
\end{equation*}%
which predicts an inverted hierarchy.

The above results provide simple expressions for the quark and lepton mass
matrices in terms of the mixing angles. The manner in which we introduced
the mixing parameters is self-consistent i.e. what we put in to construct
the mass matrices (the "input") is recovered when we try to obtain the
mixing matrices (the "output").

Another interesting result is that for the massive seesaw neutrinos only one
mass-scale appears, $M_{33}$ for normal hierarchy and $M_{11}$ for inverted
hierarchy. So the details of the seesaw mass distribution do not play a
role, which is an important advantage in the model.

The texture zeros play a crucial role in determining the physical parameters
in our model. It is well known that zeros in the mass matrices can arise
through discrete symmetries\textbf{\ }$\left[ 1,11,23,24,25,26,27,28\right]
. $\textbf{\ }The precise group structure we need in order to reproduce the
above texture zeros needs to be worked out and is being attempted currently%
\textbf{\ }$\left[ 29\right] $\textbf{.}

To conclude, we have succeeded in obtaining a simple, transparent, and
uniform framework to descibe four different pieces of data involving quarks
and leptons. We assumed the Yukawa matrices to be real and symmetric with
the dominant eigenvalue as an input. No attempt was made to do a detailed
numerical analysis to fit the data but the matrices described above are
found to give a very good descritipn of the mass eigenvalues and mixing
matrices. Apart from the seesaw particles that are essential to the
neutrinos, no new particles have been proposed--that is another big
advantage to our model.

\bigskip

\textbf{Acknowlegements}

We wish to thank Dr. Rajasekaran for several discussions on this subject and
Dr. D. P. Roy for his helpful comments. This work was supported in part by
the U. S. Departmant of Energy under Grant No. DE-FG03-94ER40837

\bigskip

\textbf{Appendix I}

We notice from (3.9) and (3.10) that, because the determinant and trace are
invariant, we have

\begin{equation}
detM_{0}^{\left( 1\right) }=detM_{diag}^{\left( 1\right) }=0  \tag{A.1}
\label{A.1}
\end{equation}
\ \ \ \ \ \ \ \ \ \ \ \ \ \ \ \ \ \ \ \ \ \ \ \ \ \ \ \ \ \ \ \ \ \ \ \ \ \
\ \ \ \ \ \ \ \ \ \ \ \ \ \ \ \ \ \ \ \ \ \ \ \ \ \ \ \ \ \ \ \ \ \ \ \ \ \
\ \ \ \ \ \ \ \ \ \ \ \ \ \ \ \ \ \ \ \ \ \ \ \ \ \ \ \ \ \ \ \ \ \ \ \ \ \
\ \ \ \ \ \ \ \ \ \ \ \ \ \ 

\begin{equation}
trM_{0}^{\left( 1\right) }=trM_{diag}^{\left( 1\right) }=1  \tag{A.2}
\label{A.2}
\end{equation}%
for simplicity we will take $M_{0}^{\left( 1\right) }$ to be real and
symmetric. Using (A.2) in equation (3.4) we obtain\ \ \ \ \ \ \ \ \ \ \ \ \
\ \ \ \ \ \ \ \ \ \ \ \ \ \ \ \ \ \ \ \ \ \ \ \ \ \ \ \ \ \ \ \ \ \ \ \ \ \
\ \ \ \ \ \ \ \ \ \ \ \ \ \ \ \ \ \ \ \ \ \ \ \ \ \ \ \ \ \ \ \ \ \ \ \ \ \
\ \ \ \ \ \ \ \ \ \ \ \ \ \ \ \ \ \ \ \ \ \ \ \ \ \ \ \ \ \ \ \ \ \ \ \ \ \
\ \ \ \ \ \ \ \ \ \ \ \ \ \ \ \ \ \ \ \ \ \ \ \ \ \ \ \ \ \ \ \ \ \ \ \ \ \
\ \ \ \ \ \ \ \ \ \ \ \ \ \ \ \ \ \ \ \ \ \ \ \ \ \ \ \ \ \ \ \ \ \ \ \ \ \
\ \ \ \ \ \ \ \ \ \ \ \ \ \ \ \ \ \ \ \ \ \ \ \ \ \ \ \ \ 

\begin{equation}
M_{0}^{(1)}=\frac{1}{2}\left[ \left( M_{0}^{(1)}\right) ^{3}+M_{0}^{(1)}%
\right]  \tag{A.3}  \label{A.3}
\end{equation}
\ \ \ \ \ \ \ \ \ \ \ \ \ \ \ \ \ \ \ \ \ \ \ \ \ \ \ \ \ \ \ \ \ \ \ \ \ \
\ \ \ \ \ \ \ \ \ \ \ \ \ \ \ \ \ \ \ \ \ \ \ \ \ \ \ \ \ \ \ \ \ \ \ \ \ \
\ \ \ \ \ \ \ \ \ \ \ \ \ \ \ \ \ \ \ \ \ \ \ \ \ \ \ \ \ \ \ \ \ \ \ \ \ \
\ \ \ \ \ \ \ \ \ \ \ \ \ \ \ \ which we will rewrite as

\begin{equation}
\frac{1}{2}\left( \mathbf{1+}M_{0}^{^{\left( 1\right) }}\right) \left( 
\mathbf{1-}M_{0}^{^{\left( 1\right) }}\right) M_{0}^{\left( 1\right) }=0 
\tag{A.4}  \label{A.4}
\end{equation}
\ \ \ \ \ \ \ \ \ \ \ \ \ \ \ \ \ \ \ \ \ \ \ \ \ \ \ \ \ \ \ \ \ \ \ \ \ \
\ \ \ \ \ \ \ \ \ \ \ \ \ \ \ \ \ \ \ \ \ \ \ \ \ \ \ \ \ \ \ \ \ \ \ \ \ \
\ \ \ \ \ \ \ \ \ \ \ \ \ \ \ \ \ \ \ \ \ \ \ \ \ \ \ \ \ \ \ \ \ \ \ \ \ \
\ \ \ \ \ \ \ \ \ \ \ \ \ \ \ \ \ \ \ \ \ \ \ \ \ \ \ \ \ \ \ \ \ \ \ \ \ \
\ \ \ \ \ \ \ \ \ \ \ \ \ \ \ \ \ \ \ \ \ \ \ \ \ \ \ \ \ \ \ \ \ \ \ \ \ \
\ \ \ \ \ \ \ \ \ \ \ \ \ \ \ \ \ \ \ \ \ \ \ \ \ \ \ \ \ \ \ \ \ \ \ \ \ 

Out of the three factors above, only $\left( \mathbf{1+}M_{0}^{^{\left(
1\right) }}\right) $has an inverse since its determinant does not vanish,

\begin{equation}
det\left( \mathbf{1+}M_{0}^{^{\left( 1\right) }}\right) =det\left( \mathbf{1+%
}M_{diag}^{^{\left( 1\right) }}\right) \neq 0\   \tag{A.5}  \label{A.5}
\end{equation}%
whereas the other two have vanishing determinants. Removing this term in
(A.4) we obtain,\ \ \ \ \ \ \ \ \ \ \ \ \ \ \ \ \ \ \ \ \ \ \ \ \ \ \ \ \ \
\ \ \ \ \ \ \ \ \ \ \ \ \ \ \ \ \ \ \ \ \ \ \ \ \ \ \ \ \ \ \ \ \ \ \ \ \ \
\ \ \ \ \ \ \ \ \ \ \ \ \ \ \ \ \ \ \ \ \ \ \ \ \ \ \ \ \ \ \ \ \ \ \ \ \ \
\ \ \ \ \ \ \ \ \ 

\begin{equation}
\left( \mathbf{1-}M_{0}^{^{\left( 1\right) }}\right) M_{0}^{\left( 1\right)
}=0  \tag{A.6}  \label{A.6}
\end{equation}
\ \ \ \ \ \ \ \ \ \ \ \ \ \ \ \ \ \ \ \ \ \ \ \ \ \ \ \ \ \ \ \ \ \ \ \ \ \
\ \ \ \ \ \ \ \ \ \ \ \ \ \ \ \ \ \ \ \ \ \ \ \ \ \ \ \ \ \ \ \ \ \ \ \ \ \
\ \ \ \ \ \ \ \ \ \ \ \ \ \ \ \ \ \ \ \ \ \ \ \ \ \ \ \ \ \ \ \ \ \ \ \ \ \
\ \ \ \ \ \ \ \ \ \ \ \ \ \ \ \ \ \ \ \ \ 

We note that a general 3x3 real, symmetric matrix with unit trace can be
written as

\begin{equation}
M_{0}^{\left( 1\right) }=\left( a+b+1\right) ^{-1}%
\begin{bmatrix}
a & d & \alpha \\ 
d & b & \beta \\ 
\alpha & \beta & 1%
\end{bmatrix}
\tag{A.7}  \label{A.7}
\end{equation}%
And, therefore,\ \ \ \ \ \ \ \ \ \ \ \ \ \ \ \ \ \ \ \ \ \ \ \ \ \ \ \ \ \ \
\ \ \ \ \ \ \ \ \ \ \ \ \ \ \ \ \ \ \ \ \ \ \ \ \ \ \ \ \ \ \ \ \ \ \ \ \ \
\ \ \ \ \ \ \ \ \ \ \ \ \ \ \ \ \ \ \ \ \ \ \ \ \ \ \ \ \ \ \ \ \ \ \ \ \ \
\ \ \ \ \ \ \ \ \ \ \ \ \ \ \ \ \ \ \ \ \ \ \ 

\begin{equation}
\left( \mathbf{1-}M_{0}^{^{\left( 1\right) }}\right) =\left( a+b+1\right)
^{-1}%
\begin{bmatrix}
1+b & -d & -\alpha \\ 
-d & 1+a & -\beta \\ 
-\alpha & -\beta & b+a%
\end{bmatrix}
\tag{A.8}  \label{A.8}
\end{equation}%
From (A.6), (A.7) and (A.8) we have\ \ \ \ \ \ \ \ \ \ \ \ \ \ \ \ \ \ \ \ \
\ \ \ \ \ \ \ \ \ \ \ \ \ \ \ \ \ \ \ \ \ \ \ \ \ \ \ \ \ \ \ \ \ \ \ \ \ \
\ \ \ \ \ \ \ \ \ \ \ \ \ \ \ \ \ \ \ \ \ \ \ \ 

\begin{equation}
\begin{bmatrix}
1+b & -d & -\alpha \\ 
-d & 1+a & -\beta \\ 
-\alpha & -\beta & b+a%
\end{bmatrix}%
\begin{bmatrix}
a & d & \alpha \\ 
d & b & \beta \\ 
\alpha & \beta & 1%
\end{bmatrix}%
=\allowbreak  \tag{A.9}  \label{A.9}
\end{equation}%
\begin{equation}
\begin{bmatrix}
-d^{2}-\alpha ^{2}+a\left( b+1\right) & -bd-\alpha \beta +d\left( b+1\right)
& -\alpha -d\beta +\alpha \left( b+1\right) \\ 
-ad-\alpha \beta +d\left( a+1\right) & -d^{2}-\beta ^{2}+b\left( a+1\right)
& -\beta -d\alpha +\beta \left( a+1\right) \\ 
-a\alpha -d\beta +\alpha \left( a+b\right) & -b\beta -d\alpha +\beta \left(
a+b\right) & a+b-\alpha ^{2}-\beta ^{2}%
\end{bmatrix}%
=0  \tag{A.10}  \label{A.10}
\end{equation}%
\ \ \ \ \ \ \ \ \ \ \ \ \ \ \ \ \ \ \ \ \ \ \ \ \ \ \ \ \ \ \ \ \ \ \ \ \ \
\ \ \ \ \ \ \ \ \ \ \ \ \ \ \ 

\bigskip

It is then easy to show that

\begin{equation}
a=\alpha ^{2}  \tag{A.11a}  \label{A.11a}
\end{equation}

\begin{equation}
b=\beta ^{2}  \tag{A.11b}  \label{A.11b}
\end{equation}

\begin{equation}
d=\alpha \beta \   \tag{A.11c}  \label{A.11c}
\end{equation}%
\ \ \ \ \ \ \ \ \ \ \ \ \ \ \ \ \ \ \ \ \ \ \ \ \ \ \ \ \ \ \ \ \ \ \ \ \ \
\ \ \ \ \ \ \ \ \ \ \ \ \ \ \ \ \ \ \ \ \ \ \ \ \ \ \ \ \ \ \ \ \ \ \ \ \ \
\ \ \ \ \ \ \ \ \ \ \ \ \ \ \ \ \ \ \ \ \ \ \ \ \ \ \ \ \ \ \ \ \ \ \ \ \ \
\ \ \ \ \ \ \ \ \ \ \ \ \ \ \ \ \ \ \ \ \ \ \ \ \ \ \ \ \ \ \ \ \ \ \ \ \ \
\ \ \ \ \ \ \ \ \ \ \ \ \ \ \ \ \ \ \ \ \ \ \ \ \ \ \ \ \ \ \ \ \ \ \ \ \ \
\ \ \ \ \ \ \ \ \ \ \ \ \ \ \ \ \ \ \ \ \ \ \ \ \ \ \ \ \ \ \ \ \ \ \ \ \ \
\ \ \ \ \ \ \ \ \ \ \ \ \ \ \ \ \ \ \ \ \ \ \ \ \ \ \ \ \ \ \ \ \ \ \ \ \ \
\ \ \ \ \ \ \ \ \ \ \ \ \ \ \ \ \ \ \ \ \ \ \ \ \ \ \ \ \ \ \ \ \ \ \ \ \ \
\ \ \ \ \ \ \ \ \ \ \ \ \ \ \ \ \ \ \ \ \ \ The matrix $M_{0}^{\left(
1\right) }$ is then given by

\begin{equation}
M_{0}^{\left( 1\right) }=\left( 1+\alpha ^{2}+\beta ^{2}\right) ^{-1}\left[ 
\begin{array}{ccc}
\alpha ^{2} & \alpha \beta & \alpha \\ 
\alpha \beta & \beta ^{2} & \beta \\ 
\alpha & \beta & 1%
\end{array}%
\right]  \tag{A.12}  \label{A.12}
\end{equation}%
If the 33-term above dominates the matrix so that\ \ \ \ \ \ \ \ \ \ \ \ \ \
\ \ \ \ \ \ \ \ \ \ \ \ \ \ \ \ \ \ \ \ \ \ \ \ \ \ \ \ \ \ \ \ \ \ \ \ \ \
\ \ \ \ \ \ \ \ \ \ \ \ \ \ \ \ \ \ \ \ \ \ \ \ \ \ \ \ \ \ \ \ \ \ \ \ \ \
\ \ \ \ \ \ \ \ \ \ \ \ \ \ \ \ \ \ \ \ \ \ \ \ \ \ \ \ \ \ \ \ \ \ 

\begin{equation}
\alpha <1\text{ }and\ \beta <1\ \   \tag{A.13}  \label{A.13}
\end{equation}%
then we have a classic hierarchy pattern in $M_{0}^{\left( 1\right) }$.\ \ 

\ \ \ \bigskip\ \ \ \ \pagebreak\ \ \ \ \ \ \ \ \ \ \ \ \ \ \ \ \ \ \ \ \ \
\ \ \ \ \ \ \ \ \ \ \ \ \ \ \ \ \ \ \ \ \ \ \ \ \ \ \ \ \ \ \ \ \ \ \ \ \ \
\ \ \ \ \ \ \ \ \ \ \ \ \ \ \ \ \ \ \ \ \ \ \ \ \ \ \ \ \ \ \ \ \ \ \ \ \ \
\ \ \ \ \ \ \ \ \ \ \ \ \ \ \ \ \ \ \ \ \ \ \ \ \ \ \ \ \ \ \ \ \ \ \ \ \ \
\ \ \ \ \ \ \ \ \ \ \ \ \ \ \ 

\textbf{Appendix II}

Equation (4.22) given by

\begin{equation}
-\dfrac{d\lambda _{b}}{dx}=\frac{1}{2}\lambda _{b}^{3}+\ \frac{1}{12}\lambda
_{t}^{2}\lambda _{b}  \tag{B.1}  \label{B.1}
\end{equation}%
can be solved analytically by first substituting expression (2.3) for $%
\lambda _{t}$ and solving

\begin{equation}
-\dfrac{dy}{dx}=\frac{1}{12}\lambda _{t}^{2}y=\frac{1}{12}\lambda
_{0t}^{2}\left( 1+\lambda _{0t}^{2}x\right) ^{-1}y  \tag{B.2}  \label{B.2}
\end{equation}%
which gives, with $C=constant$,

\begin{equation}
y=C\left( 1+\lambda _{0t}^{2}x\right) ^{-\frac{1}{12}}  \tag{B.3}
\label{B.3}
\end{equation}

If we then take

\begin{equation}
\lambda _{b}(x)=\beta (x)\left( 1+\lambda _{0t}^{2}x\right) ^{-\frac{1}{12}}
\tag{B.4}  \label{B.4}
\end{equation}%
and normalize

\begin{equation}
\lambda _{b}(0)=\beta (0)=\lambda _{0b}  \tag{B.5}  \label{B.5}
\end{equation}%
then the equation for $\beta (x)$ is obtained from (B.1) whose solution is
then given by the following complicated relation

\begin{equation}
\dfrac{1}{\beta ^{2}}=\dfrac{6}{5\lambda _{0t}^{2}}\left[ \left( 1+\lambda
_{0t}^{2}x\right) ^{-\frac{5}{6}}-1\right] +\dfrac{1}{\lambda _{0b}^{2}} 
\tag{B.6}  \label{B.6}
\end{equation}

Substituting this value of $\beta (x)$ in equation (B.4) will then give us
the complete analytic expression for $\lambda _{b}$

\ \ \ \ \ \ \ \ \ \ \ \ \ \ \ \ \ \ \ \ \ \ \ \ \ \ \ \ \ \ \ \ \ \ \ \ \ \
\ \ \ \ \ \ \ \ \ \ \ \ \ \ \ \ \ \ \ \ \ \ \ \ \ \ \ \ \ \ \ \ \ \ \ \ \ \
\ \ \ \ \ \ \ \ \ \ \ \ \ \ \ \ \ \ \ \ \ \ \ \ \ \ \ \ \ \ \ \ \ \ \ \ \ \
\ \ \ \ \ \ \ \ \ \ \ \ \ \ \ \ \ \ \ \ \ \ \ \ \ \ \ \ \ \ \ \ \ \ \ \ \ \
\ \ 

$\bigskip $\newpage

\ \ \ \ \ \ \ \ \ \ \ \ \ \ \ \ \ \ \ \ \ \ \ \ \ \ \ \ \ \ \ \ \ \ \ \ \ \
\ \ \ \ \ \ \ \ \ \ \ \ \ \ \ \ \ \ \ \ \ \ \ \ \ \ \ \ \ \ \ \ \ \ \ \ \ \
\ \ \ \ \ \ \ \ \ \ \ \ \ \ \ \ \ \ \ \ \ \ \ \ \ \ \ \ \ \ \ \ \ \ \ \ \ \
\ \ \ \ \ \ \ \ \ \ \ \ 

\textbf{References}

$\left[ 1\right] $ P. Ramond, R. G. Roberts and G. G. Ross, Nucl. Phys. 
\textbf{B406}, 19 (1993)

$\left[ 2\right] $ L. Wolfenstein, Phys. Rev. Lett. \textbf{51}, 1945
(1983)\ \ \ \ \ 

$\left[ 3\right] $\ M. H. Ahn \textit{et al, }hep-ex / 0212007

$\left[ 4\right] $ K. Eguchi \textit{et al, }hep-ex / 0212021

$\left[ 5\right] $ K. Scolderg \textit{, }hep-ph / 9905016; T. Kajita,
Invited talk at PASCOS 99,

\ \ \ \ \ Y.Fukuda \textit{et al,}\ Phys. Rev. Lett. \textbf{81}, 1562
(1998)\ 

$\left[ 6\right] $ Q. R. Ahmed \textit{et al}, Phys. Rev. Lett. \textbf{87},
071031 (2001);

\ \ \ \ A. Bandyopadpadhyay, S. Choubey, S. Goswami, and D. P. Roy,

\ \ \ \ Phys. Lett.\textbf{\ B540}, 14 (2002); V. Barger, D. Marfatia, K.
Whisnat, and

\ \ \ \ B. P. Wood, ibid. \textbf{B537}, 179 (2002); G. L. Fogli et al Phys.
Rev\textbf{\ D66},\ 

\ \ \ \ 053010 (2002); J. N. Bahcall, M. C. Gonzales-Garcia and

\ \ \ \ C. Pena-Garay, hep-ph / 0204314\textbf{\ }\ \ \ 

$\left[ 7\right] $ M. Maltoni \textit{et al, }hep-ph / 0212129; A.
Bandyopadhyay\ \textit{et al, }

\ \ \ \ \ hep-ph / 0212146 ;J. N. Bahcall \ \textit{et al }hep-ph / 0212147

$\left[ 8\right] $ M. Appollonio\ \textit{et al}\ \ Phys. Lett. \textbf{B466}%
, 415 (1999); F. Boehm \textit{et al }

\ \ \ \ \ Phys. Rev \textbf{D64},\ 112001 (2001)

$\left[ 9\right] $ See, for example, reviews by G. Altarelli and F.Ferugio,
Phys. Rep.

\ \ \ \ \ \textbf{320}, 295(1999) H. Fritzsch and Z. Xing, Pog. Part. Nucl.
Phys.

\ \ \ \ \ \textbf{45},1 (2000), H. Murayama, Int. J. Mod. Phys. \textbf{A17}%
, 3403, (2002)\textbf{\ }

$[10]$ B. Pontecorvo, JETP 34,\textbf{\ 247} (1958); Z. Maki, M. Nakagawa,
and

\ \ \ \ \ \ S. Sakata, Prog. Theo. Phys. \textbf{28}, 870 (1962)\ 

$\left[ 11\right] $\ V. Barger, M. S. Berger\ and P. Ohlmann, Phys. Rev. 
\textbf{D47}, 1093 (1993)

$\left[ 12\right] $ K.S. Babu, C. N. Leung and J. Pantaleone, Phys. Lett. 
\textbf{B319},191(1993)

\ \ \ \ \ \ P. Chankowski and Z. Pluciennick, Phys. Lett. \textbf{B316}, 312
(1993);

\ \ \ \ \ \ \ S. Antusch, M. Drees, J.Kirsten, M. Lindner and M. Ratz, Phys.
Lett.

\ \ \ \ \ \ \ \textbf{B519}, 238 (2001);S. Antusch and M.Ratz, \textbf{%
JHEP0207},059(2002)

$\left[ 13\right] $ M. Gell-Mann, P. Ramond and R. Slansky, in \textit{%
Supergarvity, }Proceedings\ 

\ \ \ \ \ \ \ of the Workshop, Stony Brook, New York 1979, edited by D. Z.
Freeman

\ \ \ \ \ \ \ and P. Van Nieuwenhuizen (North Holland, Amsterdam (1979);

\ \ \ \ \ \ \ T. Yanagida, in Proceedings of the Workshop on the Unified
Theory

\ \ \ \ \ \ \ and Baryon Number in the Universe, edited by O. Sawada and

\ \ \ \ \ \ \ A.Sugamoto (KEK, 1979); R.N. Mahopatra and G. Senjanovic,

\ \ \ \ \ \ \ Phys. Rev. Lett. \textbf{44},912 (1980)

$\left[ 14\right] $ B. R. Desai and D. P. Roy, Phys. Rev. \textbf{D58},
113007-1(1998)

$\left[ 15\right] $ H. Fritzsch, Phys. Lett \textbf{B70}, 436 (1977) H.
Fritzsch, Phys. Lett \textbf{B73},

\ \ \ \ \ \ \ 317(1978); H. Georgiand C. Jarlskog, \textit{ibid}. \textbf{89B%
}, 297 (1979) ;

\ \ \ \ \ \ \ P. Kaus and S. Meshkov, Mod. Phys.Lett. A3, 1251 (1988);

\ \ \ \ \ \ \ E. Ma, Phys. Rev. \textbf{D43}, R2761(1991)\ \ \ \ \ \ \ \ \ \
\ \ \ \ \ \ \ \ \ \ \ \ \ \ \ \ \ \ \ \ \ \ \ \ \ \ \ \ \ \ \ \ \ \ \ \ \ \
\ \ \ \ \ \ \ \ \ \ \ \ \ \ \ \ \ \ \ \ \ \ \ \ \ \ \ \ \ \ \ \ \ \ \ \ \ \
\ \ \ \ \ \ \ \ \ \ \ \ \ \ \ \ \ \ \ \ \ \ \ \ \ \ \ \ \ \ \ \ \ \ \ \ \ \
\ \ \ \ \ \ \ \ \ \ \ \ \ \ \ \ \ \ \ \ \ \ \ \ \ \ \ \ \ \ \ \ \ \ \ \ \ \
\ \ \ \ \ \ \ \ \ \ 

$\left[ 16\right] $ T. K. Kuo, S. W. Mansour and G. H. Wu. Phys. Rev. 
\textbf{D60}, 093004

\ \ \ \ \ \ \ (1999); Phys. Lett. \textbf{B467}, 116 (1999); S. H. Chin, T.
K. Kuo and

\ \ \ \ \ \ \ G. H. Wu, Phys. Rev. \textbf{D62}, 053014 (2000)\ \ \ 

$\left[ 17\right] $ B. R. Desai and A. R. Vaucher, Phys. Rev \textbf{D63},
13001-1 (2001)

$\left[ 18\right] $ P.H. Frampton, S. L. Glashow, and D. Marfatia, Phys.
Lett. \textbf{B536},

\ \ \ \ \ \ \ 79 (2002)\ 

$\left[ 19\right] $\ B.R. Desai, D.P. Roy, and A.R. Vaucher, Mod. Phys.
Lett. \textbf{A18}

\ \ \ \ \ \ \ 1355(2003)

$\left[ 20\right] $ W.A. Bardeen, C.T. Hill, and M. Lindner, Phys. Rev. 
\textbf{D41}, 1647 (1990)\ 

$\left[ 21\right] $ R. N. Mohapatra, M. K. Parida, and G. Rajasekaran hep-ph
/ 0301234

\ \ \ \ \ \ \ and references therein

$\left[ 22\right] $ Particle Data Group K.Hagiwara \textit{et al}\ \ Phys.
Rev. \textbf{D66},\ 010001 (2002)\ 

$\left[ 23\right] $ Y.Yamanaka, H. Sugawara, and S. Pakvasa, Phys. Rev.%
\textbf{\ D25},

\ \ \ \ \ \ \ 1895 (1982)

$\left[ 24\right] $\textbf{\ }E$.$Ma, Phys. Rev.\textbf{\ D43}, 2761 (1991);
N. G. Deshpande, M. Gupta and

\ \ \ \ \ \ \ P. B. Pal, ibid.\textbf{\ D45}, 953 (1992); K. S. Babu, E. Ma,
and J.W.F. Valle,

\ \ \ \ \ \ \ Phys. Lett. \textbf{B552}, 207 (2003)

$\left[ 25\right] $\textbf{\ }T. Brown, S. Pakvasa, H. Sugawara, and Y.
Yamanaka, Phys. Rev.\textbf{\ }

\ \ \ \ \ \ \ \textbf{D30}, 255 (1984)

$\left[ 26\right] $\textbf{\ }E. Ma, Phys. Rev.\textbf{\ D44}, 587 (1991),
E. Ma, ibid \textbf{D61}, 033012-1(2000),

\ \ \ \ \ \ \ E. Ma and G.Rajasekaran, ibid 113012-1 (2001)

$\left[ 27\right] $\textbf{\ }K. S. Babu and R. N. Mohapatra, Phys. Rev.
Lett. \textbf{24}, 2418 (1995)

$\left[ 28\right] $\textbf{\ }P. H. Frampton and A.Rasin, Phys. Lett. 
\textbf{B478}, 424 (2000)

$\left[ 29\right] $\textbf{\ }B.R. Desai and G. Rajasekaran (work in
progress)

\ \ 

\ 

\end{document}